\documentclass[prb,aps,showpacs,superscriptaddress,twocolumn]{revtex4}
\usepackage{amsmath}
\usepackage{amssymb,color}
\usepackage {graphicx}

\newcommand{\beq}{\begin{equation}}
\newcommand{\eeq}{\end{equation}}
\newcommand{\bea}{\begin{eqnarray}}
\newcommand{\eea}{\end{eqnarray}}

\def\k{{\bf k}}

\newcommand{\BFA}{BaFe$_2$As$_2$}

\newcommand{\LFPO}{LaFePO}
\newcommand{\KFA}{KFe$_2$As$_2$}

\begin{document}

\title{Evolution of  symmetry and structure of the gap in Fe-based superconductors with  doping and interactions}

\author{S.~Maiti}
\affiliation{Department of Physics, University of Wisconsin, Madison, Wisconsin 53706, USA}
\author{M.M.~Korshunov}
\affiliation{Department of Physics, University of Florida, Gainesville, Florida 32611, USA}
\affiliation{L.V. Kirensky Institute of Physics, Siberian Branch of Russian Academy of Sciences, 660036 Krasnoyarsk, Russia}
\author{T.A.~Maier}
\affiliation{Computer Science and Mathematics Division and Center for Nanophase Materials Sciences, Oak Ridge National Lab, Oak Ridge, TN 37831, USA}
\author{P.J.~Hirschfeld}
\affiliation{Department of Physics, University of Florida, Gainesville, Florida 32611, USA}
\author{A.V.~Chubukov}
\affiliation{Department of Physics, University of Wisconsin, Madison, Wisconsin 53706, USA}

\date{\today}


\begin{abstract}
We present a detailed study of the symmetry and structure of the
pairing gap in Fe-based superconductors (FeSC). We treat FeSC as
quasi-2D, decompose the pairing interaction in the XY plane in
$s$-wave and $d$-wave channels into contributions from scattering
between different Fermi surfaces and analyze how each scattering
evolves with doping and input parameters. We verify that each
interaction is well approximated by the lowest angular harmonics.
We
use this simplification to analyze the interplay between
the interaction with and without spin-fluctuation components, the
origin of the attraction in the $s^\pm$ and $d_{x^2-y^2}$
channels, the competition between them,
 the angular dependence of the $s^\pm$ gaps along the
electron Fermi surface, the conditions under which $s^\pm$ gap
develops nodes, and the origin of superconductivity in heavily
electron- or hole-doped systems, when only Fermi surfaces of one
type are present. We also discuss the relation between RPA and RG
approaches for FeSC.
\end{abstract}

\maketitle

\section{Introduction \label{sec:intro}}

The symmetry and the structure of the superconducting gap in
Fe-based superconductors (FeSC), and their evolution and
possible change with doping are currently subjects of intensive
debates in the condensed matter community.

The vast majority of researchers believe that superconductivity in
FeSC is of electronic origin and results from the screened Coulomb
interaction, enhanced at particular momenta due to strong magnetic
fluctuations~\cite{tom_09,kuroki,arita,graser,peter,
peter_2,Rthomale,Thomale,dhl_10,zlatko,Andrey,saurabh,bang}
 or orbital fluctuations~\cite{kontani,kontani_last,shin,orbital}. For systems with a
single Fermi surface sheet, such interaction cannot lead to a
simple $s$-wave superconductivity, but it can give rise to a
superconductivity with non-$s$-wave symmetry -- $p$-wave for
strong ferromagnetic spin fluctuations and $d$-wave for strong
antiferromagnetic spin fluctuations. In FeSC, however, the
electronic structure is more complex. The low-energy states are
formed by the hybridization of all five Fe-$d$ orbitals
 what in the band basis not only gives rise to
 multiple sheets of the Fermi surface (FS) [weakly doped FeSC contain  two electron
FSs and either two or three hole FSs],
but also leads to a complex mixing of contributions from intra- and inter-orbital terms in the interactions
  between low-energy
fermions. In this situation, in addition to the potentially
non-$s$-wave superconductivity,
 FeSC may also
develop superconductivity with an $s$-wave symmetry of the gap
even for the repulsive electron-electron interaction. For
parameters used
 for orbital interactions in most studies of FeSC,
the $s$-wave gap, averaged over the FSs, changes sign between
different FS sheets (it is commonly called the $s^\pm$ gap), and
the superconducting state competes with the spin-density-wave
(SDW) state. But for other parameters, a conventional
superconducting state, with a sign preserving $s^{++}$ gap,
becomes possible, and such a state will compete with the
charge-density-wave (CDW) state \cite{Andrey}.

The existing theoretical approaches to
pairing in FeSC can be broadly divided into two categories. One
assumes that fermions at energies smaller than a fraction of the
bandwidth can be treated as itinerant, with a moderate
self-energy~\cite{reviews},
although strong coupling effects, such as interaction-driven
renormalization of the whole bandwidth, have to be
incorporated.~\cite{singh-du,haule,lara} In the itinerant
approach, the pairing is often treated in a
BCS/Eliashberg formalism, with the interaction taken as a
combination of direct electron-electron interaction and effective
interaction mediated by collective bosonic excitations. Another
approach assumes that the system is not far from the Mott regime,
and that the pairing should be affected by the tendency towards a
Mott insulator \cite{Mott,parish,leni}.

This work falls into the first category. Already within this category,
there are several, seemingly different approaches to the pairing: the
RPA-type spin-fluctuation approach
(RPA/SF) \cite{tom_09,graser,peter,peter_2,kuroki08,graser_11,kuroki,suzuki,tom_last},
the functional RG approach (fRG) \cite{Rthomale,Thomale,fRG_KBaFeAs,dhl_10,dhl_AFESE}, and the analytic (logarithmic) RG approach based on a ``minimal'' model for the
pnictides \cite{zlatko,Andrey,saurabh}, in which the interaction in each
pairing channel is restricted to the leading angular harmonics in each
pairing channel (the leading angular harmonics approximation, or LAHA).
The very positive fact for the itinerant approach as a whole is that, so
far, the results of all these different approaches agree on the pairing
symmetry and the gap structure in hole-doped and electron-doped FeSC.
Namely, all three approaches predict that the leading pairing instability
at small/moderate dopings is in the $s$-wave channel, and the gap, averaged
over the individual FS sheets, changes sign between hole and electron
sheets (an $s^\pm$ gap). The $s^{\pm}$ gap generally varies along each of
the FSs. If the FeSC are treated as 2D systems (i.e., if the variation of
the interaction along $k_z$ is neglected), the variation of the gap is
stronger on the electron FSs, and can be large enough to create nodes. All
three approaches also predict that the gap with the nodes is more likely
in either undoped or electron-doped FeSC, while in hole-doped FeSC a
no-nodal state is more likely.

The RPA and fRG formalisms have been also applied
\cite{graser_11,das,dhl_AFESE} to study superconductivity in recently
discovered heavily electron-doped KFe$_2$Se$_2$, where only
electron FSs remain, according to recent ARPES studies
\cite{exp:AFESE,exp:AFESE_ARPES}. The results of RPA/SF and fRG
approaches for this limiting case are again in agreement -- both
predict that the gap should now have $d_{x^2-y^2}$ symmetry, i.e.,
it should change sign between the two electron FSs.
Other approaches, however, found a conventional s-wave
superconductivity in this limit, by one
reason~\cite{si_11,bernevig_11} or the other~\cite{kontani_last}.
A more complex order parameter has also been
suggested.~\cite{mazin}

The $d-$wave gap symmetry was also predicted by fRG
\cite{fRG_KBaFeAs} for heavily hole doped KFe$_2$As$_2$, where
only hole FSs are present. The RPA/SF analysis for this
material~\cite{suzuki} found attraction of comparable strength in
both s-wave and d-wave channels.

The goal of this paper is to understand in more detail  the
evolution of the $s^\pm$ gap with  hole and electron doping and the
interplay between the $s^\pm$ and $d_{x^2-y^2}$ pairing in FeSC.  The
idea is to ``decompose'' the full pairing interaction  into
contributions from  scattering processes between different FSs and
check how each process evolves with doping and input parameters.
Such an analysis has been performed within the RPA/SF formalism in
Ref.~\onlinecite{tom_09} but only for an $s$-wave interaction and
only for a fixed doping and a particular set of input parameters.

In this work, we combine the 5-band RPA/SF and LAHA approaches.
Specifically, we take the full set of interactions $\Gamma_{i,j}
(\k, \k')$ from the 5-band RPA/SF calculation as input ($i$, $j$
label different FSs, $\k$ and $\k'$ are momenta along these FSs)
and show that, for all cases that we studied, different
interactions $\Gamma_{i,j} (\k, {\bf k'})$ are well approximated
by only two angular harmonics in $s$-wave and $d_{x^2-y^2}$
channels (we choose one of the momenta such that only these two
pairing channels contribute). This leaves us with a finite number
of interaction parameters in each of the two pairing channels (the
number of independent parameters is 4 or 5, equal to the total
number of FSs). We then solve the pairing problem, obtain the
eigenfunctions (which determine the gap structure) and eigenvalues
(which are the dimensionless couplings) and analyze how both
eigenfunctions and eigenvalues evolve with the parameters. In this
work, we neglect potential new physics associated with 3D effects
and treat FeSC as quasi-2D systems, i.e., neglect the $k_z$
dependence of the quasiparticle dispersion and of the
interactions.

The key goal of our work is to understand whether superconductivity in
FeSCs  is governed by a single underlying pairing mechanism for all hole and electron dopings,
 despite that the pairing symmetry and the gap structure may change, and whether the entire variety of
  pairing states can be adequately described within the effective low-energy model with small numbers of input parameters.

The specific set of issues that we address are:

\begin{itemize}

\item
{\em What is the origin of the strong angular dependence of the
$s^\pm$ gap along the electron FSs?} The angular dependence of the
effective interaction $\Gamma_{i,j} (\k, \k')$ due to
the change in orbital character of the states on the FS, the
competition of the scattering between hole-electron and
electron-electron-sheets and the local Coulomb repulsion  are
candidates that can give rise to a strong anisotropy of the gap.

\item
{\em Are the angular dependencies of all interactions relevant for
the gap structure, or can some interactions be safely approximated
as angle-independent?} In principle, the angular dependencies of
both the electron-hole and electron-electron interactions can
affect the structure of $s^{\pm}$ gap. In the LAHA approach, we
can vary the angular dependence of each interaction by hand and
explore how it affects the gap structure.

\item
{\em Why do the $s^{\pm}$ solutions obtained within the RPA/SF and
fRG approaches have nodes for systems with two hole and two
electron FSs and no nodes for systems with three hole and two
electron FSs? Is this behavior generic, or just a trend meaning
that in both cases the $s^{\pm}$ gap is either nodal or non-nodal,
depending on the input parameters?} This issue is difficult to
address in both RPA/SF and fRG approaches as these are numerical
methods which require certain input parameters and exploring
parameter space is computationally expensive. But it can be
addressed within LAHA as one can continuously change any of the
input parameters.

\item
{\em What causes the pairing when only electron FSs are present?}
Possibilities include pairing driven by the angular dependences of
the interactions; $s-$wave pairing caused by virtual scattering to
gapped hole states -"incipient" s-wave ($s^\pm$ gap but
without hole FSs), a $d-$wave pairing (a plus-minus gap on
electron FSs) due to magnetically enhanced repulsive interaction
at momentum transfer $(\pi,\pi)$ between the electron FSs, or an
$s-$wave pairing if the interaction at $(\pi,\pi)$ is strong and
attractive.

\item {\em What causes the pairing at large hole doping, when only hole
pockets are present?} It is possible that the scattering between
hole FSs with wave vector $(\pi,\pi)$ favors  $s^\pm$ pairing
while the alternative is that Coulomb avoidance within each FS
pocket makes nodal $d$-wave the preferable symmetry.

\item
{\em How is the structure of the pairing interaction affected when
the spin-fluctuation component is added to the direct
fermion-fermion interaction?} Inclusion of the SF contribution can
affect the relative magnitudes of the interactions $\Gamma_{ij}
(\k, \k')$ between different FS sheets, and in principle
can change its angular dependence.

\end{itemize}

This last issue is relevant for understanding the comparison
between RPA/SF and RG approaches, which we pause to discuss in
some detail. These two approaches differ in the assumption of what
are the relevant energy scales for magnetism and
superconductivity. In the RG approach it is assumed that magnetism
and superconductivity are produced by the same low-energy fermions
and have to be treated on equal footing, starting from a model
with only density-density and exchange interaction between
fermions. Then the SF component of the interaction develops
together with the pairing vertex (one builds up step-by-step
``parquet'' renormalizations  simultaneously in the particle-hole
and particle-particle channels) \cite{Rthomale,Andrey,saurabh}. In
the RPA/SF approach \cite{graser,peter,peter_2,kuroki08} the
assumption is that magnetism is superior to superconductivity and
comes from fermions at energies comparable to the bandwidth, which
are above the upper edge for the RG treatment. If so, the SF
component should be included into the bare interaction, which does
not need to be further renormalized in the particle-hole channel.
In this situation, only renormalizations in the particle-particle
channel (which always come from low-energy fermions) remain
relevant, and the RG treatment becomes equivalent to the BCS
theory. There is no good justification to select a particular set
of diagrams for the renormalization of the interaction at high
energies, but in many cases the RPA approximation (which amounts
to a summation of ladder series of vertex renormalization diagrams
in the spin channel and generally accounts for a Stoner-type
instability at some $Q$) yields quite reasonable effective
interactions, particularly near a magnetic instability.

Which of the two approaches better describes FeSC is debatable.
The value of the ordered magnetic moment is quite small, at least
in the FESCs, and magnetic excitations measured by neutron
scattering die off at energies $100-150$meV. This behavior is
consistent with the idea that magnetism comes from relatively
low-energy fermions. The one-loop parquet RG equations for the
coupled flow of the vertices are rigorously justified only when
the fermionic dispersion can be approximated by a $k^2$
dependence. This also does not extend too far in energy, i.e., the
upper limit for the RG approach is also only a fraction of the
bandwidth.

But are these two approaches fundamentally different? To compare
them, we remind the reader that in the RG treatment, the flow of
magnetic and superconducting vertices remain coupled only down to
energies of order $E_F$. At energies
$E < E_F$, each vertex flows independently, and the flow of the
pairing vertex ($\Gamma_{sc}$) becomes the same as in the BCS
theory ($d \Gamma_{sc}/d \log E = \Gamma^2_{sc}$.  From this
perspective, the real result of RG as far as pairing is concerned,
is the renormalization of the pairing interaction from its bare
value to the renormalized one at $E_F$. [This is indeed only true
if the RG flow does not reach a fixed point down to $E_F$, but
this is likely to be the case for FeSC simply because $T_c$ and
$T_N$ are both small compared to $E_F \sim 0.1$eV.] It is quite
likely (although not guaranteed) that the angle-independent
components of the renormalized interactions can be reproduced by
choosing some other input parameters in the orbital basis, i.e.,
the effect of the RG flow could be absorbed into the modification
of the bare theory. The same is true for the RPA/SF approach --
the angle-independent components  of the new SF interactions
likely can be reproduced the renormalization of input parameters,
although the needed values of $U, V, J, J'$ may seem quite exotic.

The situation is a bit
more tricky for the angle dependent components of the interactions.
The RG with coupled magnetic and superconducting
vertices can be rigorously justified at weak coupling only if
angular dependence of the vertices are (i) weak and (ii) are
preserved under RG.  This is how the analytic RG flow has been
obtained~\cite{saurabh}. fRG does include some variation of the
angle dependence of interactions, but this goes beyond justifiable
logarithmic accuracy. Given that fRG and analytic
RG yield virtually identical results, it seems as if the only
effect of RG renormalizations down to $E_F$ is the rescaling of
the overall magnitudes of the interactions. If the effect of
adding the SF component to the interaction in RPA/SF approach also
predominantly gives rise to rescaling of the magnitudes of the
interactions, then the two approaches are not fundamentally
different.

This does not imply that the RPA/SF and RG approaches are
equivalent. Rather, the implication is that the outcome of
applying each of the two formalisms is the new ``bare'' theory
with new input parameters. These new parameters do differ somewhat
between RPA/SF and RG, but the difference should not matter much
if the pairing symmetry and the gap structure are quite robust
with respect to parameter variations. This is another issue that
one can straightforwardly verify using the LAHA formalism which
allows one to continuously change the parameters. At this stage
the reader might wish to peek into Sec. VI (Conclusions), where
the above mentioned questions are answered within the scope of
this work.

This paper is organized as follows. In the
Sec.~\ref{sec:RPAandLAHA} we briefly discuss the RPA/SF and LAHA
formalisms and outline the computational procedure. In
Sec.~\ref{sec:FormOfGamma} we discuss how the gap symmetry and
structure are affected by the SF component of the interaction. In
Sec.~\ref{sec:Doping} we show the results for weakly
electron-doped and hole-doped FeSC, and in
Sec.~\ref{sec:Overdoping} we discuss strongly electron- and
hole-doped FeSC, which contain only electron or hole pockets,
respectively. We present our conclusions in
Sec.~\ref{sec:Conclusions}.
A short summary of this work is presented in
Ref~\onlinecite{maiti_short}

\section{The RPA/SF and LAHA formalisms and the computational procedure \label{sec:RPAandLAHA}}

We first briefly describe the RPA/SF and LAHA approaches and then
outline the computational procedure.

\subsection{RPA/SF formalism}

The approach and its application to FeSC have been discussed in
detail in several recent publications \cite{graser,peter,peter_2},
so we will be brief. The point of departure for RPA/SF is a
5-orbital model with intra-orbital and inter-orbital hopping
integrals and density-density (Hubbard) and exchange intra-orbital
and inter-orbital interactions, given by
\begin{eqnarray}
H_{int} & = & U \sum_{f,s} n_{f,s\uparrow} n_{f,s\downarrow} +
\sum_{f,s,t \neq s} \left( \frac{V}{2} n_{fs} n_{ft} - \frac{J}{2} \vec{S}_{fs}\cdot \vec{S}_{ft} \right) \nonumber \\
& + & \frac{J'}{2}  \sum_{f,s,t \neq s}  \sum_\sigma
c_{fs\sigma}^\dagger c_{fs\bar{\sigma}}^\dagger c_{ft\bar{\sigma}}
c_{ft\sigma}\label{eq:multiorb_Hubbard}
\end{eqnarray}
where $c_{fs\sigma}$ is the annihilation operator for electron on
lattice site $f$ with orbital index $s$ and spin $\sigma$, $n_{fs}
=  n_{f,s\uparrow} + n_{f,s\downarrow}$, $n_{fs\sigma} =
c_{fs\sigma}^\dagger c_{fs\sigma}$ is the density operator,
$\vec{S}_{fs} = (1/2) c^\dagger_{fs\alpha} {\vec
\sigma}_{\alpha\beta} c_{fs\beta}$ is the spin operator, and
$\bar{\sigma} = -\sigma$. We have designated different symbols for
the intra-orbital
 Coulomb interaction $U$,
interorbital Coulomb interaction $V$, interorbital exchange $J$
and ``pair hopping'' term $J'$ for generality, but note that if
they are generated from a single spin-rotational invariant
two-body term they are related by $J'=J/2$ and $V=U-5J/4$.
The model parameters for hopping integrals (36 total) are obtained
from the fit to density functional theory (DFT) band structure
\cite{ref:Cao}.

The SF component of the interaction is now obtained by summing up
second and higher-order ladder diagrams in matrix orbital
formalism. The total interaction (the sum of the direct, first
order term and SF contribution) is then converted from orbital to
band basis by dressing it by matrix elements associated with the
hybridization of five Fe orbitals. The end result of this procedure
for the purposes of the analysis of superconductivity is the
effective BCS-type Hamiltonian in the band description
\beq
{\cal H} = \sum_{i,\k} \epsilon_{i} (\k) c^\dagger_{i \k} c_{i \k} +
\sum_{i,j, \k, \k'} \Gamma_{ij} (\k,\k') c^\dagger_{i \k} c^\dagger_{i -\k} c_{j \k'}c_{j -\k'}
\label{1}
\eeq
The quadratic term describes low-energy excitations near hole and
electron FSs, labeled by $i$ and $j$, and the interaction term describes
the scattering of a pair $(k\uparrow, -k\downarrow)$ on the FS $i$ to
a pair $(-k'\uparrow, k'\downarrow)$ on the FS $j$. The
effective singlet interaction $\Gamma_{ij}(\k,\k')$ is then given by
\begin{eqnarray}
{\Gamma}_{ij} (\k,\k') & = & \sum_{s,t,p,q} a_{\nu_i}^{t*}(-\k)
a_{\nu_i}^{s*}(\k)
\mathrm{Re}\left[{\Gamma}_{st}^{pq} (\k,\k',0) \right] \nonumber \\
& \times & a_{\nu_j}^{p}(\k')  a_{\nu_j}^{q}(-\k'),
\end{eqnarray}
with
\begin{eqnarray}
{\Gamma}_{st}^{pq} (\k,\k',\omega) &=& \left[\frac{1}{2}
U^s+\frac{1}{2} U^c +\frac{3}{2} U^s
\chi_1^{RPA}  (\k-\k',\omega) U^s \right. \nonumber\\
 &-&  \left. \frac{1}{2} U^c  \chi_0^{RPA}  (\k-\k',\omega)
U^c  \right]_{ps}^{tq}.
\label{eq:fullGamma}
\end{eqnarray}
Here the $\chi_1^{RPA}$ and  $\chi_0^{RPA}$  describe the spin-fluctuation
contribution and orbital
(charge)-fluctuation contribution, respectively, $a_{\nu_j}^p$ is the matrix
element connecting orbital $p$ with the band $\nu_j$ on FS $j$,
and the matrices $U^c$ and $U^s$ contain interaction parameters
from Eq.~(\ref{eq:multiorb_Hubbard}) as described in
Ref.~\onlinecite{graser}.  We will call an approximation to the
total interaction which includes only the two first-order terms in
Eq.~(\ref{eq:fullGamma}) as constant or non-spin-fluctuation
(NSF), while the total interaction includes the third and fourth
spin-fluctuation terms (SF) as well. In principle this interaction
includes also charge/orbital fluctuations via $U^c$, but as these
are negligible for realistic parameters, we use the simpler ``spin
fluctuation'' designation.

Throughout this paper, we consider the pairing in the unfolded
Brillouin zone, leaving aside the issue of possible changes of the
gap due to folding. These are particularly important in the case
of the 122 systems with $I4/mmm$ symmetry \cite{mazin,kontani}. In
the unfolded zone, the 2D electronic structure of weakly and
moderately electron-doped FeSC consists of two near-circular FSs
centered at $\Gamma$ point ($\k=0$) and two elliptical
electron FSs centered at $X$ and $Y$ points, $\k = (0,\pi)$
and  $\k = (\pi,0)$, respectively. For hole-doped and some
undoped FeSC, there exists also another, third,  hole FS located
at $M$ point  $\k = (\pi,\pi)$.  In the folded zone, all
three hole FSs are centered at $(0,0)$, and the two electron FSs
move to $(\pi,\pi)$ and are hybridized through the coupling via a
pnictogen.

The interaction $\Gamma_{ij} (\k,\k')$ contains all pairing
components for tetragonal ($D_{4h}$) lattice symmetry: $A_{1g}$
($s$-wave), $B_{1g}$ ($d_{x^2-y^2}$), $B_{2g}$ ($d_{xy}$) and
$A_{2g}$ ($g$-wave).  The $s$-wave gap $\Delta_s (k_x,k_y)$ is
symmetric under $k_x \to \pm k_x,k_y$, the $d_{x^2-y^2}$ gap
$\Delta_{d_{x^2-y^2}} (k_x,k_y)$ changes sign under $k_x \to k_y$,
and so on. We focus here on $s$- and $d_{x^2-y^2}$-wave
symmetries.


%

With these considerations in mind, the BCS gap equation then
becomes the eigenvalue problem
\beq
-\sum_j \oint_{C_j} \frac{dk'_{\parallel}}{2\pi} \frac{1}{2\pi v_F (\k'_F)} \Gamma_{ij} (\k_F,\k'_F) \Delta_{\alpha,j} (\k'_F) = \lambda_\alpha \Delta_{\alpha,i} (\k_F),
\label{1a}
\eeq
where $\alpha$ is either $s$ or $d_{x^2-y^2}$.  For a circular FS,
Eq.~\ref{1a} is simplified to
\beq
-\sum_j \int_0^{2\pi} \frac{d\psi'}{2\pi} N_{F,j}\Gamma_{ij} (\psi, \psi') \Delta_{\alpha,j} (\psi') = \lambda_\alpha \Delta_{\alpha,i} (\psi)
\label{1aa}
\eeq
where $N_{F,j} = m_j/2\pi$ is the density of states at the  FS
$j$, and $\psi$ and $\phi'$ are the angles along the FSs $i$ and
$j$, respectively.

Eqs.~\ref{1a} and~\ref{1aa} are integral equations which in
general can be solved only numerically. Taking $M$ points on each
FS, one obtains $M$ eigenfunctions and $M$ different $\lambda$'s
in each of the two pairing channels. For $s$-wave, some of
eigenfunctions correspond to an $s^{++}$ gap, while others
correspond to an $s^\pm$ gap. The eigenfunction corresponding to
the largest positive $\lambda_\alpha$ describes the pairing state
immediately below $T_c$.

\subsection{LAHA formalism}

The generic idea of LAHA approximation is to model $\Gamma_{ij} (\k, \k')$ by a rather simple function of the two
momenta, such that the gap equation can be solved {\it and
analyzed} analytically.  In cuprates, numerous groups approximated
$d_{x^2-y^2}$ gap by the first harmonic $\cos{k_x} - \cos{k_y}$ ($\cos 2 \phi$ for a circular FS) and neglected higher harmonics
like $\cos{6 \phi}$, $\cos 10 \phi$ etc. The smallness of $\cos
(4n+2) \phi$ terms with $n \geq 1$ does not follow from any
underlying principle, but numerically $\cos 2 \phi$ approximation
works rather well, at least at and above optimal doping.

Such an approximation should generally work even better for FeSC
because all FSs are small and even electron ones are almost
circular.  By analogy with the cuprates, one may try to
approximate $s$-wave eigenfunction by a constant along each FS,
and approximate $d_{x^2-y^2}$ gaps by $\cos 2 \phi$.  There
is a caveat, however --- such approximation is only valid for the
gaps along hole FSs which are centered at the points along $k_x =
\pm k_y$ [i.e., at  $k=(0,0)$ and $(\pm \pi,\pm \pi)$]. Electron
FSs are centered at $X$ and $Y$ points, which by itself are not
$k_x \to \pm k_y$  symmetric. As a result,  some of $s$-wave gap
functions, like $\cos k_x + \cos k_y$ behave as $\pm \cos 2\phi$
along electron FSs, while some of $d$-wave gap functions like
$\cos k_x - \cos k_y$ are approximated by constants on the two
electron FSs.  In the latter case, the only ``memory'' about
$d$-wave is that the sign of a constant changes between the two
electron FSs.

The implication of this result is that, within LAHA,
angle-independent and $\cos 2 \phi$ terms must appear together in
both $s$-wave and $d$-wave components of the interactions, and
with comparable  magnitudes.  A simple analysis then shows
that the form of the interaction depends on whether it involves
hole or electron FSs.  For the interaction between fermions on a
hole FS, in LAHA
\beq
\Gamma_{hh} (\phi, \phi') = A_{hh} + {\tilde A}_{hh} \cos 2\phi \cos 2\phi'\label{2}
\eeq
where $\phi$ and $\phi'$ are the angles along a hole FS (measured
relative to the $k_x$ axis), and $A$ and ${\tilde A}$ terms are
$s$-wave and $d$-wave components, respectively. For the
interaction between fermions from a hole and an electron FSs,
\beq
\Gamma_{eh} (\phi, \theta) = A_{eh} \left(1 + 2 \alpha  \cos
2 \theta \right) + {\tilde A}_{eh} \cos 2\phi \left(1 + 2 {\tilde \alpha} \cos 2\phi' \right)
\label{3}
\eeq
where $\theta$ is the angle along an electron FS (again, measured
relative to the $k_x$ axis). Finally, for the interaction between
fermions from an electron FS, we have in LAHA
\bea
&&\Gamma_{ee} (\theta, \theta') = A_{ee} \left[1 + 2 \alpha
\left(\cos 2 \theta + \cos 2 \theta'\right) + 4 \beta
\cos 2 \theta \cos 2 \theta'\right] \nonumber \\
&& + {\tilde A}_{ee}  \left[1 + 2 {\tilde \alpha}
\left(\cos 2 \theta + \cos 2 \theta'\right) + 4 {\tilde \beta}
\cos 2 \theta \cos 2 \theta'\right]
\label{4}
\eea

The $s$-wave and $d$-wave components look identical, but they
transform differently between intra and inter-pocket interactions
involving the electron FSs.

Below we present the full LAHA result for $\Gamma_{ij} (\k_F,
\k'_F)$ for the case when the FS consists of two hole and two
electron pockets. The extension of the case of three hole FSs is
straightforward. We have
\begin{widetext}
\bea
\label{eq:interactions}
\Gamma_{h_1h_1}(\phi,\phi')&=&U_{h_1h_1}+\tilde{U}_{h_1h_1} \cos2\phi
\cos2\phi'\\\nonumber
\Gamma_{h_2h_2}(\phi,\phi')&=&U_{h_2h_2}+\tilde{U}_{h_2h_2} \cos2\phi
 \cos2\phi'\\\nonumber
\Gamma_{h_1h_2}(\phi,\phi')&=&U_{h_1h_2}+\tilde{U}_{h_1h_2}~\cos2\phi
 \cos2\phi'\\\nonumber
 \Gamma_{h_1e_1}(\phi,\theta)&=&U_{h_1e}(1+2\alpha_{h_1e}
\cos2\theta)+\tilde{U}_{h_1e} (1+2{\tilde \alpha}_{h_1e}
 \cos2\theta) \cos2\phi\\\nonumber
\Gamma_{h_1e_2}(\phi,\theta)&=&U_{h_1e}(1-2\alpha_{h_1e}
\cos2\theta)+\tilde{U}_{h_1e} (-1+2{\tilde \alpha}_{h_1e}
 \cos2\theta) \cos2\phi\\\nonumber
 \Gamma_{h_2e_1}(\phi,\theta)&=&U_{h_2e}(1+2\alpha_{h_2e}
\cos2\theta)+\tilde{U}_{h_2e} (1+2{\tilde \alpha}_{h_2e}
 \cos2\theta) \cos2\phi\\\nonumber
\Gamma_{h_2e_2}(\phi,\theta)&=&U_{h_2e}(1-2\alpha_{h_2e}
\cos2\theta)+\tilde{U}_{h_2e} (-1+2{\tilde \alpha}_{h_2e}
 \cos2\theta) \cos2\phi\\\nonumber
 \Gamma_{e_1e_1}(\theta,\theta')&=& U_{ee} \left[1 + 2 \alpha_{ee}
(\cos2\theta+ \cos2\theta') + 4 \beta_{ee} \cos2\theta \cos2\theta'\right] +
{\tilde U}_{ee} \left[1 + 2 {\tilde \alpha}_{ee}
(\cos2\theta+ \cos2\theta') + 4 {\tilde \beta}_{ee} \cos2\theta \cos2\theta'\right] \\\nonumber
\Gamma_{e_2e_2}(\theta,\theta')&=& U_{ee} \left[1 - 2 \alpha_{ee}
(\cos2\theta+ \cos2\theta') + 4 \beta_{ee} \cos2\theta \cos2\theta'\right] +
{\tilde U}_{ee} \left[1 - 2 {\tilde \alpha}_{ee}
(\cos2\theta+ \cos2\theta') + 4 {\tilde \beta}_{ee} \cos2\theta \cos2\theta'\right] \\\nonumber
\Gamma_{e_1e_2}(\theta,\theta')&=& U_{ee} \left[1 + 2 \alpha_{ee}
(\cos2\theta - \cos2\theta') - 4 \beta_{ee} \cos2\theta \cos2\theta'\right] +
{\tilde U}_{ee} \left[-1 - 2 {\tilde \alpha}_{ee}
(\cos2\theta - \cos2\theta') + 4 {\tilde \beta}_{ee} \cos2\theta \cos2\theta'\right] \\\nonumber
\eea
\end{widetext}

Intra-pocket $s$-wave components  $U_{h_i h_i}$ and  $U_{ee}$
represent the total strength of intra- and inter-orbital Coulomb
(Hubbard) interaction  and are positive (repulsive). The signs of
$U_{h_i e}$ are determined by the relative magnitudes of
intra-orbital $U$ and inter-orbital $V$ terms. For a toy model of two orbitals which hybridize to give
one hole and one electron band,  $U_{h e} > 0$  if intra-orbital
Hubbard repulsion $U$ exceeds inter-orbital
$V$, and $U_{h e} <0$
for $V > U$~\cite{Andrey}.
 It is generally expected that the intra-orbital
interaction is the largest, and below we assume that $U_{h e}$ is
positive. If, however,
$V > U$, the interaction between electrons and holes is attractive
in which case it favors a conventional $s$-wave superconductivity~\cite{kontani}.
The signs of $d$-wave components of the interactions are
determined by a more subtle balance between different $U$s and
$J$s.

The equation for $\Delta_{\alpha_i}$ and $\lambda_\alpha$ is the
same as Eq.~\ref{1aa}. However, now we don't have to solve
integral equation on the gap because for $\Gamma_{i j}$ given by
Eq.~\ref{eq:interactions}, $s$-wave and $d$-wave gaps have only four
components each.  For $s$-wave gap we have
\bea
&&\Delta_{h_1} (\phi) =\Delta_{h_1},~ \Delta_{e_1} (\theta) =\Delta_{e} + {\bar \Delta}_e \cos 2\theta \\
&&\Delta_{h_2} (\phi) =\Delta_{h_2}, ~ \Delta_{e_2} (\theta) =\Delta_{e} - {\bar \Delta}_e \cos 2\theta \nonumber
\label{5}
\eea
and for $d$-wave gap we have
\bea
&&\Delta_{h_1} (\phi) = \Delta_{h_1} \cos 2 \phi,~ \Delta_{e_1} (\theta) =\Delta_{e} + {\bar \Delta}_e \cos 2\theta \\
&&\Delta_{h_2} (\phi) =  \Delta_{h_2} \cos 2 \phi, ~ \Delta_{e_2} (\theta) = -\Delta_{e} + {\bar \Delta}_e \cos 2\theta \nonumber
\label{6}
\eea

To obtain $\lambda_{\alpha}$ and eigenfunctions  we  need to write
down and diagonalize $4\times 4$ matrix gap equations. For this
purpose, it is convenient to introduce bare pairing vertices
$\Delta^{(0)}_i$ and solve the set of conventional ladder
equations for the full $\Delta^{(0)}_i$. For the $s$-wave solution we
obtain
\begin{widetext}
\bea
\label{eq:SC 2H_2E}
\left(
\begin {array}{cccc}
 1+u_{h_1h_1} L & u_{h_1h_2} L & 2u_{h_1e} L &  2\alpha_{h_1e} u_{h_1e} L\\
 u_{h_1h_2} L & 1+u_{h_2h_2} L & 2u_{h_2e} L &  2\alpha_{h_2e} u_{h_2e} L\\
 u_{h_1e} L & u_{h_2e} L& 1+2u_{ee} L  &  2\alpha_{ee} u_{ee}  L\\
 2 \alpha_{h_1e} u_{h_1e} L & 2\alpha_{h_2e} u_{h_2e} L& 4 \alpha_{ee} u_{ee} L
 & 1+ 4 \beta_{ee} u_{ee}  L\\
\end{array}
\right) \left(
\begin {array}{c}
  \Delta_{h_1}\\
  \Delta_{h_2}\\
  \Delta_{e}\\
  \bar{\Delta}_{e}\\
\end{array}
\right) =
\left(
\begin {array}{c}
  \Delta^{0}_{h_1}\\
  \Delta^{0}_{h_2}\\
  \Delta^{0}_{e}\\
  \bar{\Delta}^{0}_{e}\\
\end{array}
\right)
\eea
where $u_{ij} = U_{i j} (N_{F,i} N_{F,j})^{1/2}$ are dimensionless
interactions, $L \sim ln\left( \frac{E_F}{T_c} \right)$ ($T_c$ is
measured in units of energy). The eigenvalue $\lambda_s$ and the
corresponding eigenfunctions are obtained by diagonalizing this
matrix equation and casting the result as
\beq
\label{eq:lambda_def}
(1-[\underline{\underline{\lambda_{s}}}]
L)(\Delta_{h_1},~\Delta_{h_2},~ \Delta_{e},~ \bar{\Delta}_{e})^t=
(\Delta^{(0)}_{h_1},~\Delta^{(0)}_{h_2},~ \Delta^{(0)}_{e},~
\bar{\Delta}^{(0)}_{e})^t\eeq For $d$-wave gap function we obtain
\bea \label{eq:SC 2H_2Ed} \left(
\begin {array}{cccc}
 1+\frac{\tilde{u}_{h_1h_1}}{2} L & \frac{\tilde{u}_{h_1h_2}}{2} L & 2{\tilde u}_{h_1e} L &  2{\tilde \alpha}_{h_1e} {\tilde u}_{h_1e} L\\
 \frac{\tilde{u}_{h_1h_2}}{2} L & 1+\frac{\tilde{u}_{h_2h_2}}{2} L & 2\tilde{u}_{h_2e} L &  2{\tilde \alpha}_{h_2e} \tilde{u}_{h_2e} L\\
 \frac{{\tilde u}_{h_1e}}{2} L & \frac{\tilde{u}_{h_2e}}{2} L& 1+2{\tilde u}_{ee} L  &  2 {\tilde \alpha}_{ee} \tilde{u}_{ee}  L\\
 {\tilde \alpha}_{h_1e} {\tilde u}_{h_1e} L & {\tilde \alpha}_{h_2e} \tilde{u}_{h_2e} L& 4 {\tilde \alpha}_{ee} \tilde{u}_{ee}  L  & 1+ 4 {\tilde \beta}_{ee}\tilde{u}_{ee}  L\\
\end{array}
\right) \left(
\begin {array}{c}
  \Delta_{h_1}\\
  \Delta_{h_2}\\
  \Delta_{e}\\
  \bar{\Delta}_{e}\\
\end{array}
\right) =
\left(
\begin {array}{c}
  \Delta^{0}_{h_1}\\
  \Delta^{0}_{h_2}\\
  \Delta^{0}_{e}\\
  \bar{\Delta}^{0}_{e}\\
\end{array}
\right)
\eea
\end{widetext}
Diagonalizing this set, we obtain $\lambda_d$ and the
corresponding eigenfunctions.

We emphasize that the equations on $\lambda_s$ and $\lambda_d$ are
4th order algebraic equations, hence there are effectively only
four parameters which determine the couplings and the gap
structure in each channel. This is the same number as the minimal
number of interaction parameters in the original orbital model.

If the two hole FSs can be treated as equal (i.e., $u_{h_ih_j} =
u_{hh}$, $u_{h_1e} = u_{h_2e} = u_{he}$ and $\alpha_{h_1e} =
\alpha_{h_2e} = \alpha$, which actually is quite consistent with
the fits to RPA/SF, see below), then $\Delta_{h_1} = \Delta_{h_2}
=\Delta_h$ and the gap equation reduces to $3 \times 3$ set which
can be very easily analyzed analytically. The equation for the
$s$-wave gap in this case is
\begin{widetext}
\bea
\label{eq:SC 2H_2E_3}
\left(
\begin {array}{ccc}
 1+ 2u_{hh} L & 2u_{he} L &  2 \alpha_{he} u_{he} L\\
 2u_{he} L & 1+2u_{ee} L  &  2  \alpha_{ee}u_{ee} L\\
 4 \alpha_{he} u_{he} L & 4  \alpha_{ee} u_{ee} L  & 1+ 4 \beta_{ee} u_{ee} L\\
\end{array}
\right) \left(
\begin {array}{c}
  \Delta_{h}\\
  \Delta_{e}\\
  \bar{\Delta}_{e}\\
\end{array}
\right) =
\left(
\begin {array}{c}
  \Delta^{0}_{h}\\
  \Delta^{0}_{e}\\
  \bar{\Delta}^{0}_{e}\\
\end{array}
\right)
\eea

For the $d$-wave gap, we have
\bea
\label{eq:SC 2H_2E_3_d}
\left(
\begin {array}{ccc}
 1+ {\tilde u}_{hh} L & 2{\tilde u}_{he} L &  2 {\tilde \alpha}_{he} {\tilde u}u_{he} L\\
 {\tilde u}_{he} L & 1+2u_{ee} L  &  2  {\tilde \alpha}_{ee} {\tilde u}_{ee} L\\
 2 {\tilde \alpha}_{he} {\tilde u}_{he} L & 4  {\tilde \alpha}_{ee} {\tilde u}_{ee} L  & 1+ 4  {\tilde \beta}_{ee} {\tilde u}_{ee} L\\
\end{array}
\right) \left(
\begin {array}{c}
  \Delta_{h}\\
  \Delta_{e}\\
  \bar{\Delta}_{e}\\
\end{array}
\right) =
\left(
\begin {array}{c}
  \Delta^{0}_{h}\\
  \Delta^{0}_{e}\\
  \bar{\Delta}^{0}_{e}\\
\end{array}
\right)
\eea
\end{widetext}

For the case of three hole FSs, we have to introduce three
different $\Delta_{h_i}$. The gap equations in $s$-wave and
$d$-wave channels become $5\times 5$ sets. Still, they can be very
easily analyzed.

\subsection{General Considerations}

Before we proceed further, it is instructive to take a more
careful look at the $3 \times 3$ sets Eqs.~\ref{eq:SC 2H_2E_3}
and~\ref{eq:SC 2H_2E_3_d} to illustrate issues which we outlined
in the introduction. Consider first the $s$-wave gap equation.
Suppose momentarily that all interactions are angle-independent,
i.e., $\alpha_{he}$, $\alpha_{ee}$ and $\beta_{ee}=0$. Then we
have three solutions
\bea \lambda_s = \left\{
 \begin{array}{l}
 0 \\
 - \frac{u_{hh} + u_{ee}}{2} - \sqrt{\left(\frac{u_{hh}-u_{ee}}{2}\right)^2 + u_{he}^2} \\
 -\frac{u_{hh} + u_{ee}}{2} + \sqrt{\left(\frac{u_{hh}-u_{ee}}{2}\right)^2 + u_{he}^2}
 \end{array}
\right. \label{l_1} \eea A positive (attractive) $\lambda_s$
emerges only when $u^2_{eh} > u_{ee} u_{hh}$, i.e., when
inter-pocket pair hopping interaction term exceeds intra-pocket
repulsion. The eigenfunction corresponding to the positive
$\lambda_s$  is a sign-changing $s^{\pm}$ gap: $\Delta_e = -
\Delta_h$, the one corresponding to the negative $\lambda_s$ is a
conventional $s^{++}$ gap: $\Delta_e = \Delta_h$.

Suppose next that the angular dependence of the interaction is
present. Then $\pm \cos 2 \theta$ components of the gaps along
electron FSs become non-zero. Solving the cubic equation for
$\lambda_s$, we now find three non-zero solutions. When $u^2_{eh}
> u_{ee} u_{hh}$, the solution with the largest $\lambda_s$
gradually evolves from the one which already existed for constant
interactions. When $\alpha_{he}$, $\alpha_{ee}$,  and $\beta_{ee}$
increase, the $\cos 2\theta$ component of ${\bar \Delta}_e$ grows,
and at some point gets larger than $\Delta_e$, and the gap develop
nodes. This is one scenario. Another one comes from the analysis
of the region $u^2_{eh} < u_{ee} u_{hh}$, where no $s^{\pm}$
solution was possible without angular dependence of the
interactions. In that case, one of $\lambda_s$ was zero. When the
angular dependence is included, this $\lambda_s$ becomes non-zero,
and its sign is determined by the sign of
\beq
S = u_{ee} u_{hh}
\left(\alpha^2_{ee} - \beta_{ee} \right) + u^2_{eh}
\left(\alpha^2_{eh} + \beta_{ee} -2 \alpha_{ee} \alpha_{eh}\right)
\label{7}
\eeq
When $S > 0$, $\lambda_s >0$, i.e., the system
develops an attraction in $s$-wave channel exclusively because of
the angular dependence of the interactions. The eigenfunction
corresponding to such $\lambda_s$ has nodes even when the angular
dependence of the interactions are weak. The physics picture is
that the system finds a way to minimize the effect of strong
intra-pocket repulsion $u_{ee}$ by inflating $\cos 2\theta$
components of the gaps along the electron FSs, because these
components do not couple to angle-independent component of the
interaction.

Note that the sign of $S$ is predominantly determined by the
interplay between the angular dependence of electron-hole and
electron-electron interactions. When $\alpha_{eh} \gg \alpha_{ee}$,
or when $\alpha^2_{ee} > \beta_{ee}$, $S >0$ even when $u_{ee}
u_{hh} \gg u^2_{eh}$, i.e., $s^\pm$ superconductivity with nodes
develops despite the fact that intra-pocket repulsion is the
strongest. In particular, $S$ is definitely positive if only
electron-hole interaction has momentum dependence, i.e., if
$\alpha_{ee} = \beta_{ee} =0$. In this situation, $S =
u^2_{eh}\alpha^2_{eh} > 0$.

Consider next the limit when hole FSs are absent and only electron
ones are present. At the first thought, $s$-wave pairing is
impossible. On a more careful look, however, we find from
Eq.~(\ref{eq:SC 2H_2E_3}) that, even if we set $u_{eh} = 0$, one
of $\lambda_s$ for $s$-wave pairing is still positive if
$\beta_{ee} < \alpha^2_{ee}$, no matter how small angular
dependence of electron-electron interaction is. The eigenfunction
for this solution again has nodes. The physics reasoning is the
same as in the case we just considered: the angle-independent part
of the interaction is repulsive, but the system finds a way to
overcome this strong repulsion by inflating $\cos 2 \theta$
components of the gaps along the electron FSs.

We next turn to the $d$-wave gap equation~( Eq. \ref{eq:SC
2H_2E_3_d}). The generic reasoning parallels the one for $s$-wave
case, namely for interactions independent on $\cos 2 \theta$
attractive $d$-wave solution exists when ${\tilde u}^2_{eh} >
{\tilde u}_{ee} {\tilde u}_{hh}$, and for angle-dependent
interaction one of $\lambda_d$ is positive (attractive) even if
this condition is not satisfied, but ${\tilde S}$, which is a
$d$-wave analog of $S$ from Eq.~(\ref{7}), is positive.  There is,
however, one crucial distinction with the $s$-wave case: the
$d$-wave interactions ${\tilde u}_{hh}$ and ${\tilde u}_{ee}$ are
not necessary positive.  In particular, ${\tilde u}_{ee}$ is the
difference between angle-independent components of intra-pocket
and inter-pocket interactions between electron pockets. Once the
inter-pocket interaction is larger (e.g., when magnetic
fluctuations are peaked at $(\pi,\pi)$), ${\tilde u}_{ee} <0$, and
the system develops an attraction in the $d$-wave channel, even
when $d$-wave electron-hole interaction is weak.  This is
particularly relevant for the case when only electron FSs are
present, i.e., within BCS approximation ${\tilde u}_{eh}$ can be
set to zero. If ${\tilde u}_{ee} <0$ in this case, $d$-wave
solution emerges, with the sign-changing gap on the two electron
FSs. An alternative possibility is that ${\tilde u}_{ee}
>0$, but ${\tilde \alpha}^2_{ee} > {\tilde \beta}_{ee}$, and
$d$-wave attraction is produced by angle-dependent part of
$d$-wave interaction. In this situation, $\cos 2 \theta$ component
of the $d$-wave gap is large, and the gap has nodes on the two
electron FSs.

\subsection{Computational Procedure}

We use $\Gamma_{ij} (\k_F, \k'_F)$ obtained in the
RPA/SF approach as inputs and fit their functional forms by
Eq.~\ref{eq:interactions}. This gives us $U_{ij}$, $\alpha_{ij}$ and
$\beta_{ij}$. We assume for simplicity that $N_{F,j}$ are the same
for all FSs, convert $U_{ij}$ into $u_{ij} = N_F U_{ij}$ and
analyze $4 \times 4$ and $5 \times 5$ gap equations. We then vary
parameters and check how robust the solutions are. The approach
can be  extended to the case when $N_{F,j}$ depends on $j$, but
this dependence very likely does not change the physics.

We analyzed six different sets of parameters and several different
doping levels which correspond to either electron-doping or hole doping.
The parameters are presented in Table~\ref{tab:parameters}.  The
results for all sets of parameters are quite similar, and below we
show the results only for representative cases. We also analyzed the case of large electron doping, when there are no hole FSs.
\begin{table}[htp]
\caption{Six sets of parameters used in comparison of RPA/SF and
LAHA. We used different values of the chemical potential $\mu$ for
each set ($\mu=-0.30$ to $+0.30$), the range covers hole
doping(negative) to electron doping (positive).}
\label{tab:parameters}
\begin{ruledtabular}
\begin{tabular}{cccc}
set \# & U & J & V\\
\hline
set 1&1.67&0.21&1.46 \\
set 2&1.0&0.25&0.69\\
set 3&1.2&0.0&1.2\\
set 4&1.0&0.9&-1.25\\
set 5&1.0&4.0&-4.0\\
set 6&1.0&0.9&-0.7\\
\end{tabular}
\end{ruledtabular}
\end{table}
\begin{figure*}[htp]
$\begin{array}{cccc}
\includegraphics[width=1.8in]{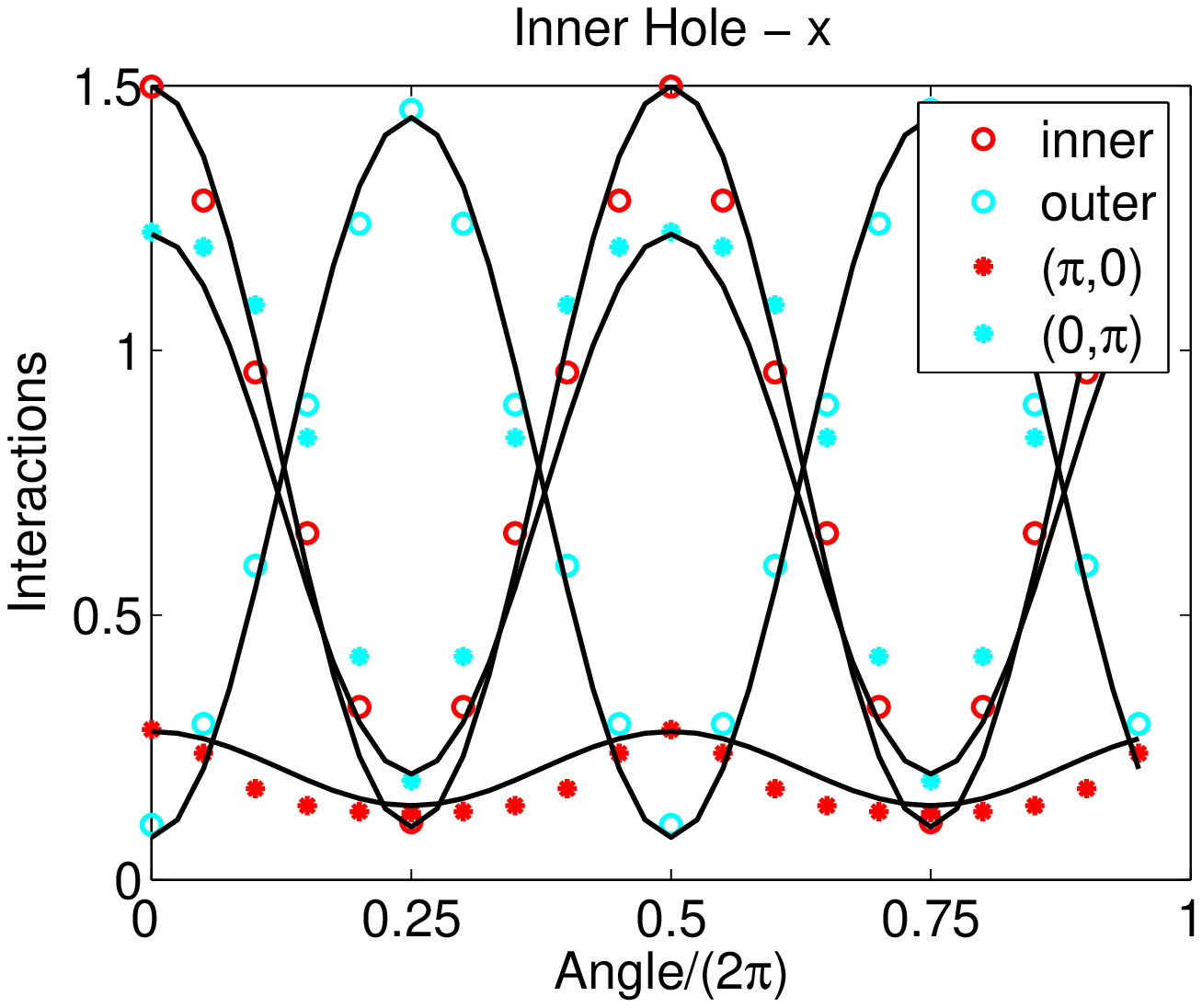}&
\includegraphics[width=1.8in]{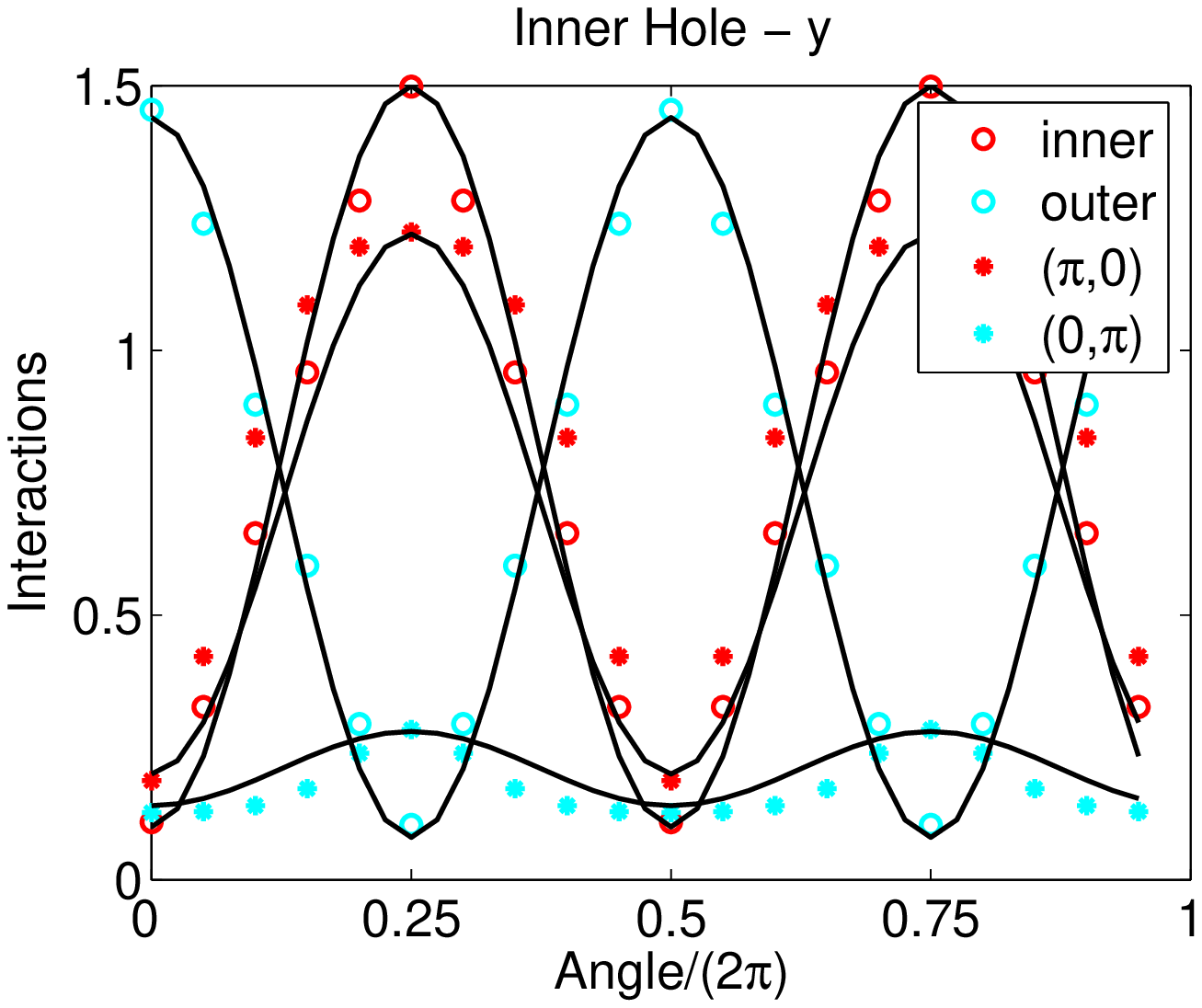}&
\includegraphics[width=1.8in]{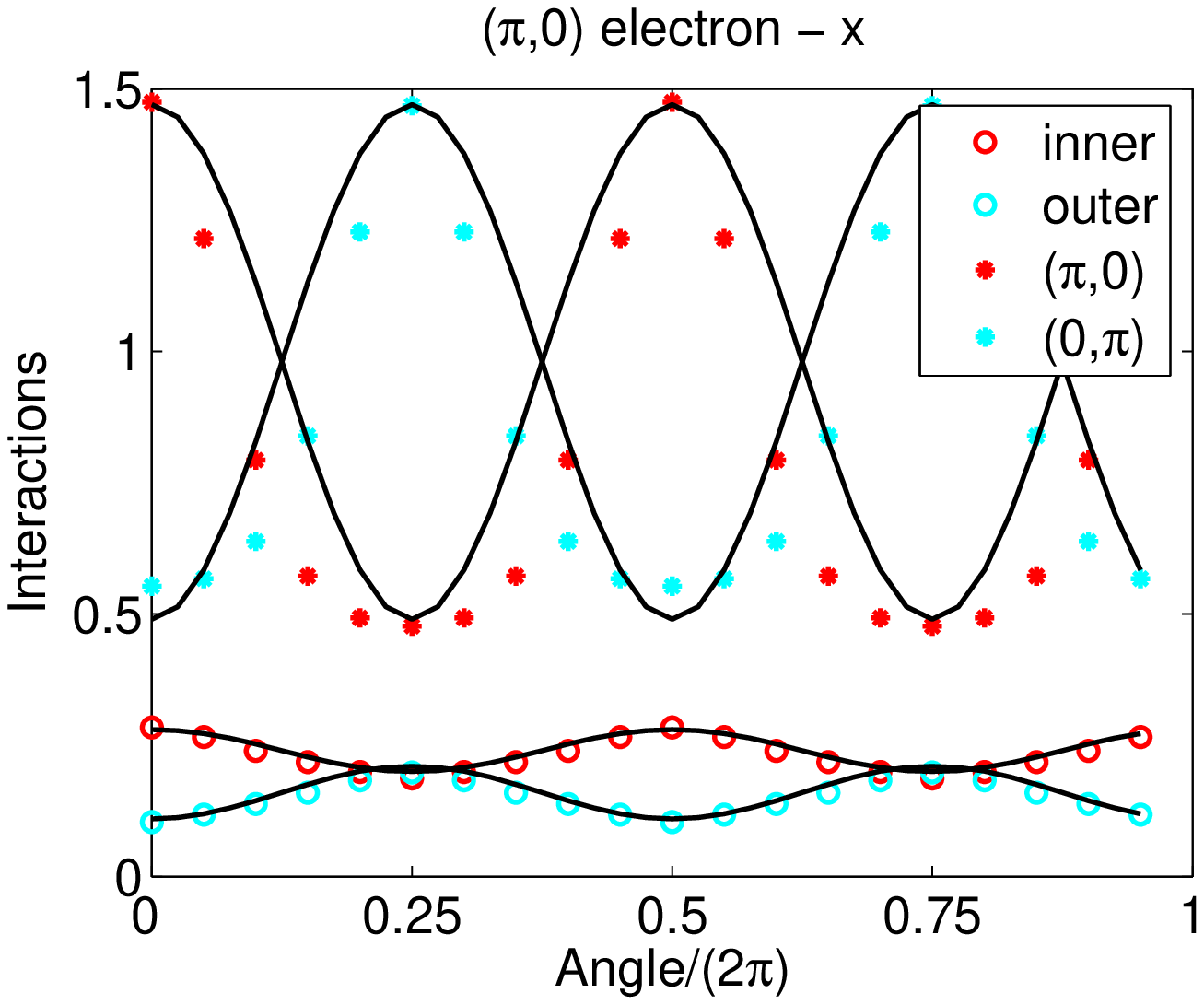}&
\includegraphics[width=1.8in]{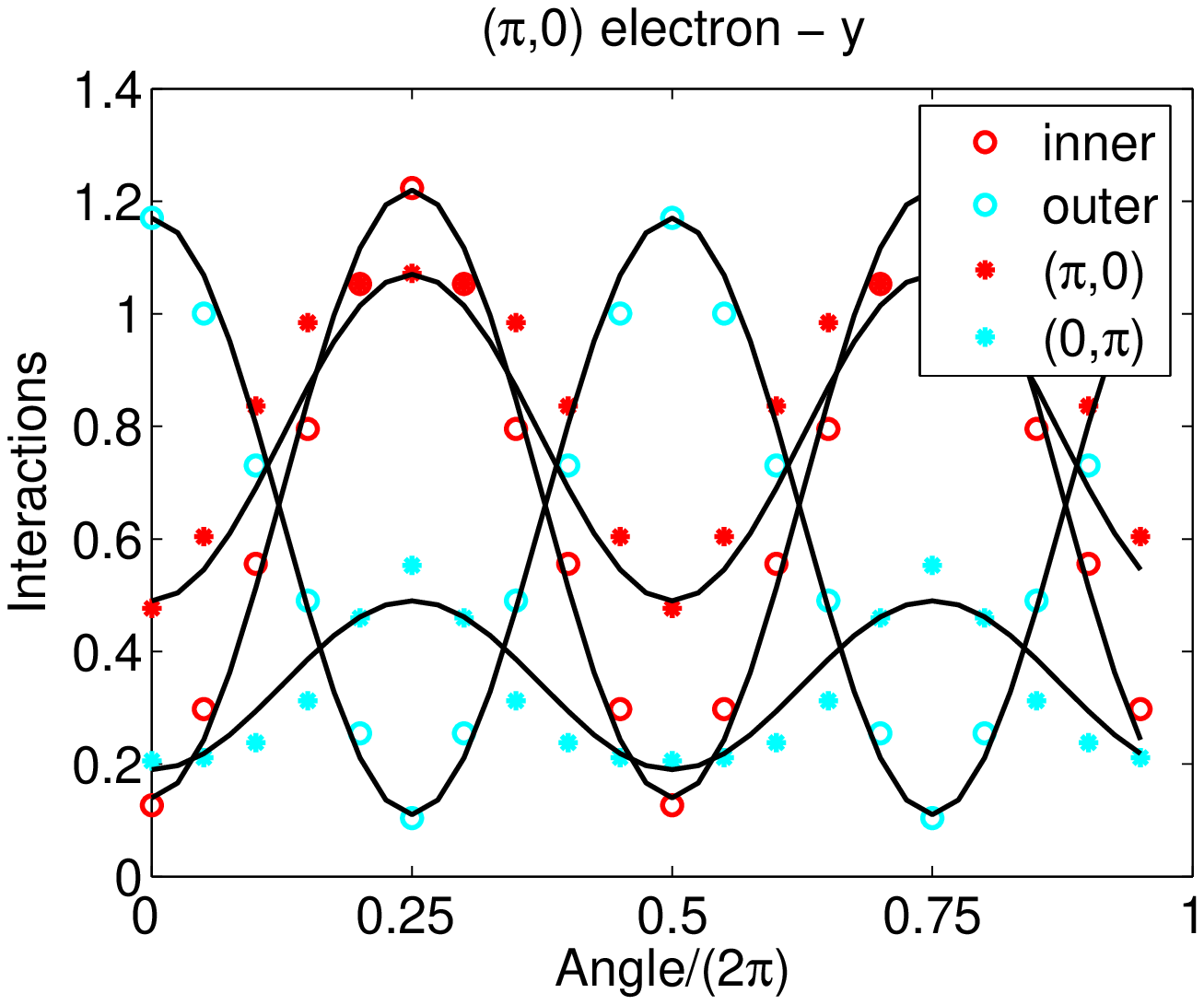}
\end{array}$
\caption{\label{fig:plots_wo_RPA} Fits of the actual interactions
$\Gamma_{ij} (\k_F, \k'_F)$ by LAHA. Symbols
represent interactions computed numerically for the 5-orbital model using
LDA band structure, solid lines are fits using
Eq.~\ref{eq:interactions}. The fit is for the set \#1 for bare
interactions (no SF component). The chemical potential is $\mu =
0.08$ which corresponds to electron doping.
We set $\k_F$ in
$\Gamma_{ij} (\k_F, \k'_F)$ to be either along $x$ or
along $y$ direction on a given FS (its location is specified in
the title on top of each figure) and varied $\k'_F$ along
each of FSs. The angle $\theta'$ is measured relative to $k_x$. The
fit is amazingly good.}
\end{figure*}
%


\begin{figure*}[t]
$\begin{array}{cccc}
\includegraphics[width=1.8in]{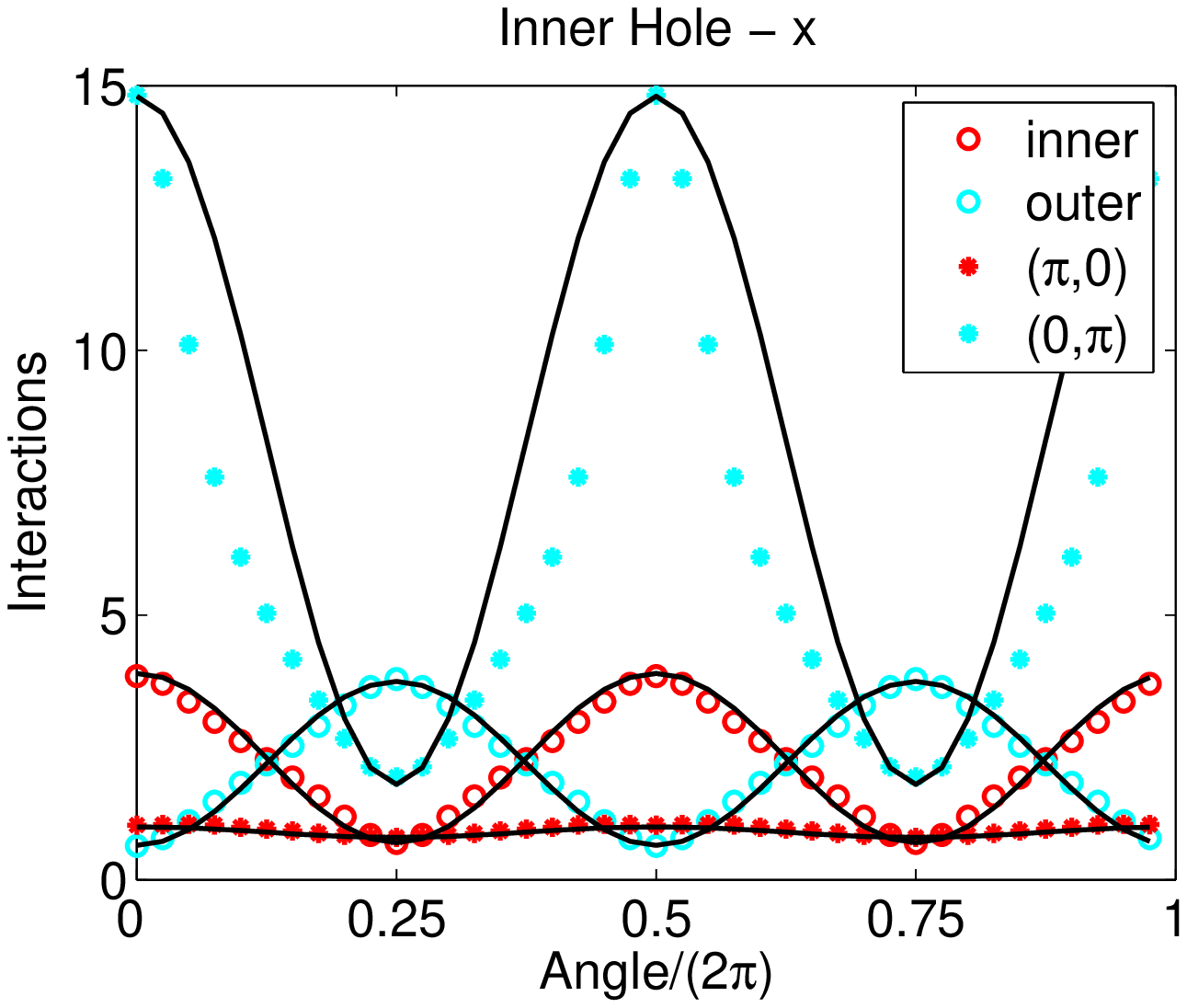}&
\includegraphics[width=1.8in]{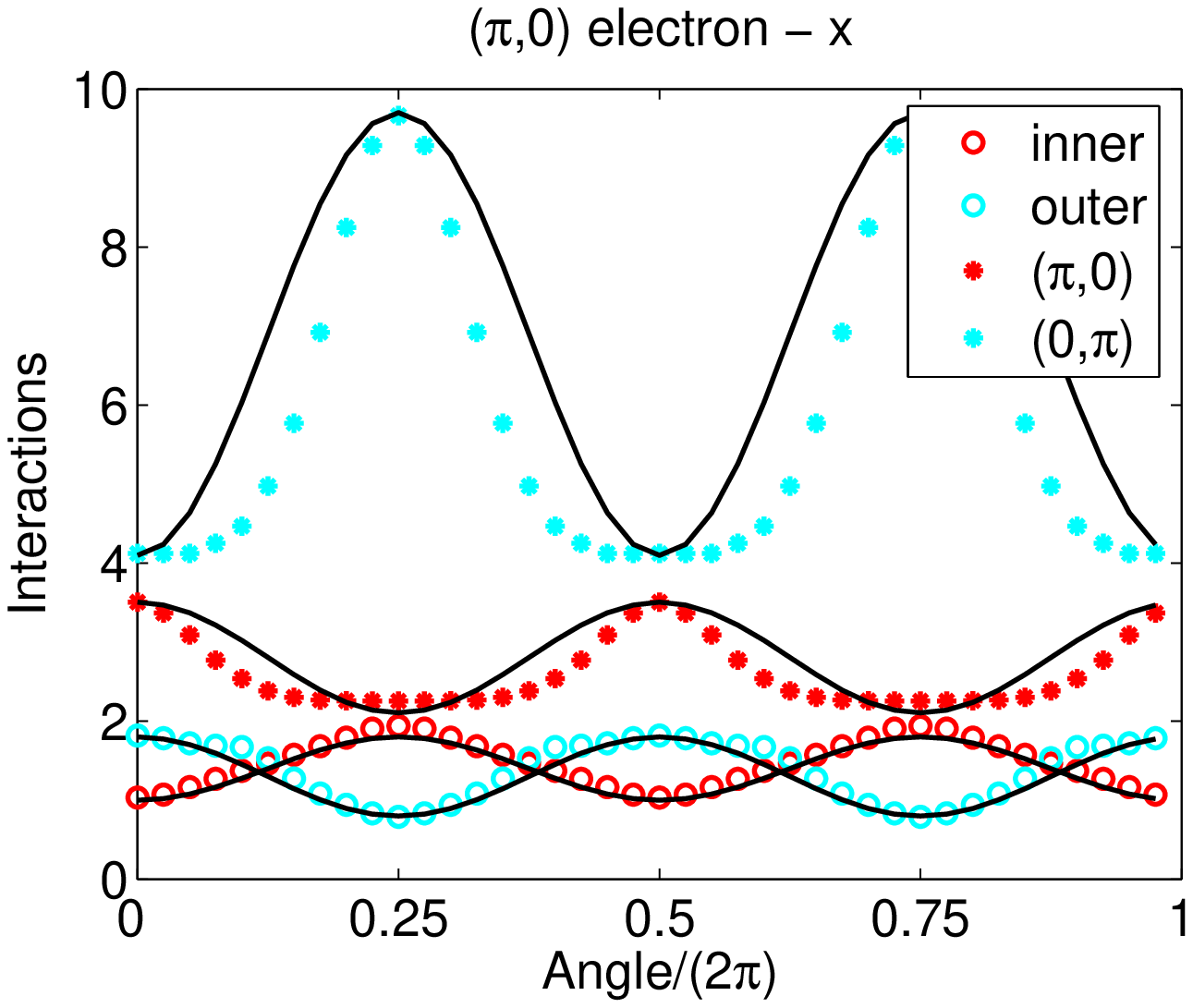}&
\includegraphics[width=1.8in]{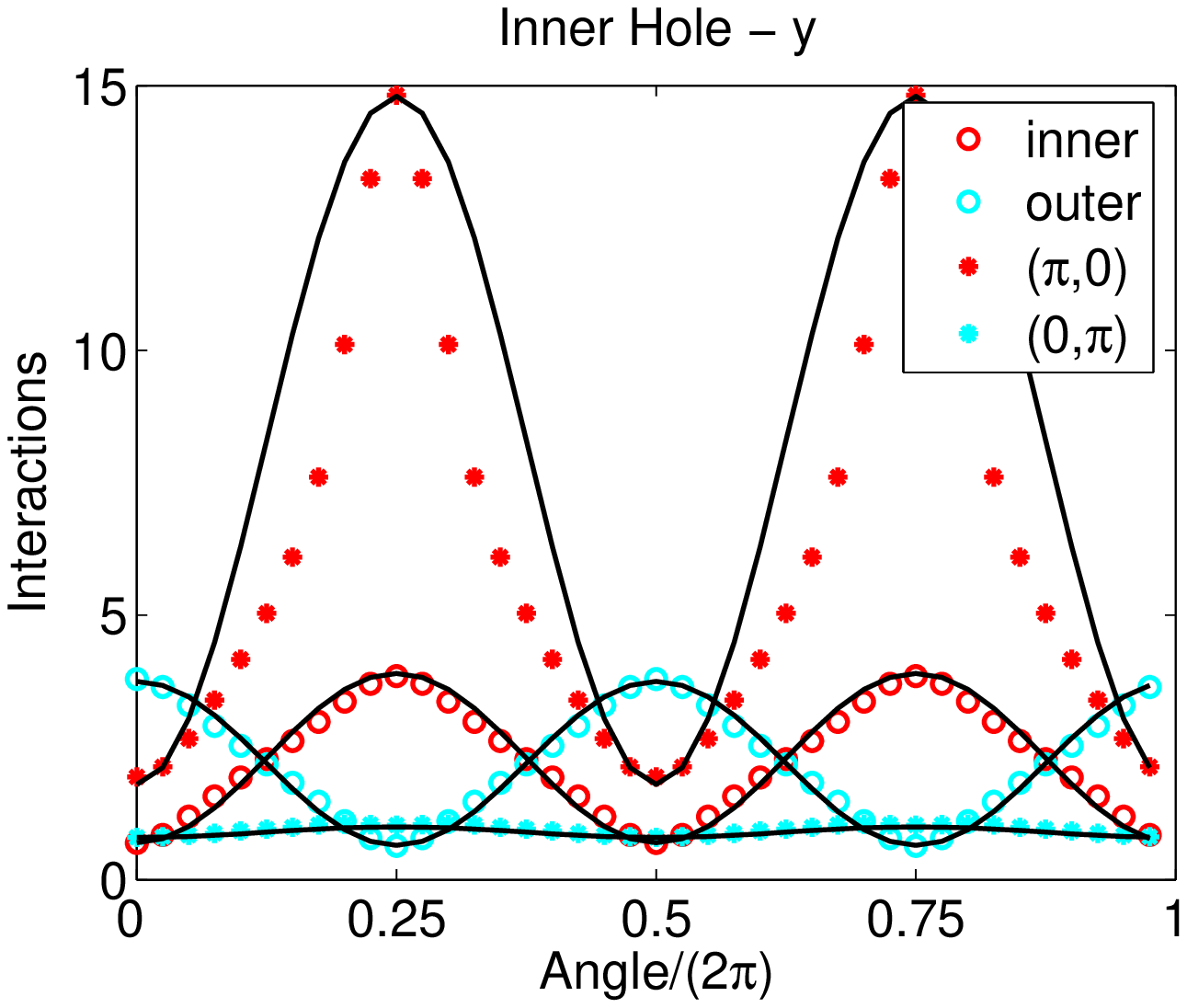}&
\includegraphics[width=1.8in]{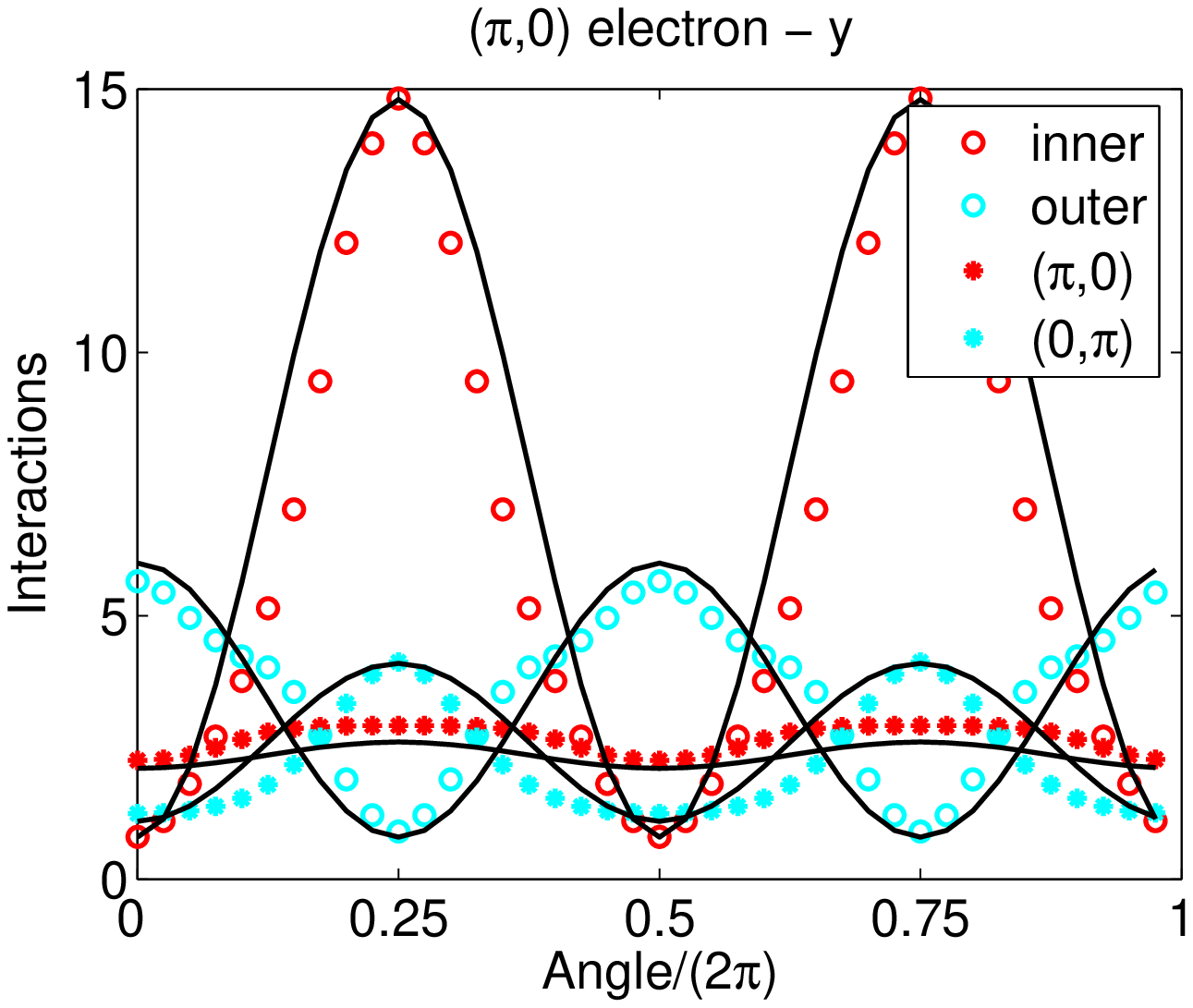}
\end{array}$
\caption{\label{fig:fits_tom} The same as in
Fig.~\ref{fig:plots_wo_RPA} but for the full interaction, with SF
component. The fit is again amazingly good. We verified that the
slight discrepancies are removed if in LAHA we add a $4\theta$
harmonics to the interaction.}
\end{figure*}

The use of LAHA is meaningful only if the fit of the actual
$\Gamma_{ij} (\k_F, \k'_F)$ by Eq.~\ref{eq:interactions} is
accurate enough. In Figs.~\ref{fig:plots_wo_RPA}
and~\ref{fig:fits_tom} we show fits for the representative case of
$\mu = 0.08$ (electron doping with $n_e=6.12$, while undoped case
is $n_e=6$) and interactions from the set \#1 ($U=1.67$,
$J=J'=0.21$, $V =1.41$) considered in Ref.~\onlinecite{tom_09}.
We set one of momenta to be either along $x$ or along $y$ axis on
one of the four FSs and vary the other momentum along all four
FSs. The parameters for $s$-wave and $d$-wave interactions
extracted from the fit are presented in Table~\ref{tab:s-set6}.

The original $\Gamma_{ij} (\k_F, \k'_F)$ are in the units of eV.
Within our approximation of an equal $N_{F,j}$ on all FSs, the
dimensionless $u_{ij}$ are $\Gamma_{ij} (\k_F, \k'_F)$ times the
rescaling factor. This factor affects the overall scale of the
eigenvalues $\lambda$, but doesn't affect the sign and relative
magnitudes of $\lambda_s$ and $\lambda_d$, which we need. For
simplicity, we then set this rescaling factor to be equal to one.

\begin{table*}[htp]
\caption{$s$- and $d$- wave parameters for the set \#1 with
$\mu=0.08$. Here and henceforth NSF and SF mean the bare
interaction without SF component and the full interaction with SF
component, respectively.} \label{tab:s-set6}
\begin{ruledtabular}
\begin{tabular}{cccccccccccccc}
$s-$wave&$u_{h_1h_1}$&$u_{h_2h_2}$&$u_{h_1h_2}$&$u_{h_1e}$&$\alpha_{h_1 e}$&$u_{h_2e}$&$\alpha_{h_2 e}$&$u_{ee}$&$\alpha_{ee}$&$\beta_{ee}$\\
\hline
NSF&$0.8$&$0.76$&$0.78$&$0.46$&$-0.24$&$0.4$&$-0.30$&0.77&0.14&0.09\\
SF&$2.27$&$2.13$&$2.22$&$4.65$&$-0.34$&$2.29$&$-0.22$&3.67&0.15&0.04\\
\hline
$d$-wave&$\tilde{u}_{h_1 h_1}$&$\tilde{u}_{h_2 h_2}$&$\tilde{u}_{h_1 h_2}$&$\tilde{u}_{h_1 e}$&$\tilde{\alpha}_{h_1 e}$&$\tilde{u}_{h_2 e}$&$\tilde{\alpha}_{h_2 e}$&$\tilde{u}_{ee}$&$\tilde{\alpha}_{ee}$&$\tilde{\beta}_{ee}$\\
\hline
NSF&$0.7$&$0.66$&$-0.68$&$-0.25$&$-0.58$&$0.24$&$-0.42$&0.11&-0.5&0.25\\
SF&$1.50$&$1.40$&$-1.50$&$-3.73$&$-0.44$&$1.44$&$-0.32$&1.03&-0.49&-0.02\\
\end{tabular}
\end{ruledtabular}
\end{table*}


We see that the fits are quite good. For the case of no SF term
(NSF), all  $\Gamma_{ij} (\k_F, \k'_F)$ are fitted well by
Eq.~\ref{eq:interactions}. For the full interaction (i.e. SF
included), all $\Gamma_{ij} (\k_F, \k'_F)$  are fitted well, the
small discrepancies can be
cured by including cos$4\theta$ terms into the interactions.

The fits for other parameters are quite similar. For example in
Figs.~\ref{fig:SF_0p05} -~\ref{fig:SF_m0p05} below we show the
``best'' and the ``worst'' fits of interactions for the set \#2
for three different values of $\mu$. For positive $\mu$, there are
4 FSs, and for negative $\mu$, there is an additional, 5th hole
FS. The fits are not perfect, but are quite good for all practical
purposes.

We continue below with the set \#1 for the discussion on how SF
contribution affects the pairing symmetry and the gap structure.
The trend is the same for all other sets. And just for a change we
will look at set \#2 for the discussion of how gap symmetry and
structure change with doping. We do so to limit the number of
figures and not overwhelm the readers. We will point out in the
text if differences arise.

\begin{figure*}[t]
$\begin{array}{cccc}
\includegraphics[width=1.8in]{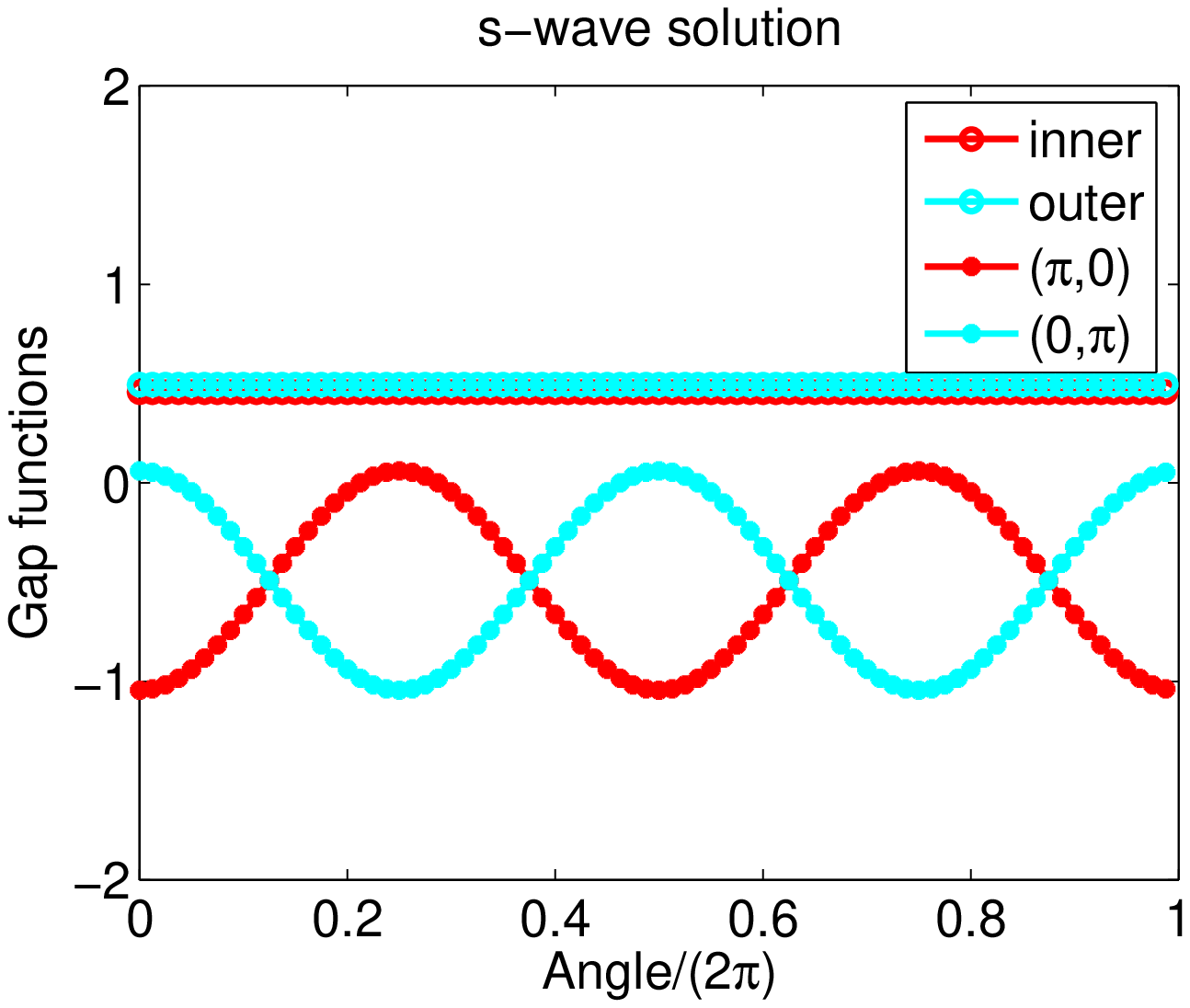}&
\includegraphics[width=1.8in]{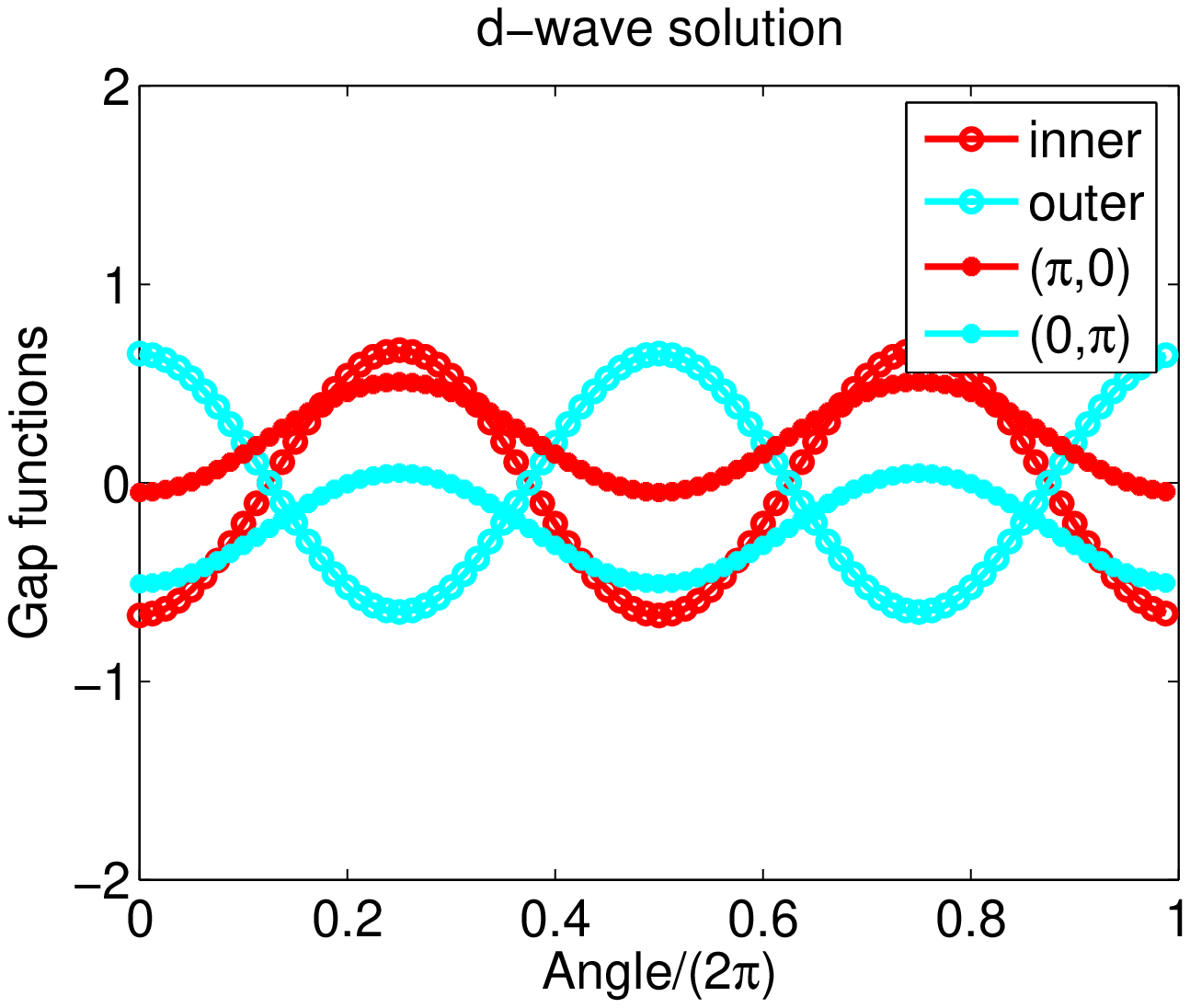}&
\includegraphics[width=1.8in]{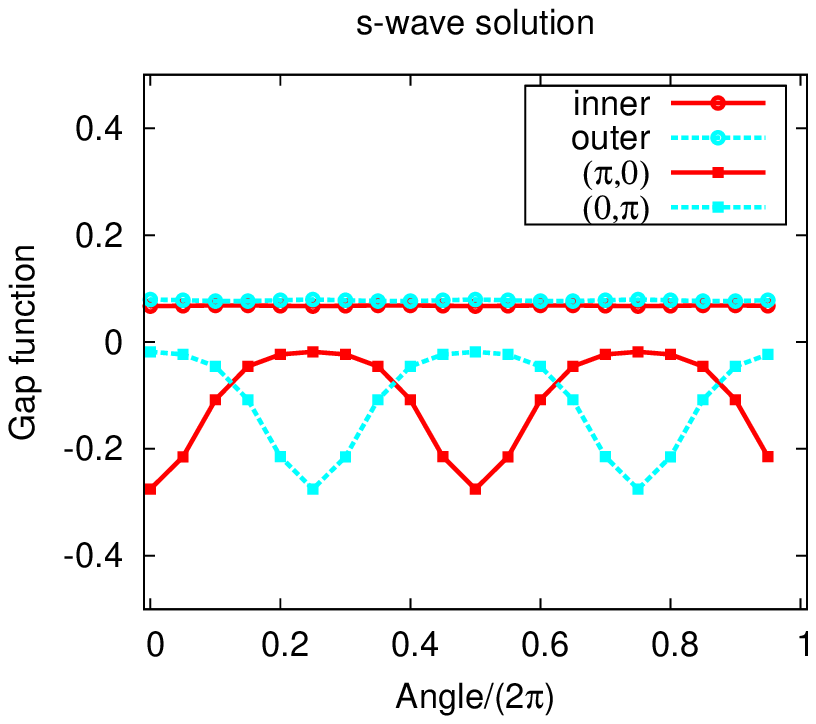}&
\includegraphics[width=1.8in]{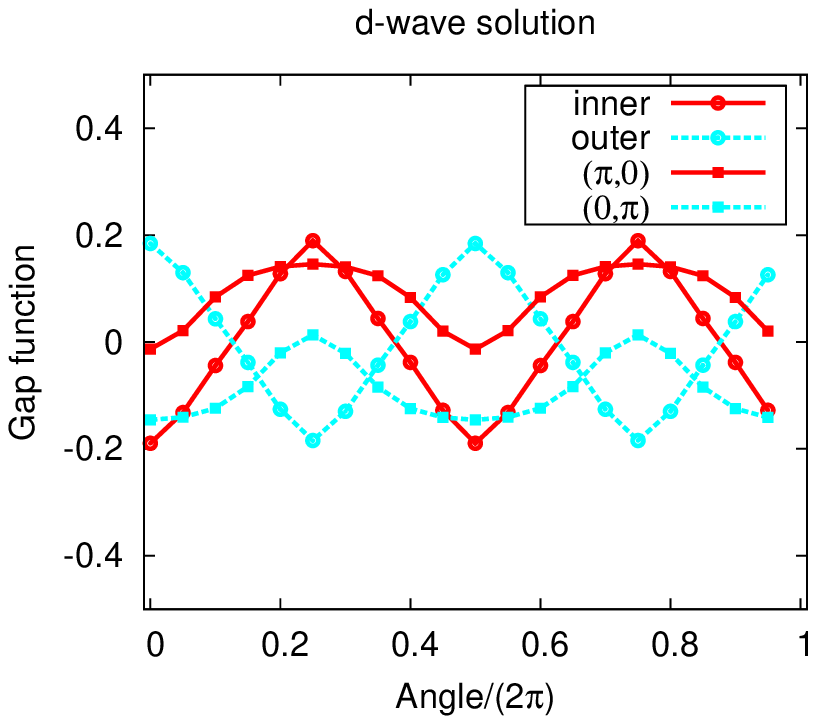}\\
\includegraphics[width=1.8in]{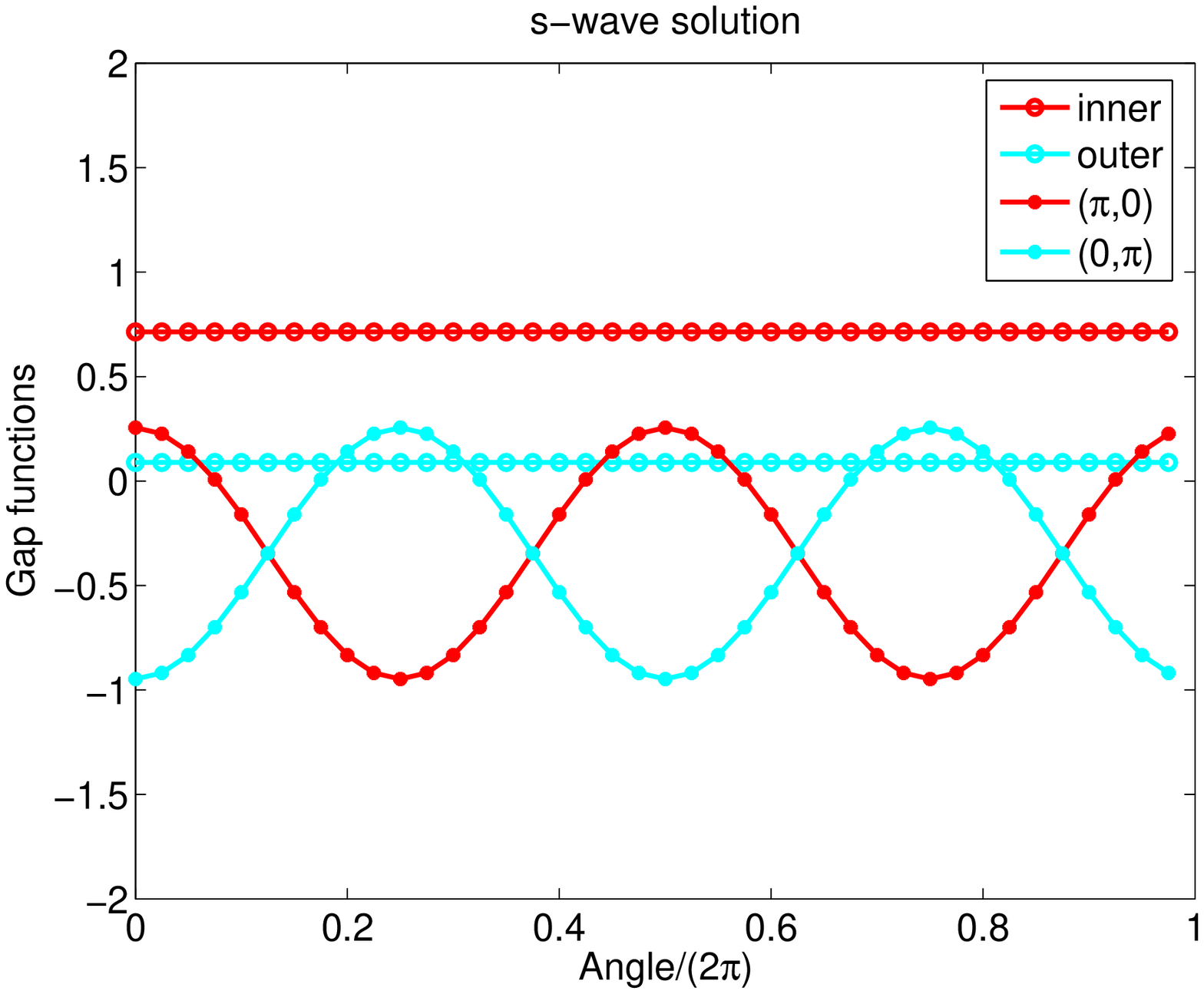}&
\includegraphics[width=1.8in]{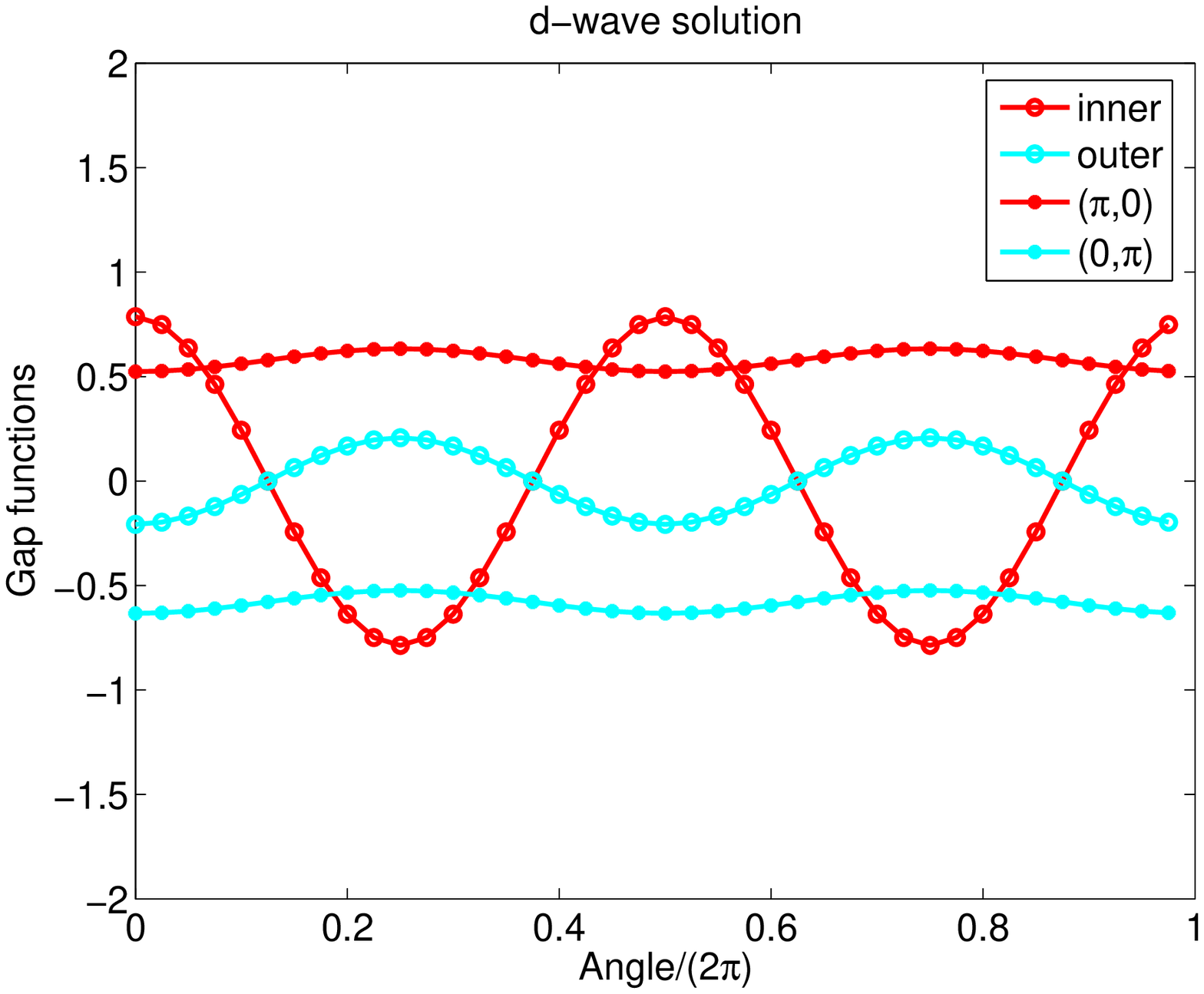}&
\includegraphics[width=1.6in]{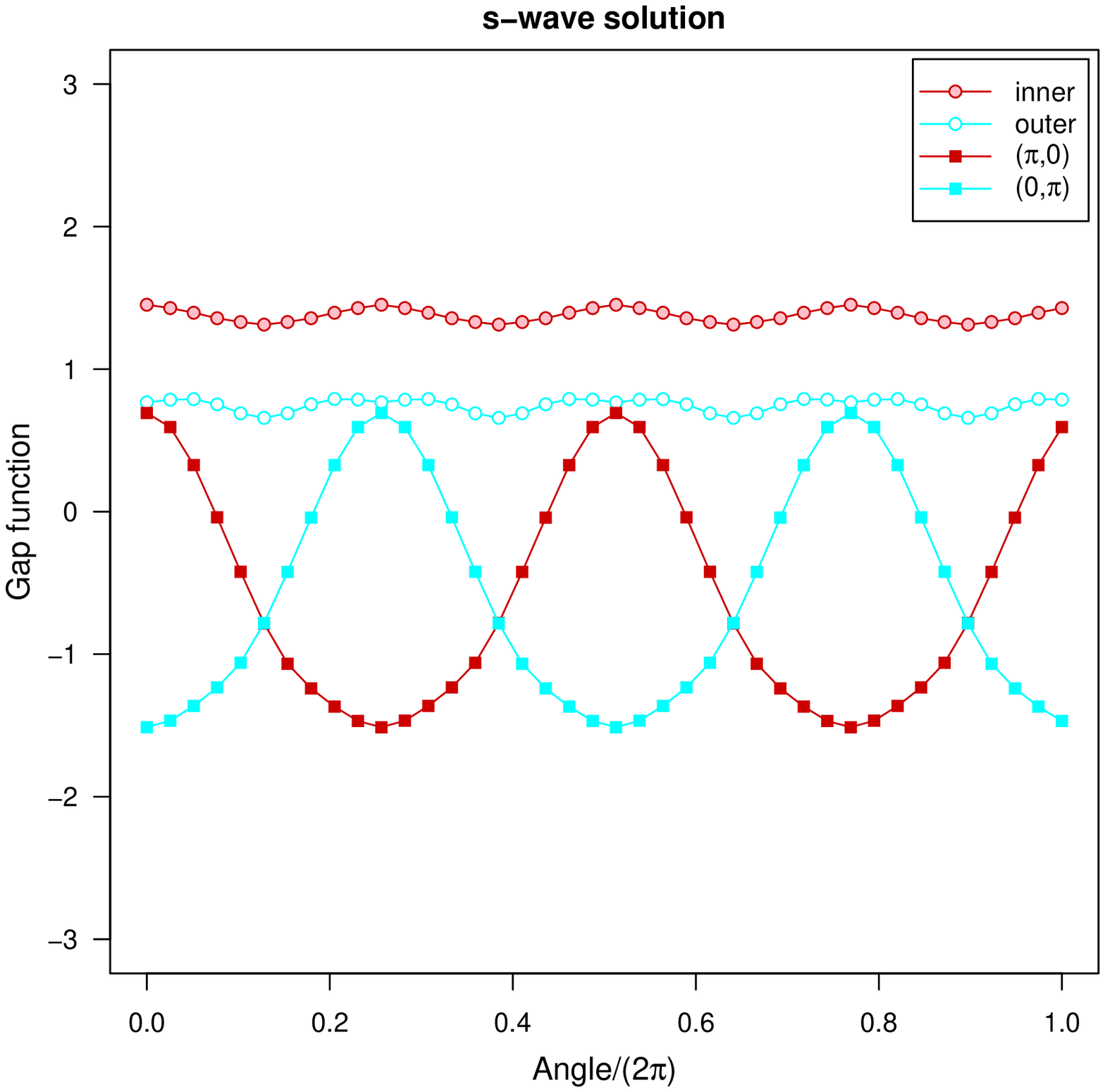}&
\includegraphics[width=1.6in]{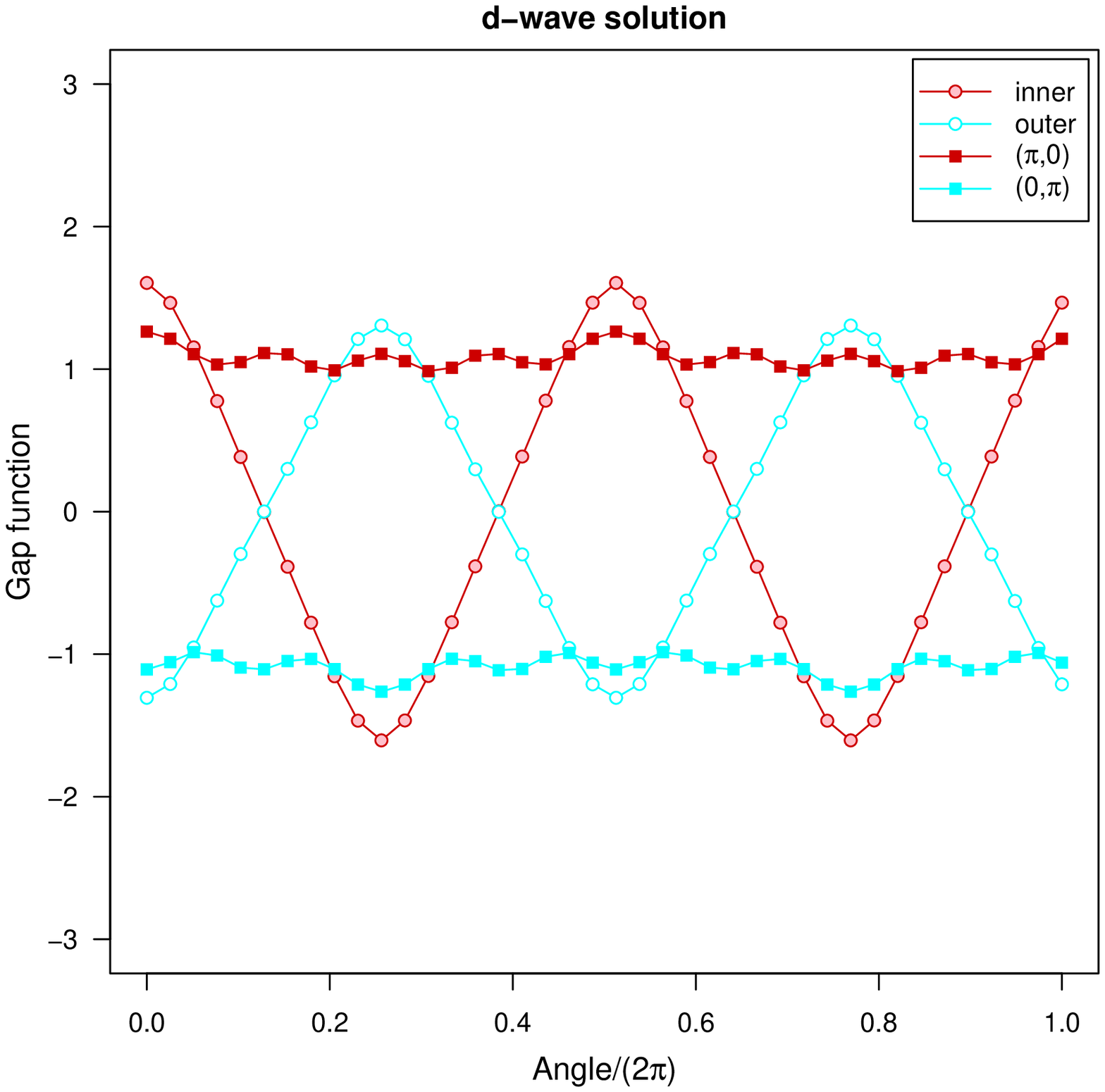}
\end{array}$
\caption{\label{fig:gaps_wo_RPA}. Top: $s$- and $d-$wave solutions
obtained by applying LAHA for the gap structure (left pair) for
the set \#1 (NSF) with $\mu=0.08$. $s$- and $d-$wave solutions
obtained numerically from the 5-orbital model based on the LDA
band structure(right pair) are also shown. There are no solutions
with positive $\lambda$, so we show the solutions with the smallest negative $\lambda$ in
both channels.  The agreement of the gap structure with LAHA
results is quite good. For LAHA, the $s$-wave solution has
$\lambda_{s}=-1.02$ and $d$-wave solution has $\lambda_{d}=-0.99$
($\lambda_s/\lambda_d \approx 1.0$).  $\lambda_s$ and $\lambda_d$
obtained from the numerical approach are negative, and their ratio
is $\lambda_s/\lambda_d = 0.6$. Bottom: The same, but now with the
SF component of interaction included. The couplings are now
positive: $\lambda_s=4.6$ and $\lambda_d=4.8$
($\lambda_s/\lambda_d \approx 0.96$) for LAHA and $\approx 1.1$
for RPA/SF calculations for the 5-orbital model \cite{tom_09}. }
\end{figure*}

\section{Sensitivity of the gap structure to the form of $\Gamma_{ij}(\k_F, \k'_F)$ \label{sec:FormOfGamma}}

We first use the parameters extracted from the fit for the set \#1
and solve the $4\times 4$  gap equation within LAHA. The results
for the case of no SF component are shown in
Fig.~\ref{fig:gaps_wo_RPA}.  For comparison, the gaps obtained by
the full numerical solution within RPA are also presented.  The
LAHA and RPA solutions are almost equivalent, which is another
indication that the LAHA fit works quite well.   The results for
the full interaction, with SF component, are also presented in
Fig.~\ref{fig:gaps_wo_RPA} along with the numerical solution for
$s$-wave gap within RPA/SF. The LAHA and RPA solutions again agree
very well. The only real difference between RPA/SF and LAHA is
that the value of the gap along the outer hole FS is somewhat
larger in RPA/SF. Like we said, this is the consequence of the
fact that to fit $\Gamma_{h_2,e} (\phi,\theta)$ in LAHA one needs
to add $cos 4 \theta$ components. Once we include these
components, the gap along the outer hole FS goes up, bringing the
LAHA result even closer to RPA/SF.

We see therefore that LAHA approximation works quite well both for
the bare interaction and the full interaction with SF
contribution.  We verified that the near equivalence between gaps
obtained within RPA/SF and LAHA holds for all other sets of
parameters from Table~\ref{tab:parameters}. This gives confidence that the physics
can be understood by analyzing $4\times 4$ $s$-wave and $d$-wave
gap equations within LAHA.  We begin with $s$-wave case.

\subsection{$s$-wave case}

\subsubsection{Modification of $\Gamma_{ij} (\k_F, \k'_F)$ by spin fluctuations}

We first take a closer look at Table~\ref{tab:s-set6}.
Comparing the values of $u_{h_1h_1}, u_{h_2h_2}, u_{h_1h_2},
u_{h_1e}, u_{h_2e}$, and $u_{ee}$ we see two trends. First,  once
SF contribution is added,  there is overall enhancement, roughly
by a factor of three,  for {\it all} interactions, including
interactions within hole pockets. On top of this overall
enhancement, there is another effect -- electron-hole interactions
$u_{he}$  further increase compared to $u_{hh}$ and $u_{ee}$. This
additional increase is by factor of $3-4$, such that the total
increase of $u_{h_1e}$  is by a factor of $10$ [the increase of
$u_{h_2e}$ is a bit smaller].

We attribute the overall increase to the ``bare'' SF interaction
term $u^2 \chi_0 (q)$ and the additional increase to the relative
enhancement of the RPA susceptibility $\chi (Q)/\chi_0 (Q)$ near
the momentum transfer $Q = (\pi,0)$ or $(0,\pi)$. The reasoning is
based on the comparison with the results of  Ref.
\onlinecite{graser} for the bare and RPA-renormalized spin
susceptibility. First,  the product $u_{ij} \chi_0 (\k_i - \k_j)$
remains roughly constant in $k-$space, if we use $u_{ij}$ from
Table~\ref{tab:s-set6}. Second, the RPA-renormalized $\chi (Q)$ is
3-4 times larger than $\chi_0 (Q)$, while $\chi (0) \approx \chi_0
(0)$. (Ref. \onlinecite{graser}). This logic also applies to the
interaction between electron pockets, for which  $u_{ee}$ is the
average between intra-pocket interaction, for which $\chi = \chi
(0) \approx \chi_0 (0)$, and inter-pocket interaction for which
$\chi = \chi (\pi,\pi) \approx 2.4 \chi_0 (\pi,\pi)$.  The total
increase of $u_{ee}$ is then expected to be $1 + 2((1+2.4)/2)
=4.4$, and we see from the Table~\ref{tab:s-set6} that $u_{ee}$
increases by  quite similar factor of $4$.

Compare next the angular parts $\alpha_{he}$, $\alpha_{ee}$ and
$\beta_{ee}$. We see from Table~\ref{tab:s-set6} that
$\alpha_{he}$ and $\alpha_{ee}$ do not change much.  The term
$\beta_{ee}$ does change and becomes 2.5 times smaller in the
presence of the SF component. However, we will show in the next
section that the gap structure is insensitive to the change of
$\beta_{ee}$ and does not change much even if we set
$\beta_{ee}=0$ (see Fig.~\ref{fig:tom_change}).

Neglecting the change of $\beta_{ee}$, we conclude that the SF
contribution to the pairing interaction increases the overall
magnitude of $\Gamma_{ij}$ and additionally increases the
magnitudes of electron-hole interactions ($u_{h_1e}$ and
$u_{h_2e}$ terms) and, to lesser extent, of electron-electron
interaction (the $u_{ee}$ term), but doesn't substantially modify
the relevant angular dependence of the electron-hole interaction.
The overall increase of the pairing interaction does not affect
the gap structure, hence the only true effect of the SF term is
the increase of electron-hole interactions compared to hole-hole
and electron-electron interactions.

As we said, controlled RG flow of the couplings gives rise to
exactly the same effect --  angular dependence are preserved
during the flow, but  the relative magnitude of electron-hole
interaction increases. From this perspective,  RG and SF
approaches, although formally different, describe very similar
physics.

\subsubsection{Effect of the angular dependencies of electron-hole and electron-electron interactions}

We now analyze explicitly how sensitive is the gap structure to
various angular-dependent components of $\Gamma_{ij}$.  Within
LAHA, we can easily change angular dependence of the interactions
and check the consequences. First, we verify how sensitive is the
solution for the gap to the change of the angular component of
electron-hole interaction.

\begin{figure*}[t]
$\begin{array}{ccc}
\includegraphics[width=2in]{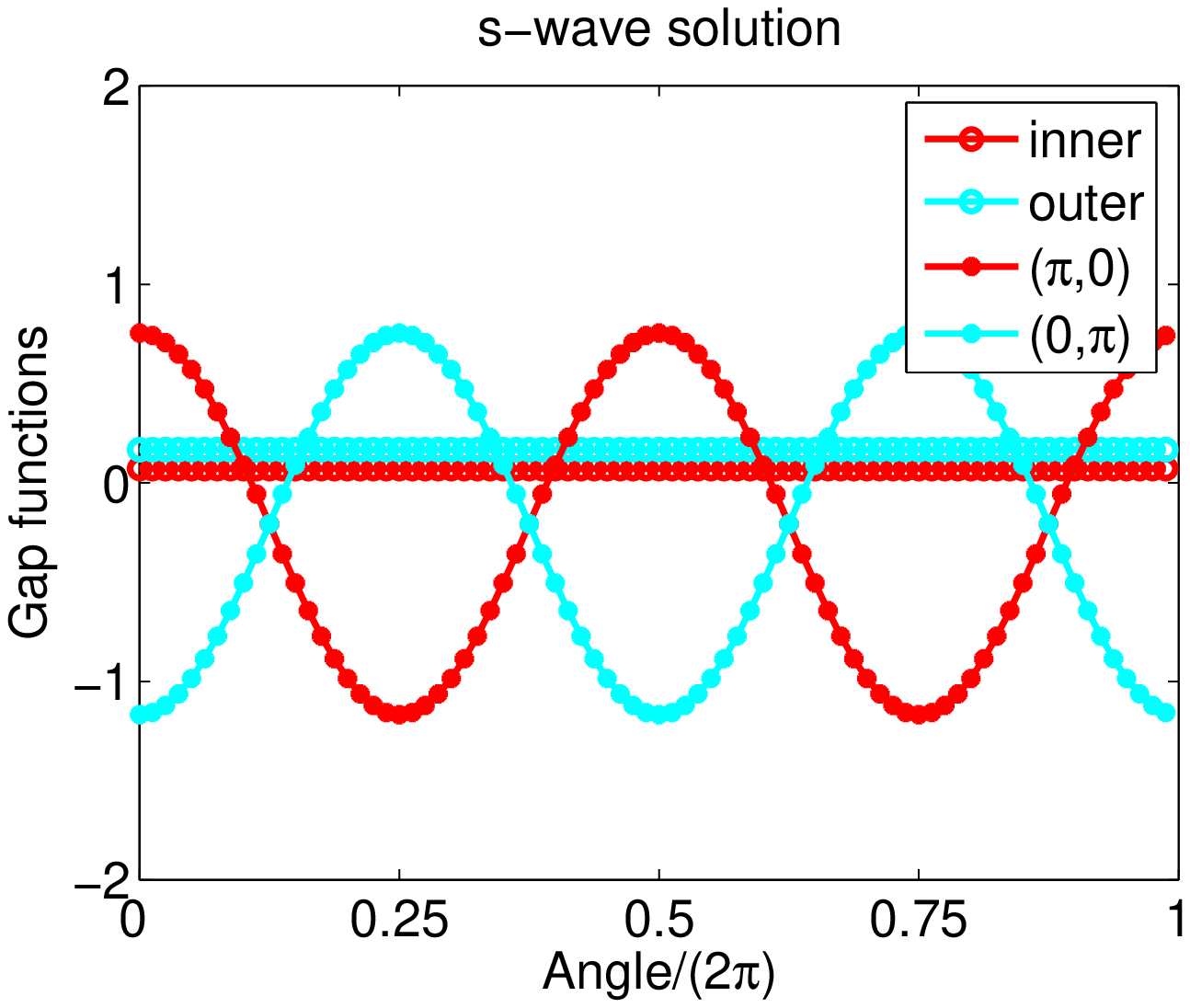}&
\includegraphics[width=2in]{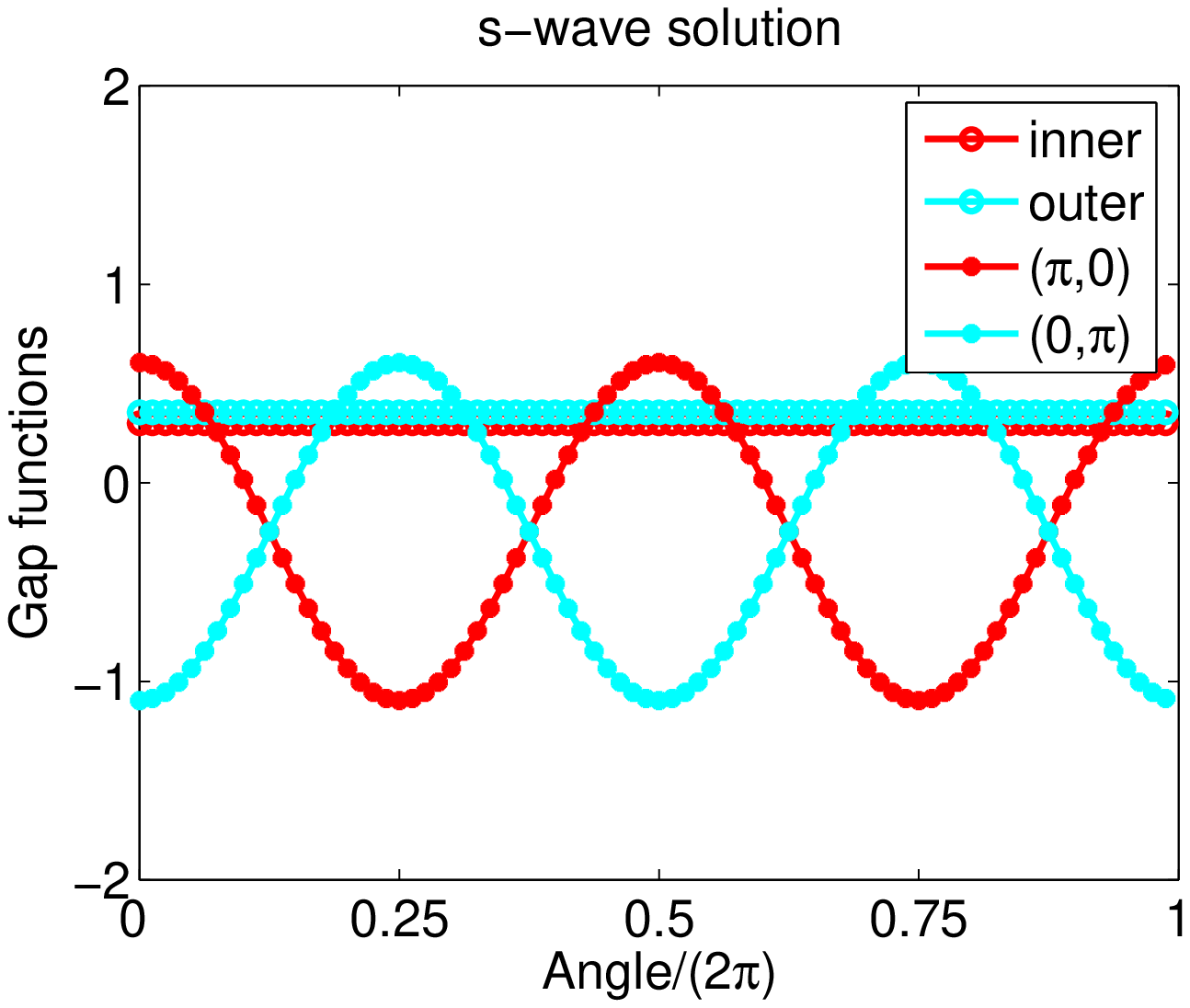}&
\includegraphics[width=2in]{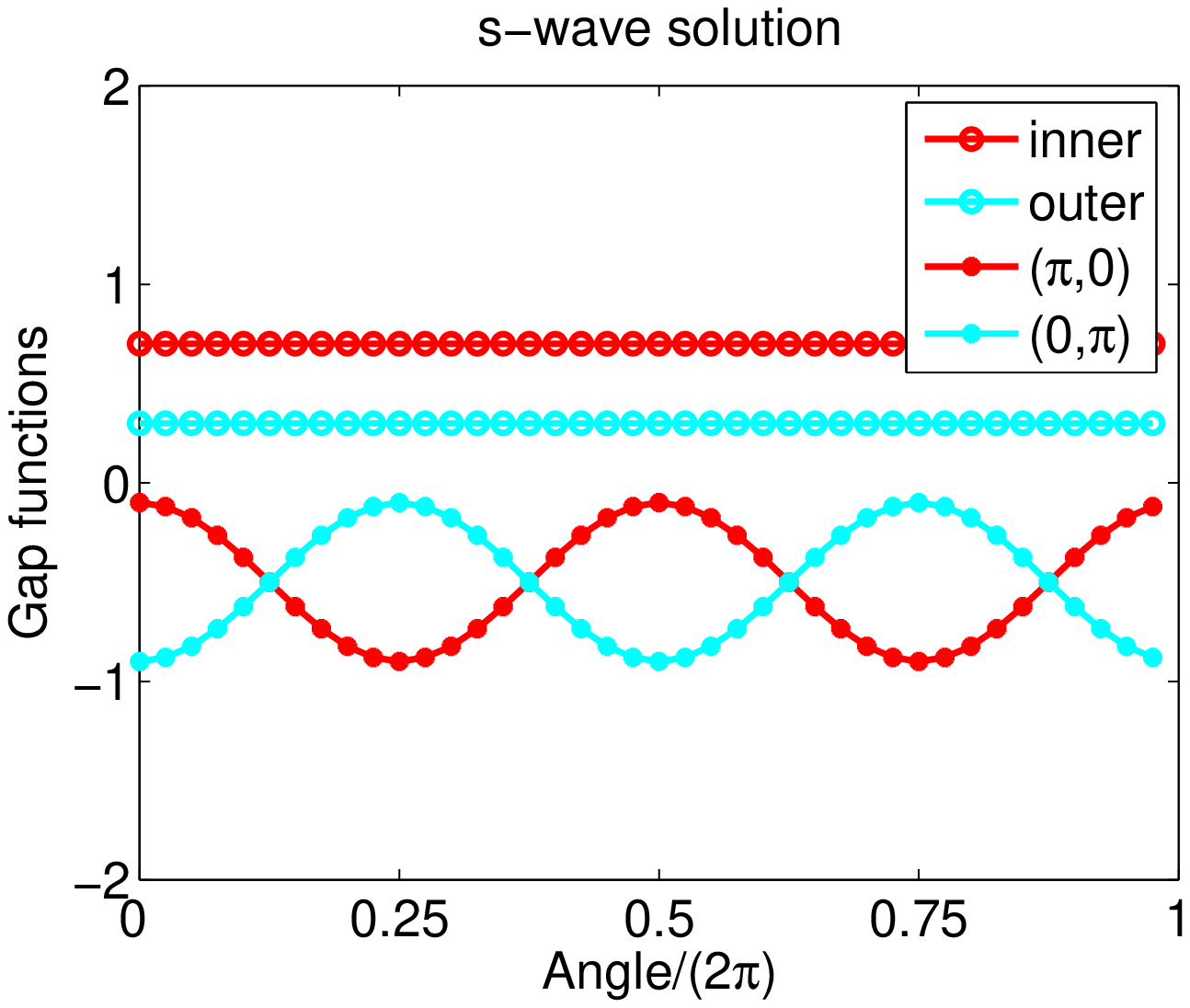}
\end{array}$
\caption{\label{fig:Max_change}Effect of electron-hole
interactions: $s$-wave solutions for the set \#1 (NSF) for
$\mu=0.08$ with the angular parts of electron-hole interactions
($\alpha_{h_ie}$) set to zero and increased by a factor of 2,
respectively (left and center). The values of $\lambda_s$ are
-0.16 and 0.226, positive $\lambda_s $ is for enhanced
$\alpha_{h_i e}$. In the right figure, we introduce additional
enhancement of the overall factors of the electron-hole
interactions by $\sim4$ and this causes nodes in the gap to
disappear.}
\end{figure*}

\begin{figure*}[t]
$\begin{array}{ccc}
\includegraphics[width=2in]{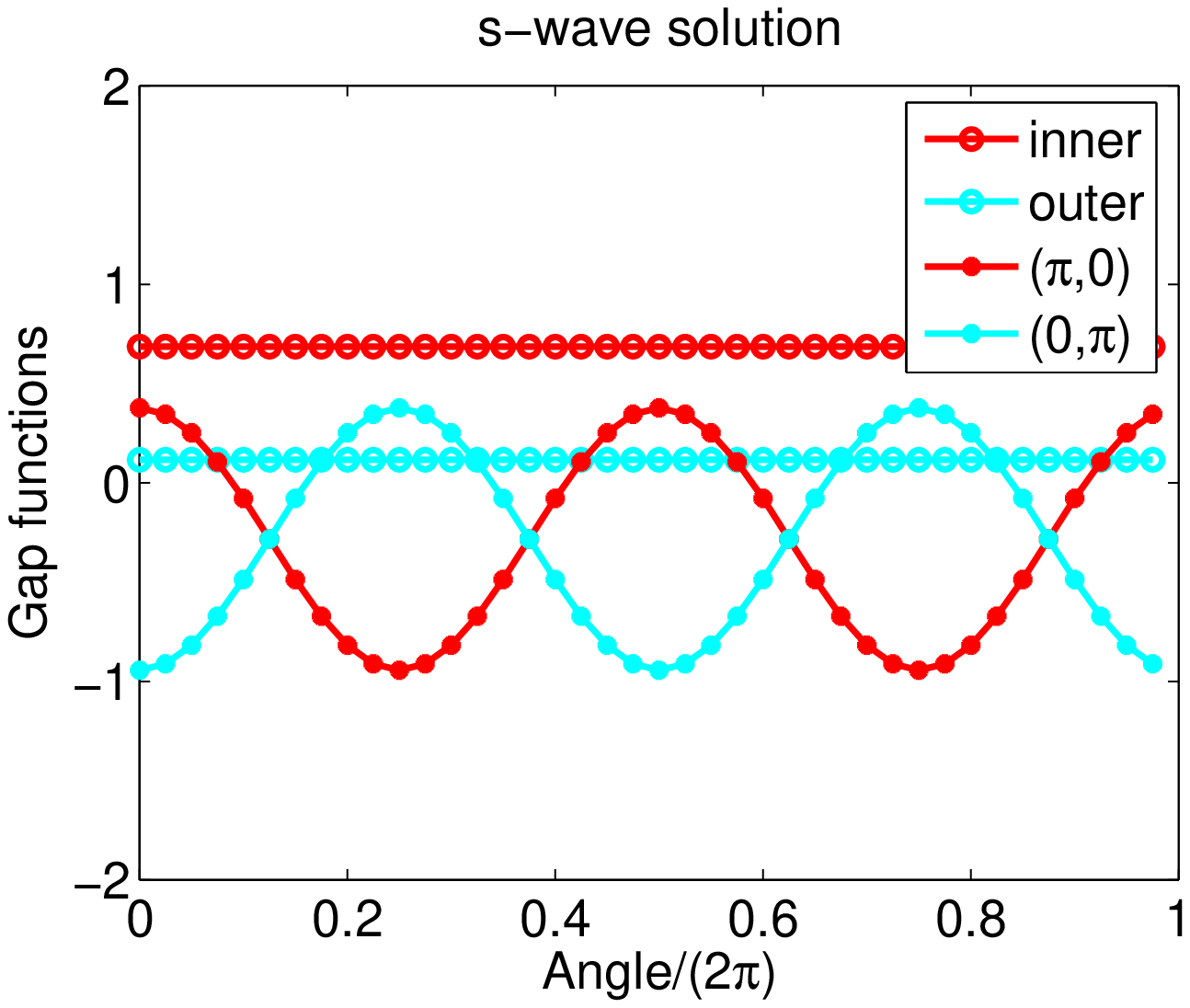}&
\includegraphics[width=2in]{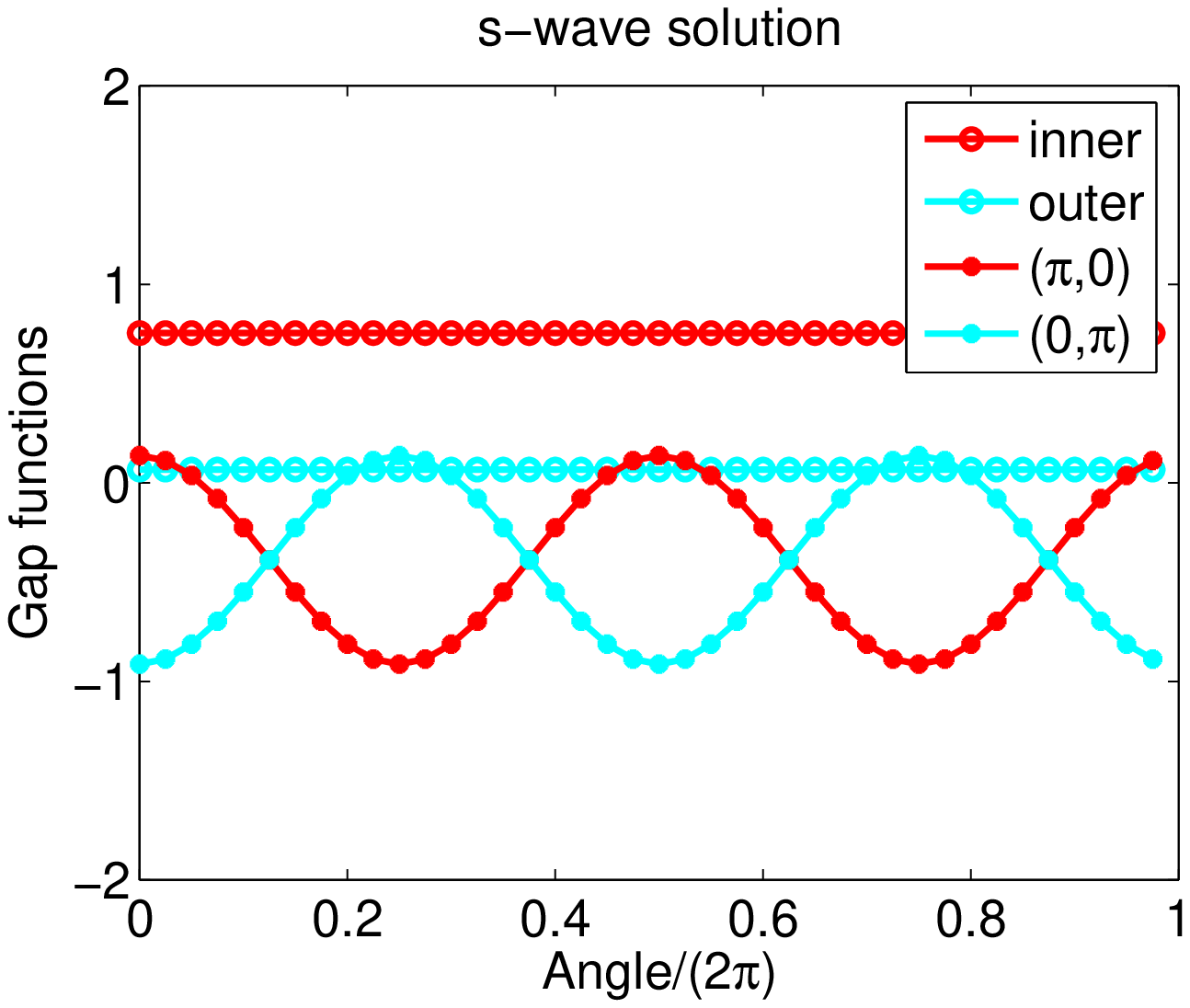}&
\includegraphics[width=2in]{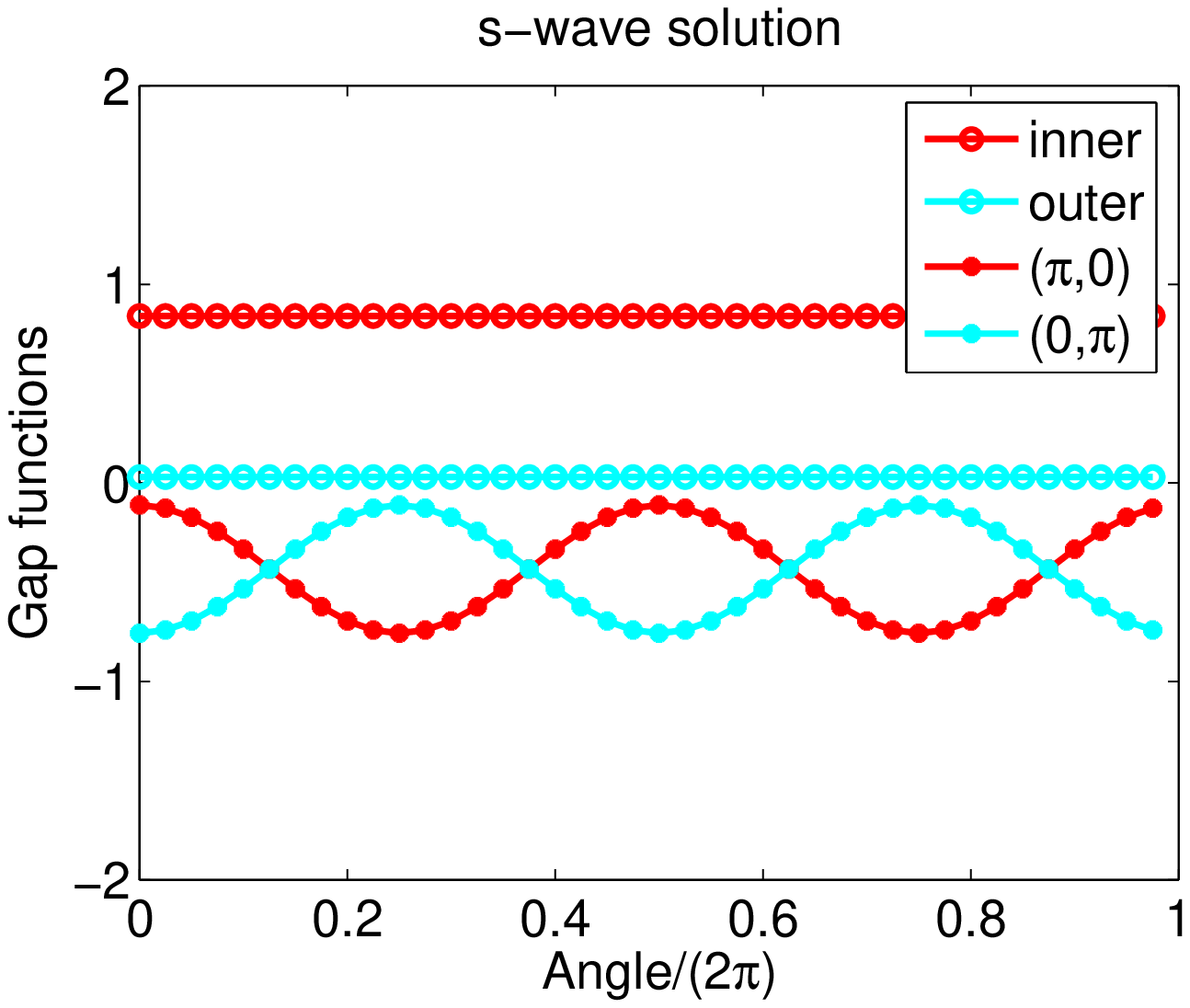}\\
\includegraphics[width=2in]{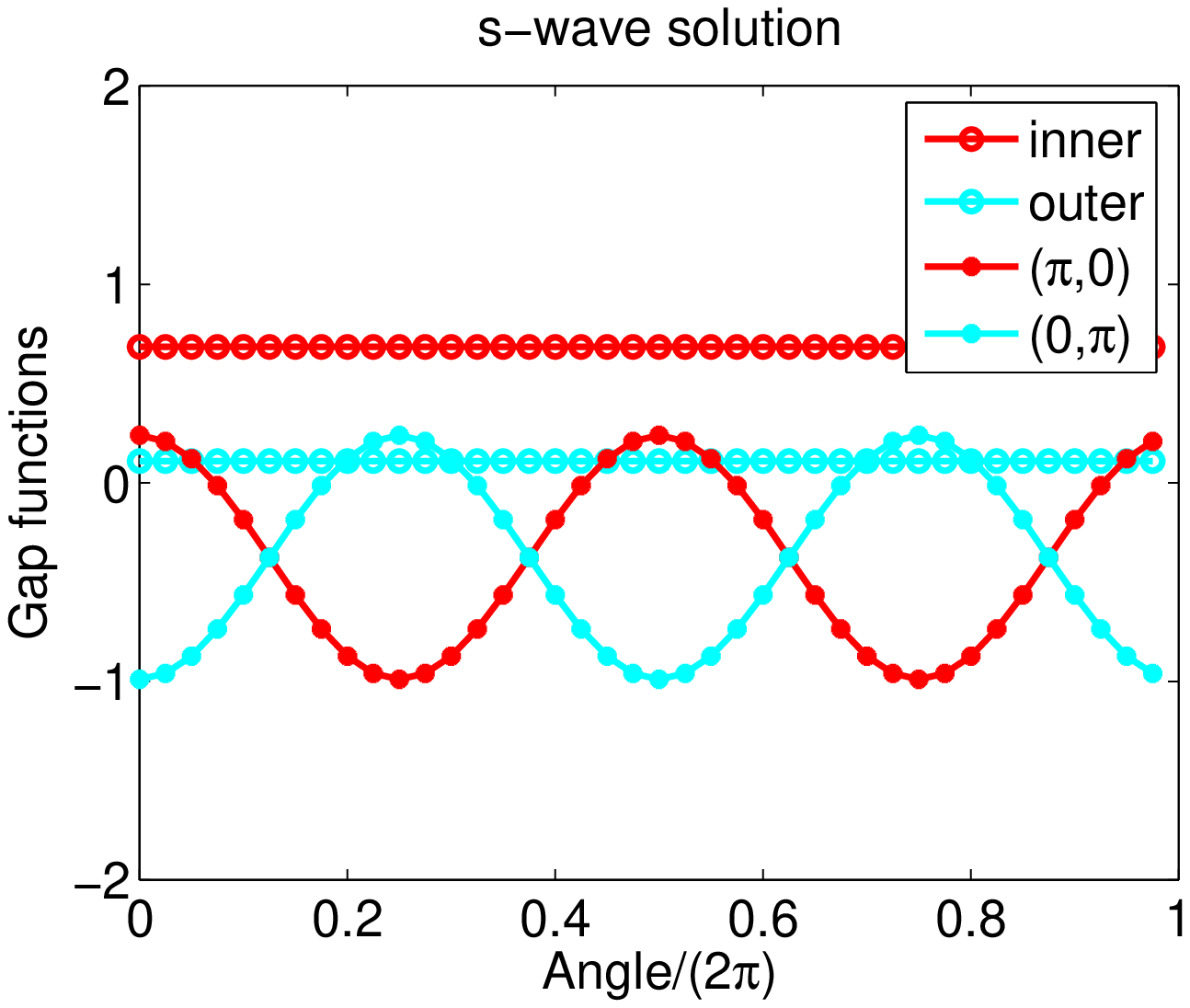}&
\includegraphics[width=2in]{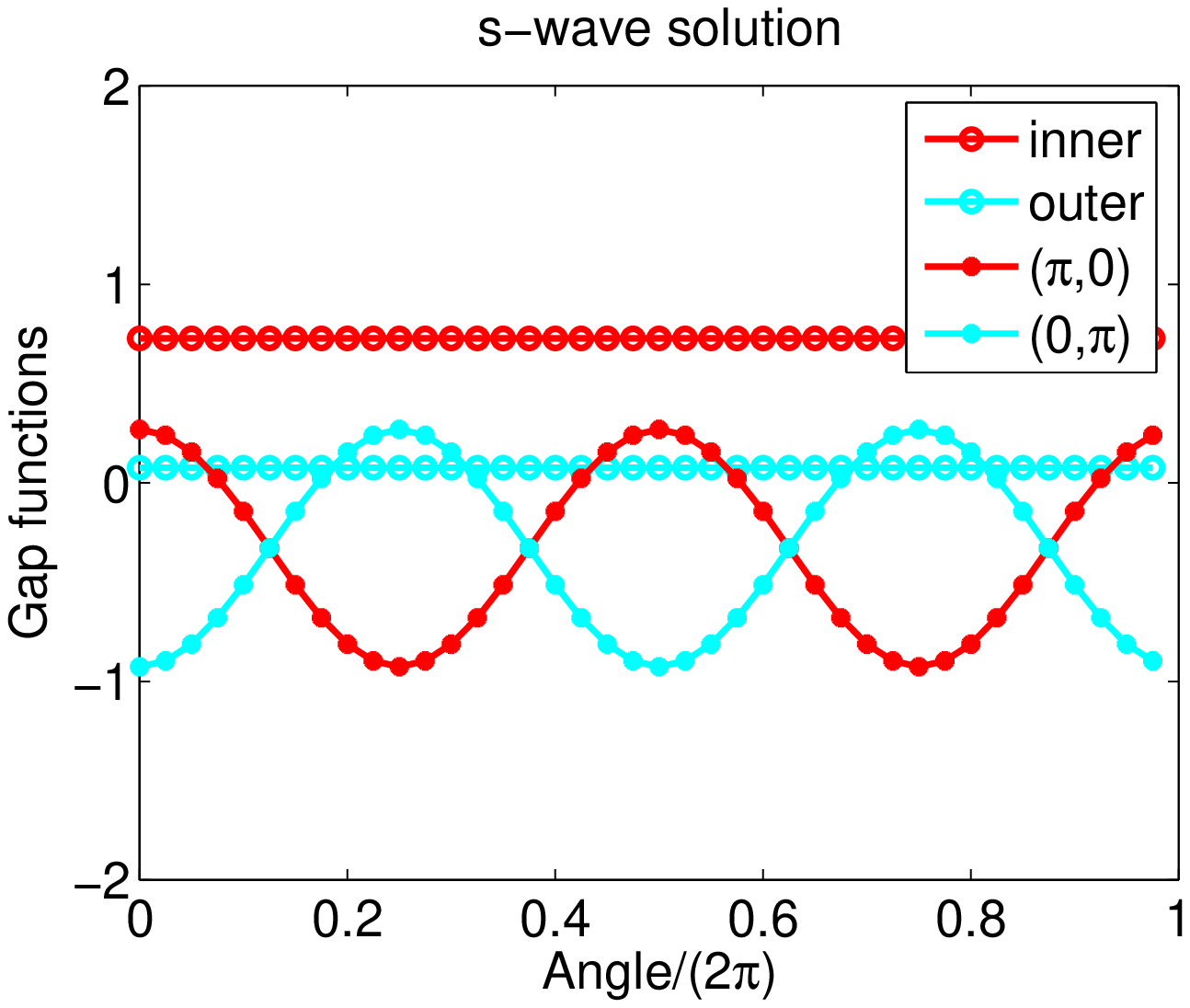}&
\includegraphics[width=2in]{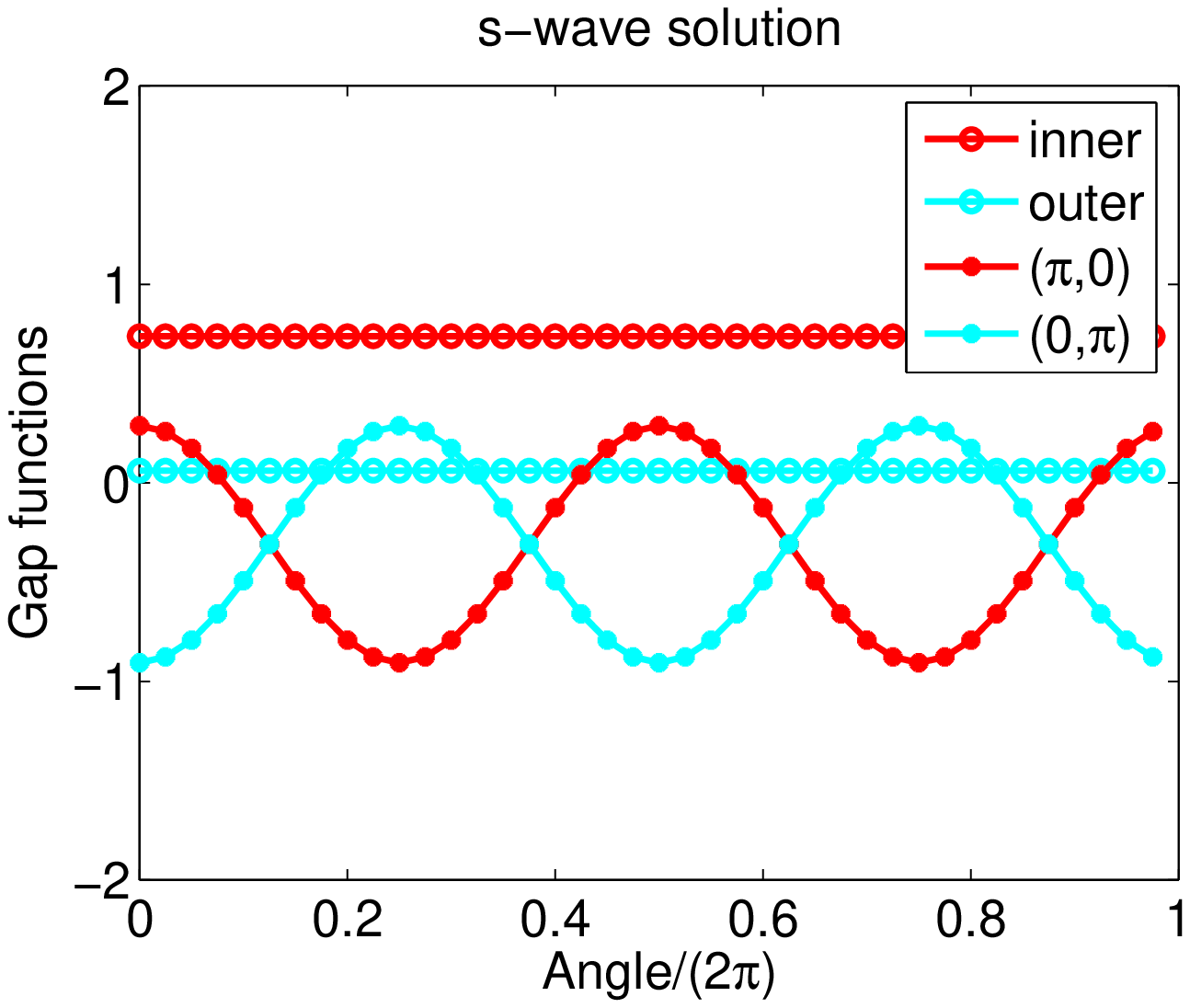}
\end{array}$
\caption{\label{fig:tom_change} The dependence of the gap
structure for $s$-wave solution  on the strength of
angle-dependent parts of the interactions. We used set  \#1(SF)
for $\mu=0.08$. The solution for the original $\alpha_{h_ie}$  is
shown in Fig. \ref{fig:gaps_wo_RPA}. Top panel: the gap structure
for the cases when the angular parts of electron-hole interaction
($\alpha_{h_ie}$) are increased by factor of 2, reduced by factor
of 2, and set to zero, while the angular parts of
electron-electron interaction  ($\alpha_{ee}$ and $\beta_{ee}$)
are kept at original values. The values of $\lambda_s$ are 7.3,
3.4, and 2.5 respectively. The angular dependence do change, and,
in particular, the nodes along the hole FS disappear when we set
$\alpha_{h_ie} =0$. Bottom panel: the same but now we keep
$\alpha_{h_ie}$ at original value and change the angular parts of
the electron-electron interactions ($\alpha_{ee}$ and $\beta_{ee}$
terms):  increase them by 2, reduce by 2, and set to zero,  The
$\lambda$'s are 4.5,4.7 and 4.0 respectively. We see that this has
very little effect on the structure of the gap.}
\end{figure*}

In Figs.~\ref{fig:Max_change} and~\ref{fig:tom_change} we show the
results for the $s^\pm$ gap for different values of
$\alpha_{h_ie}$. Comparing these figures with
Fig.~\ref{fig:gaps_wo_RPA}, we see that the effect of
$\alpha_{h_ie}$ on the gap is different for bare and full
interaction. For the bare interaction, originally there is no
$s$-wave solution with positive $\lambda_s$, but it appears once
we increase $\alpha_{h_ie}$ above a certain threshold. This can be
easily understood by  analyzing $4\times 4$ gap equation: for bare
interaction intra-pocket repulsions $u_{h_ih_j}$ and $u_{ee}$ are
stronger than inter-pocket $u_{h_ie}$, hence the pairing can only
be induced by angle-dependent components of the interaction, when
the factor $S$,  given by Eq.~(\ref{7}), becomes positive. For
original parameters $S <0$, but once we increase $\alpha_{h_ie}$,
$S$ eventually changes sign and the solution with $\lambda_s >0$
appears. The gap function for this induced solution  has strong
oscillations along electron FSs and has accidental nodes.

Consider next the full interaction with SF component. $u_{h_ie}$
are now enhanced,  and the solution with $\lambda_s>0$ exists even
if we set $\alpha_{h_ie} =0$ (see Fig.~\ref{fig:tom_change}). The
only difference between the solutions with small and larger
$\alpha_{h_ie}$ is that in the first case $s^\pm$ gap has no nodes
on the two electron FSs. We see therefore that for full
interaction   the role of $\alpha_{h_ie}$ is merely to modify
already existing $s^\pm$ solution and add angular variation to the
gap along the two electron FSs.

We next analyze how sensitive is the solution for the gap to the
change of the angular component of electron-electron interaction
$(\alpha_{ee}$ and $\beta_{ee}$ terms). In Fig.
\ref{fig:tom_change}, bottom panel, we show the results for the
gaps obtained with $\alpha_{ee}$ and $\beta_{ee}$ increased by
two, reduced by two, and set to zero. Comparing this figure with
Fig.~\ref{fig:gaps_wo_RPA}(bottom panel) we see that the changes
in $\alpha_{ee}$ and $\beta_{ee}$ lead to very little changes in
the gap structure (and the $\lambda$s). The gap remains very much
the same as in Fig.~\ref{fig:gaps_wo_RPA}, even if we set
$\alpha_{ee}$ and $\beta_{ee}$ to zero.  The implication is that,
for full interaction,  the gap structure is determined by the
angular dependence of electron-hole interaction, while
electron-electron interactions can be we approximated by
angle-independent $u_{ee}$ term.

\subsubsection*{Nodal vs non-nodal $s$-wave gap for 4 FSs}

Previous studies of the gap structure for 4 FSs within RPA/SF
formalism yielded  the gap with accidental nodes along the
electron FS~\cite{tom_09,graser}. The $s^{\pm}$ solution without
nodes only appears for an electronic structure which contains an
additional, 5th hole FS centered at $(\pi,\pi)$. The results of
the previous subsection imply that this result is the likely
outcome, but is non-universal:  for the full interaction the gap
can be either nodeless or with nodes depending on the magnitude of
$\alpha_{he}$ and on the interplay between $u^2_{he}$ and $u_{ee}
u_{hh}$ (for simplicity, we set two hole FSs to be equal).  When
$u^2_{he} < (u_{ee}u_{hh})$, the only possibility is $s^{\pm}$
pairing with nodes along electron FSs, in the opposite case the
solution can be either nodal or nodeless depending on
$\alpha_{he}$.  The larger $u^2_{he}/(u_{ee}u_{hh})$ is, the
larger $\alpha_{he}$ one needs to make SC nodeless.

To further illustrate this, in Fig.~\ref{fig:Max_change}(right) we
show the solution of $4\times4$ set for the gap  for the case when
we increase $u_{h_1e}$ and $u_{h_2e}$ keeping all other
parameters, including  $\alpha_{h_ie}$, fixed. We see that the
nodes on electron FSs disappear for substantially large $u_{h_1e}$
and $u_{h_2e}$.


In RPA/SF analysis, a way to further increase $u_{h_1e}$ compared
to other parameters is to bring the system even closer to AFM
instability. For the parameter set we are dealing with, one needs
to increase $u_{h_ie}$ by quite substantial amount, so a fine
tuning to AFM instability is required to eliminate the nodes. But
still, no-nodal solution definitely exists in some range of
parameters.

\subsection{$d$-wave case}

The consideration for the $d$-wave case proceeds similar to the
$s$-wave case. The $d$-wave pairing again  can be due to
attraction coming from angle-independent parts of the interaction,
or it can be induced by angle-dependent parts of the interactions,
even when $\lambda_d <0$ in the absence of angle-dependent terms.
We refrain from discussing all cases as in many respects the
analysis for $d$-wave pairing parallels the one for the $s$-wave
case. We point out, however, that the negative $\lambda_d$ for the
case of bare interaction (NSF) is much smaller in magnitude than
the coupling in the $s$-wave channel, i.e., there is ``less
repulsion'' in the $d$-wave channel. As a result, it takes less to
convert $d$-wave repulsion into attraction. This can be achieved
by shifting the values ${\tilde \alpha}_{h_ie}$,   ${\tilde
\alpha}_{re}$, and  ${\tilde \beta}_{ee}$. Once SF piece is added
to the interaction all  $d$-wave components increase, but, just as
for $s$-wave case, the largest increase is for electron-hole
interaction.  This increase is large enough such that the gap
equation develops a solution with $\lambda_d >0$ even if we set
angular components of electron-hole and electron-electron
interactions to zero. In the latter case, the gaps along the two
electron FSs are $\pm \Delta_e$.  When we don't set
angle-dependent components to zero, electron gaps acquire $\cos 2
\theta$ components, but, as we see in Fig. \ref{fig:gaps_wo_RPA},
this component is quite small.

We also note that the values of $\lambda_s$ and $\lambda_d$ for
the full interaction are quite close. For the set \#1 which we are
discussing in this section, we found that, within LAHA,
$\lambda_d$ is actually a bit larger than $\lambda_s$, for the
parameters extracted from the fit, but $\lambda_d$ and $\lambda_s$
are very close and $\lambda_s$ becomes larger already after a
small change of parameters.  The solution for the gap within RPA
also shows that, for this set, $\lambda_s$ and $\lambda_d$ are
very close, but it yields $\lambda_s \geq \lambda_d$. In any
event, however, $s$-wave coupling $\lambda_s$ definitely becomes
stronger than $\lambda_d$ once the system comes close enough to an
antiferromagnetic instability.

\section{Doping evolution of $s$-wave and $d$-wave gap functions \label{sec:Doping}}

To understand the evolution with doping, we choose set \#2 for
definiteness and consider the gap structure for positive and
negative values of the chemical potential.  Positive $\mu$
corresponds to electron doping and negative $\mu$ correspond to
hole doping.  For positive $\mu$, electronic structure consists of
two hole and two electron pockets, and the size of hole pockets
gets smaller as $\mu$ increases.  For negative $\mu$, the
electronic structure contains an additional, $5$th hole pocket,
centered at $(\pi,\pi)$.  As before we shall denote the electron
pocket at $(\pi,0)$ as $e_1$, the electron pocket at $(0,\pi)$ as
$e_2$, the inner hole pocket at $(0,0)$ as $h_1$, the outer hole
pocket at $(0,0)$ as $h_2$ and the hole pocket at $(\pi,\pi)$ when
present as $h_3$. In this section we only consider moderate
doping, when both hole and electron FSs are present. We consider
the limiting case of large electron and hole dopings in the next
section. For simplicity, we only present results for full
interaction with the SF component.

\begin{figure*}[htp]
$\begin{array}{ccc}
\includegraphics[width=2.4in]{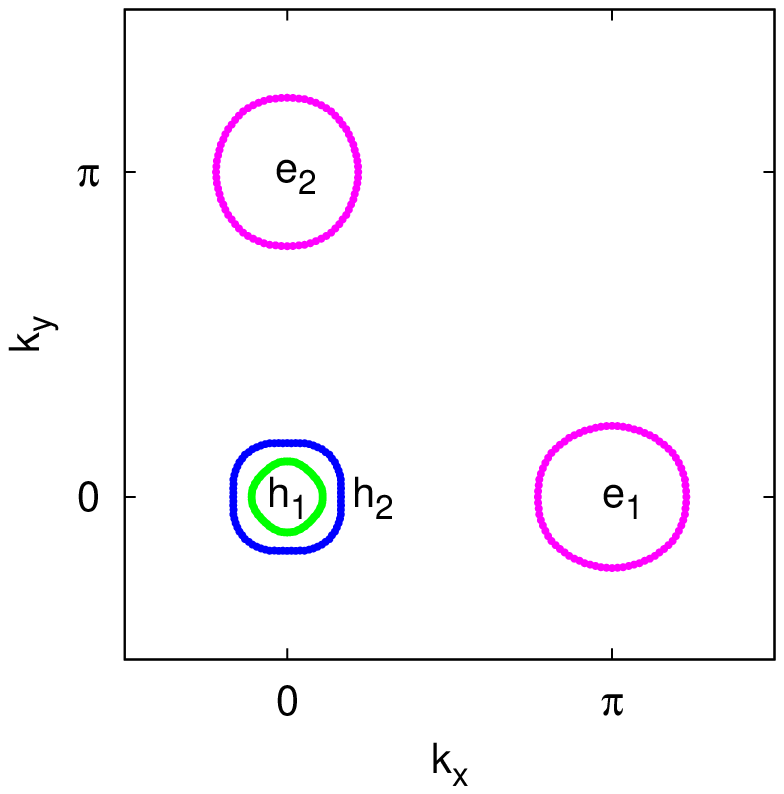}&
\includegraphics[width=2.4in]{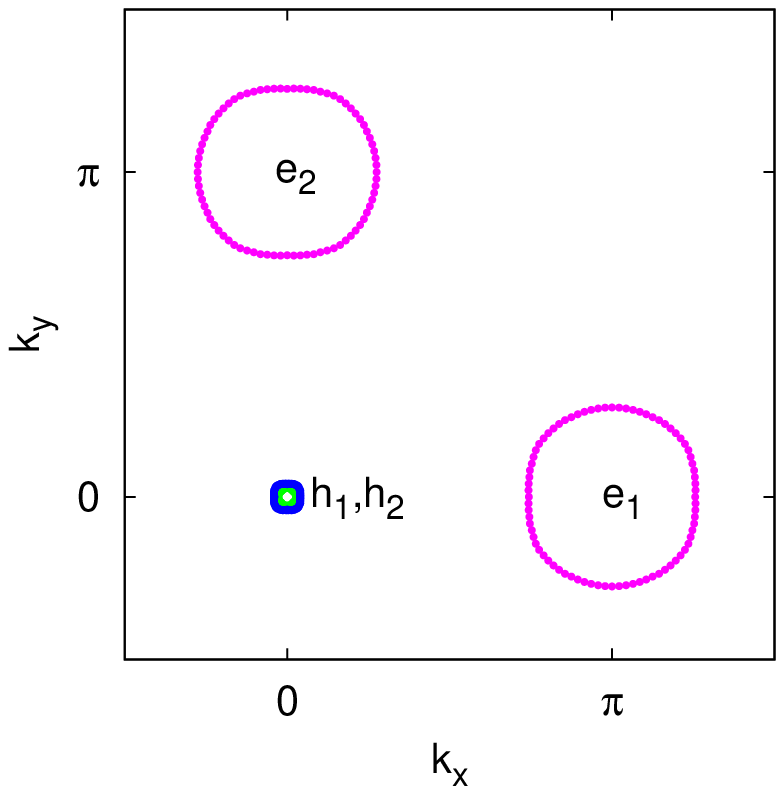}&
\includegraphics[width=2.4in]{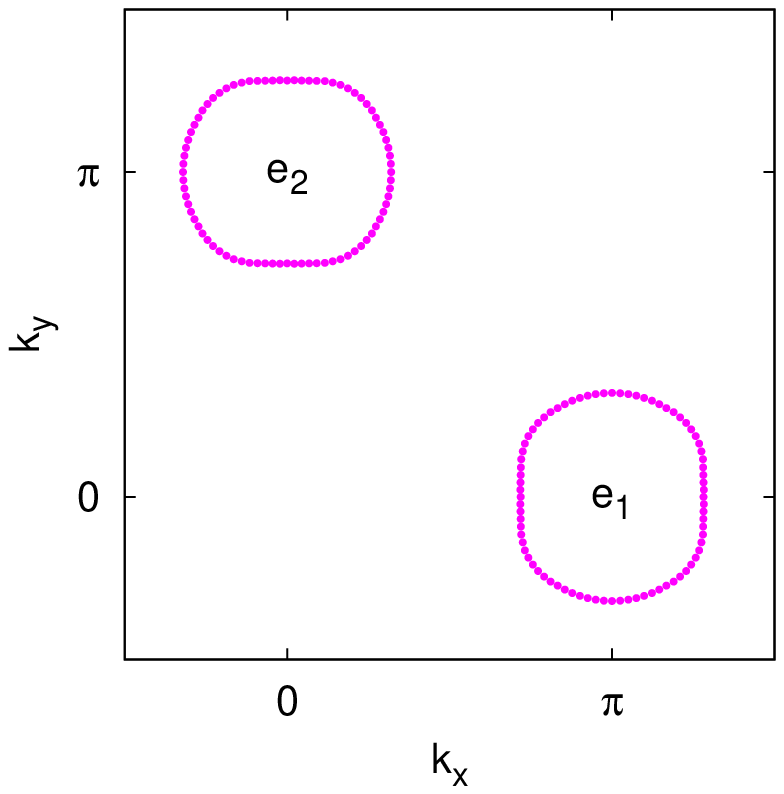}\\
\includegraphics[width=2.4in]{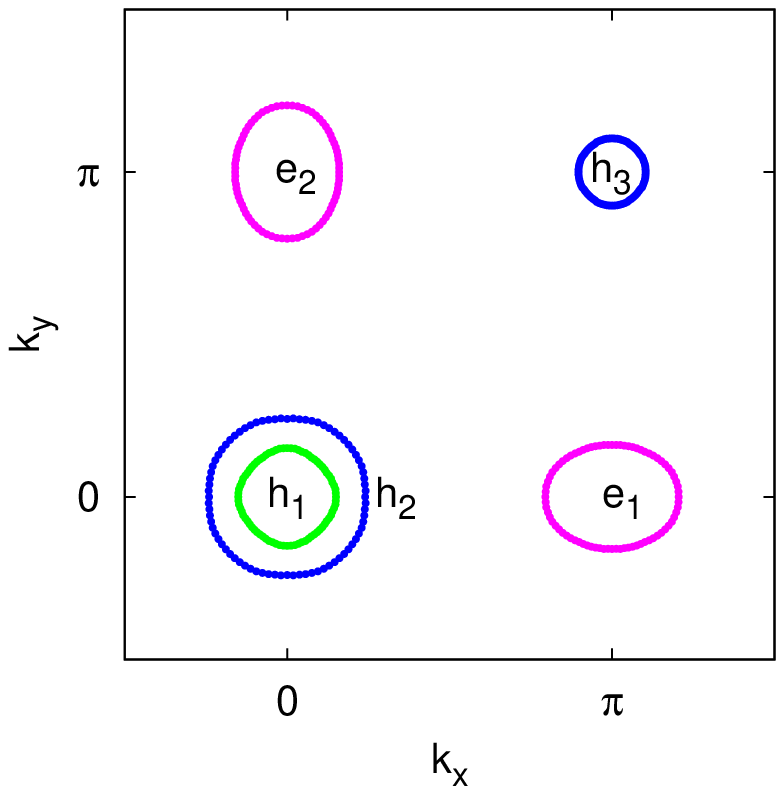}&
\includegraphics[width=2.4in]{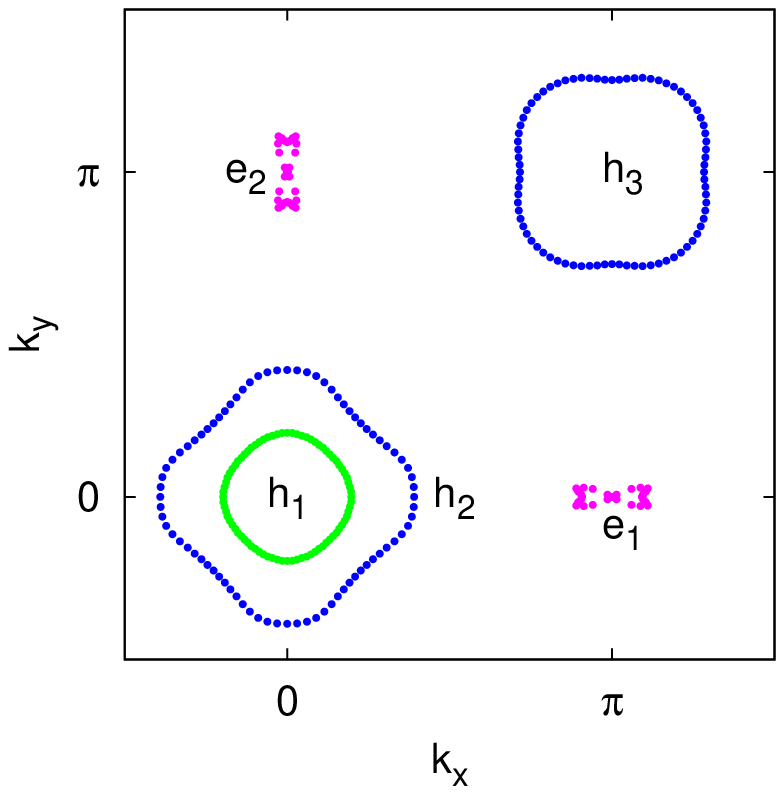}&
\includegraphics[width=2.4in]{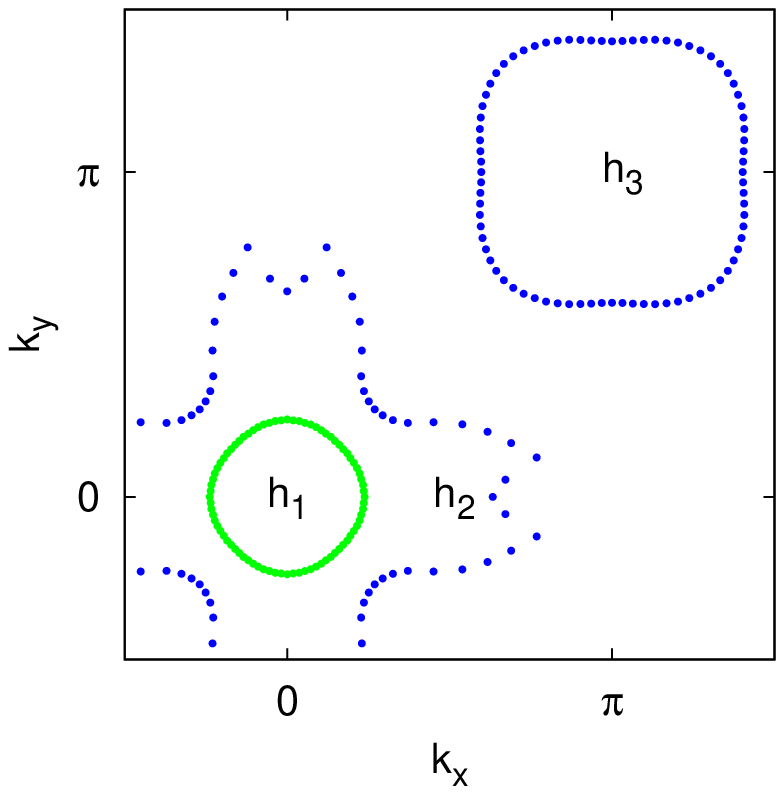}
\end{array}$
\caption{\label{fig:FS_mu_0.05and0.18} (Top: left to right) FSs
for increasing electron doping with $\mu=0.05,~0.18,~0.30$
respectively. $\mu=0.18$ represents the electron doping at which
the hole FS almost disappear, while $\mu=0.30$ represents extreme
electron doping where the hole FS completely disappear.
(Bottom-left to right) FSs for increasing hole doping with
$\mu=-0.05,~-0.18,~-0.30$ respectively. Here electron FSs almost
disappear at $\mu=-0.18$ and completely disappear at $\mu=-0.30$}
\end{figure*}

\begin{table*}[htp]
\caption{$s$- and $d$-wave parameters for the set \#2 with
$\mu=0.05$} \label{tab:3}
\begin{ruledtabular}
\begin{tabular}{cccccccccccccc}
$s$-wave&$u_{h_1h_1}$&$u_{h_2h_2}$&$u_{h_1h_2}$&$u_{h_1e}$&$\alpha_{h_1 e}$&$u_{h_2e}$&$\alpha_{h_2 e}$&$u_{ee}$&$\alpha_{ee}$&$\beta_{ee}$\\
\hline
NSF&$0.51$&$0.50$&$0.50$&$0.32$&$-0.19$&$0.28$&$-0.24$&0.50&0.11&0.08\\
SF&$0.80$&$0.79$&$0.79$&$0.79$&$-0.19$&$0.67$&$-0.19$&0.91&0.05&0.05\\
\hline
$d$-wave&$\tilde{u}_{h_1 h_1}$&$\tilde{u}_{h_2 h_2}$&$\tilde{u}_{h_1 h_2}$&$\tilde{u}_{h_1 e}$&$\tilde{\alpha}_{h_1 e}$&$\tilde{u}_{h_2 e}$&$\tilde{\alpha}_{h_2 e}$&$\tilde{u}_{ee}$&$\tilde{\alpha}_{ee}$&$\tilde{\beta}_{ee}$\\
\hline
NSF&$0.38$&$0.36$&$-0.37$&$-0.14$&$-0.60$&$0.14$&$-0.60$&0.05&-0.6&0.35\\
SF&$0.50$&$0.49$&$-0.50$&$-0.39$&$-0.46$&$0.30$&$-0.47$&-0.04&1.5&-0.69\\
\end{tabular}
\end{ruledtabular}
\end{table*}
\begin{table*}[htp]
\caption{$s$- and $d$-wave parameters for the set \#2 with
$\mu=0.18$.} \label{tab:1}
\begin{ruledtabular}
\begin{tabular}{cccccccccccccc}
$s$-wave&$u_{h_1h_1}$&$u_{h_2h_2}$&$u_{h_1h_2}$&$u_{h_1e}$&$\alpha_{h_1 e}$&$u_{h_2e}$&$\alpha_{h_2 e}$&$u_{ee}$&$\alpha_{ee}$&$\beta_{ee}$\\
\hline
NSF&$0.56$&$0.55$&$0.56$&$0.30$&$-.23$&$0.29$&$-0.23$&0.52&0.12&0.06\\
SF&$0.75$&$0.75$&$0.74$&$0.67$&$-0.19$&$0.67$&$-0.20$&0.88&0.10&0.05\\
\hline
$d$-wave&$\tilde{u}_{h_1 h_1}$&$\tilde{u}_{h_2 h_2}$&$\tilde{u}_{h_1 h_2}$&$\tilde{u}_{h_1 e}$&$\tilde{\alpha}_{h_1 e}$&$\tilde{u}_{h_2 e}$&$\tilde{\alpha}_{h_2 e}$&$\tilde{u}_{ee}$&$\tilde{\alpha}_{ee}$&$\tilde{\beta}_{ee}$\\
\hline
NSF&$0.43$&$0.43$&$-0.43$&$-0.14$&$-0.62$&$0.14$&$-0.62$&0.04&-0.625&0.44\\
SF&$0.51$&$0.51$&$-0.51$&$-0.32$&$-0.50$&$0.32$&$-0.50$&-0.05&0.9&-0.6
\end{tabular}
\end{ruledtabular}
\end{table*}
\begin{table*}[htp]
\caption{$s$- and $d$-wave parameters for the set \#2 with
$\mu=-0.05$.} \label{tab:5}
\begin{ruledtabular}
\begin{tabular}{cccccccccccccccc}
$s$-wave&$u_{h_1h_1}$&$u_{h_2h_2}$&$u_{h_3h_3}$&$u_{h_1h_2}$&$u_{h_1
h_3}$&$u_{h_2 h_3}$&$u_{h_1e}$&$\alpha_{h_1
e}$&$u_{h_2e}$&$\alpha_{h_2 e}$&$u_{h_3 e}$&$\alpha_{h_3
e}$&$u_{ee}$&$\alpha_{ee}$&$\beta_{ee}$\\
\hline
NSF&$0.50$&$0.49$&$1.00$&$0.49$&$0.20$&$0.13$&$0.34$&-0.18&0.30&-0.25&$0.61$&$0.26$&0.49&0.11&0.10\\
SF&$0.86$&$0.96$&$1.83$&$0.89$&$0.45$&$0.32$&$0.92$&-0.18&0.79&-0.21&$1.5$&$0.21$&1.00&0.11&0.08\\
\hline $d$-wave&$\tilde{u}_{h_1 h_1}$&$\tilde{u}_{h_2
h_2}$&$\tilde{u}_{h_3 h_3}$&$\tilde{u}_{h_1 h_2}$&$\tilde{u}_{h_1
h_3}$&$\tilde{u}_{h_2 h_3}$&$\tilde{u}_{h_1
e}$&$\tilde{\alpha}_{h_1 e}$&$\tilde{u}_{h_2
e}$&$\tilde{\alpha}_{h_2 e}$&$\tilde{u}_{h_3
e}$&$\tilde{\alpha}_{h_3
e}$&$\tilde{u}_{ee}$&$\tilde{\alpha}_{ee}$&$\tilde{\beta}_{ee}$\\
\hline
NSF&$0.36$&$0.36$&$0$&$-0.36$&$0$&$-0$&$-0.15$&-0.58&0.15&-0.58&$-0$&$-0.58$&0.06&-0.58&0.33\\
SF&$0.51$&$0.61$&$0.01$&$-0.56$&$0.00$&$-0.01$&$-0.45$&-0.48&0.39&-0.43&$0.02$&$0.90$&0.07&-1.00&0.46
\end{tabular}
\end{ruledtabular}
\end{table*}
\begin{table*}[htp]
\caption{$s$- and $d$-wave parameters for $\mu=-0.18$. For
technical reasons we used $U=0.9$, $J=0$, and $V =0.9$.}
\label{tab:5_1}
\begin{ruledtabular}
\begin{tabular}{cccccccccccccccc}
$s$-wave&$u_{h_1h_1}$&$u_{h_2h_2}$&$u_{h_3h_3}$&$u_{h_1h_2}$&$u_{h_1
h_3}$&$u_{h_2 h_3}$&$u_{h_1e}$&$\alpha_{h_1
e}$&$u_{h_2e}$&$\alpha_{h_2 e}$&$u_{h_3 e}$&$\alpha_{h_3
e}$&$u_{ee}$&$\alpha_{ee}$&$\beta_{ee}$\\
\hline
NSF&$0.37$&$0.37$&$0.74$&$0.36$&$0.04$&$0.10$&$0.40$&$0.0$&$0.40$&$0.0$&$0.04$&$0.0$&$0.44$&$0.0$&$0.0$\\
SF&$0.75$&$2.02$&$17.2$&$0.98$&$-0.08$&$0.41$&$1.36$&0.08&2.86&0.02&$0.31$&$-0.01$&1.40&0.01&0.06\\
\hline $d$-wave&$\tilde{u}_{h_1 h_1}$&$\tilde{u}_{h_2
h_2}$&$\tilde{u}_{h_3 h_3}$&$\tilde{u}_{h_1 h_2}$&$\tilde{u}_{h_1
h_3}$&$\tilde{u}_{h_2 h_3}$&$\tilde{u}_{h_1
e}$&$\tilde{\alpha}_{h_1 e}$&$\tilde{u}_{h_2
e}$&$\tilde{\alpha}_{h_2 e}$&$\tilde{u}_{h_3
e}$&$\tilde{\alpha}_{h_3
e}$&$\tilde{u}_{ee}$&$\tilde{\alpha}_{ee}$&$\tilde{\beta}_{ee}$\\
\hline
NSF&$0.36$&$0.36$&$0$&$-0.36$&$-0.04$&$0.04$&$-0.40$&-0.0&0.40&-0.0&$0.04$&$-0.0$&0.44&-0.0&0.0\\
SF&$0.70$&$1.94$&$13.6$&$-0.94$&$0.04$&$0.33$&$-1.32$&0.0&2.85&0.02&$0.26$&$0.02$&1.45&0.01&0.04
\end{tabular}
\end{ruledtabular}
\end{table*}


\subsection{Electron doping}

\begin{figure*}[htp]
$\begin{array}{cc}
\includegraphics[width=2in]{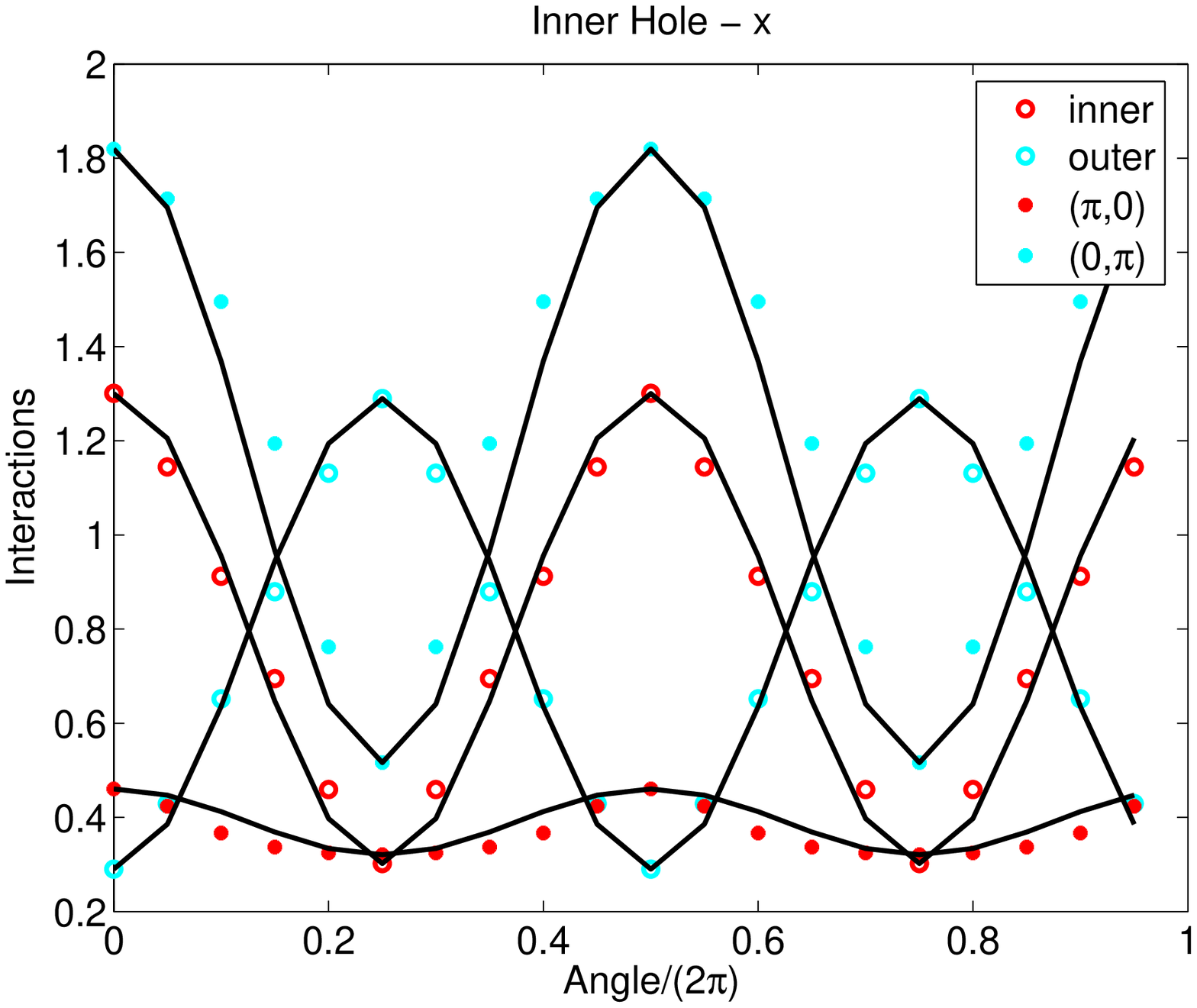}&
\includegraphics[width=2in]{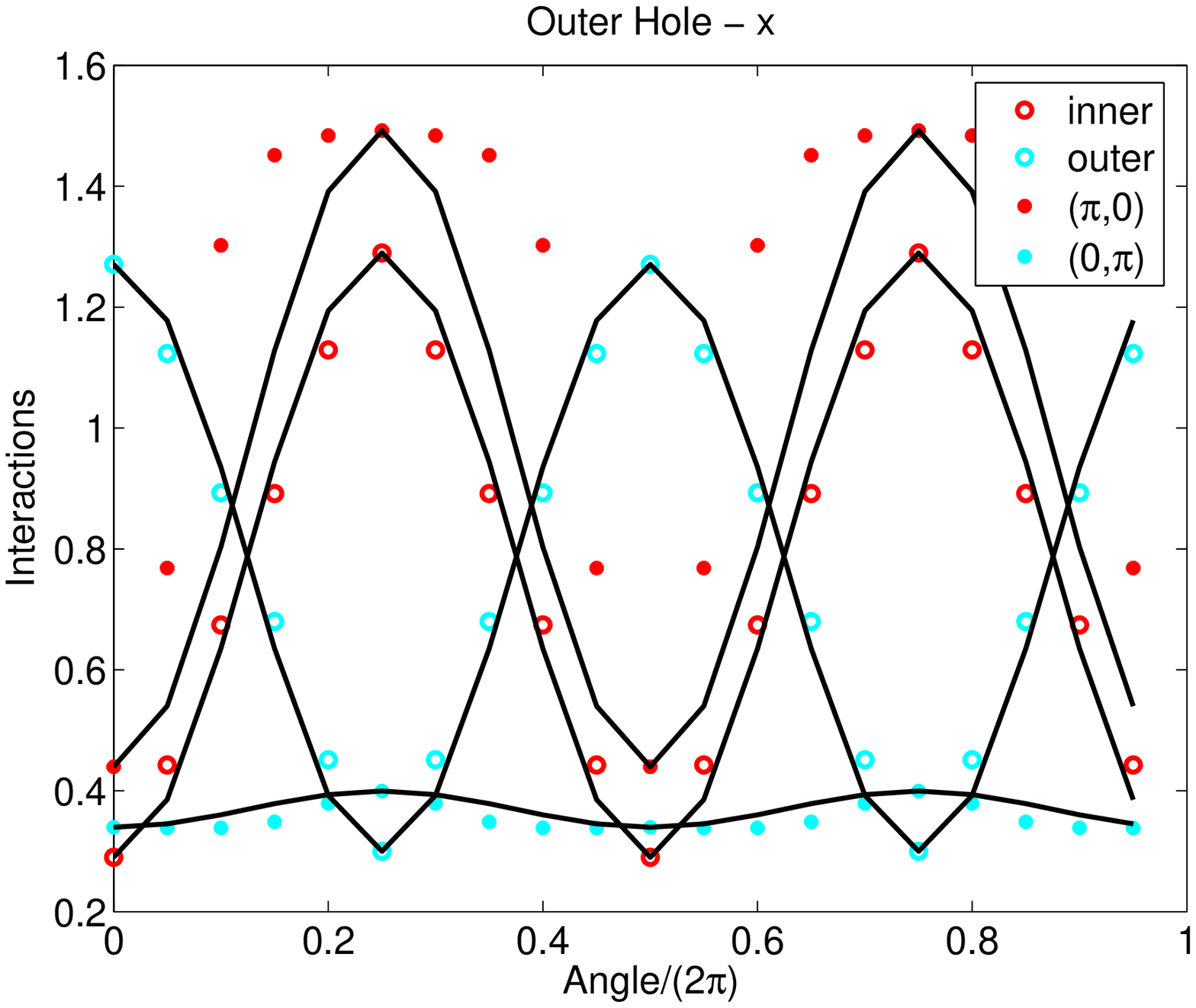}\\
\includegraphics[width=2in]{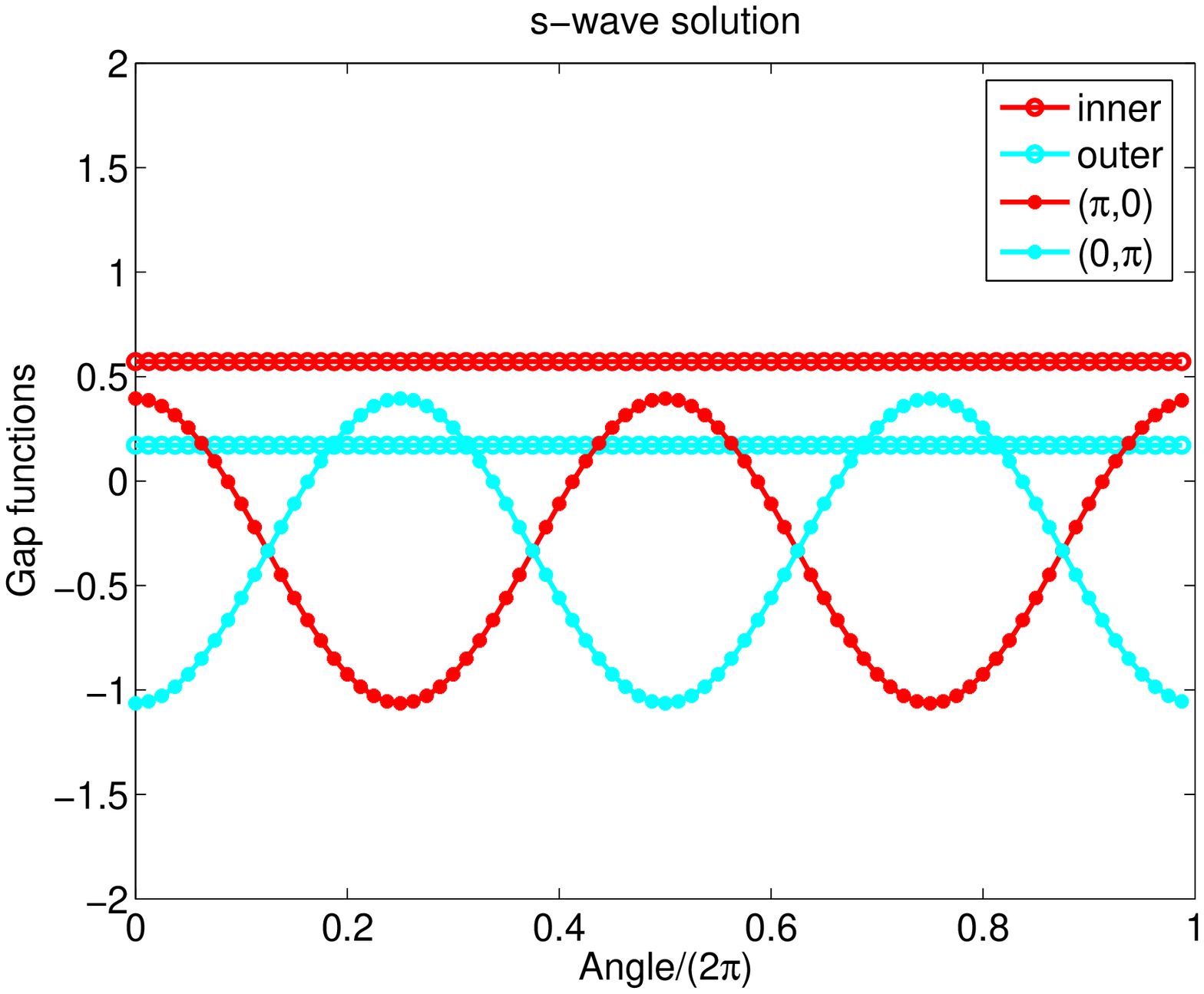}&
\includegraphics[width=2in]{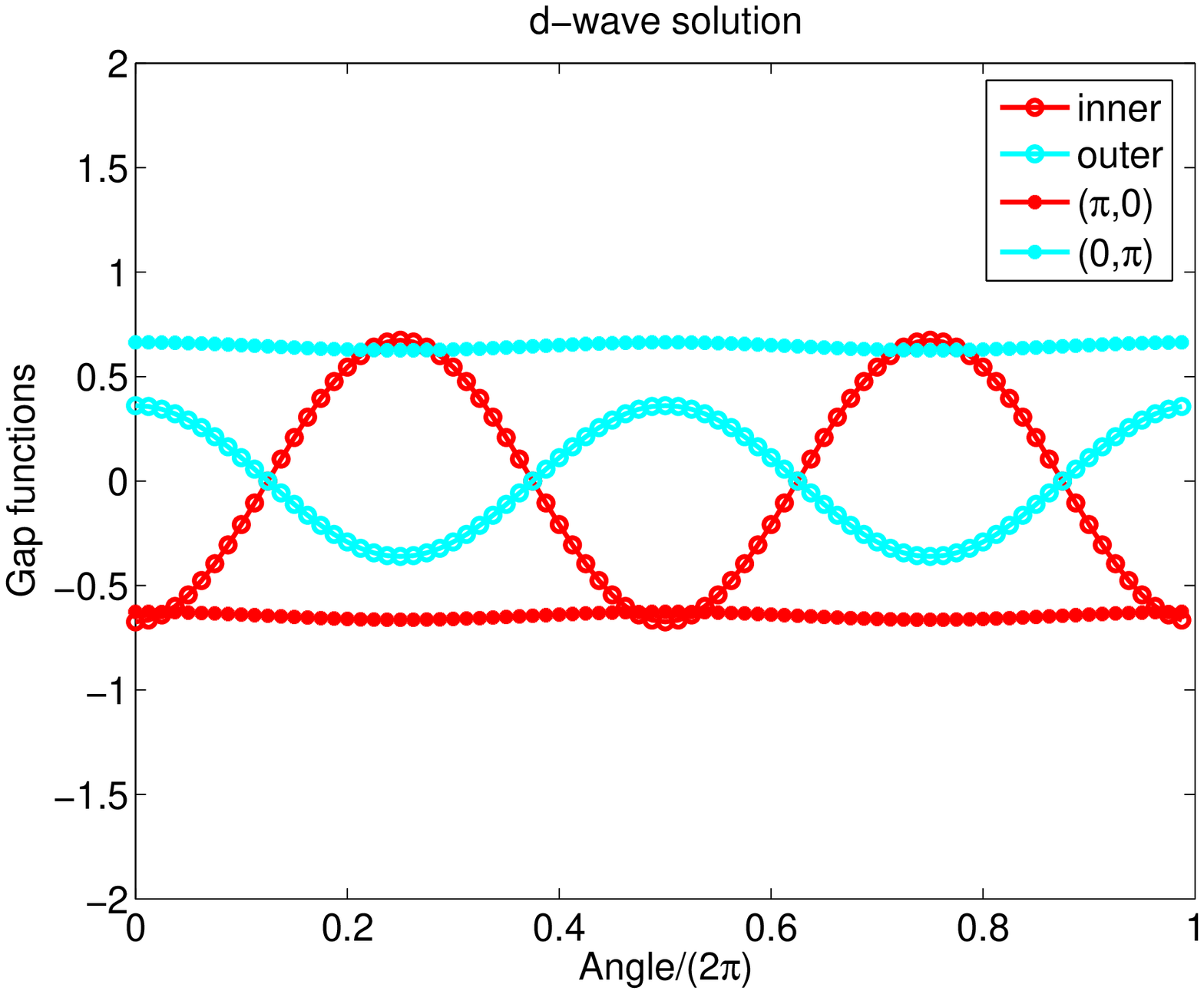}
\end{array}$
\caption{\label{fig:SF_0p05}
Top panel. The `best' and `worst' LAHA fits of the interactions
$\Gamma(\k_F, \k'_F)$ obtained from RPA/SF calculations for parameter
set \#2(SF) for $\mu=0.05$. Bottom panel: Gap-functions in
$s$- and $d$-channels obtained in the LAHA.
$\lambda_s = 0.25$, $\lambda_d = 0.37$. Note that $\lambda_d$ is somewhat larger.}
\end{figure*}

We considered several values of electron doping (positive $\mu$).
Below we present the results for the representative case of
$\mu=0.05$ ($n_e=6.09$). The FS is presented in
Fig.~\ref{fig:FS_mu_0.05and0.18} In Table~\ref{tab:3} we present
LAHA $s$-wave and $d$-wave interaction components from
Eq.~(\ref{eq:interactions}), obtained by fitting RPA/SF results.
The actual fits are presented in Fig.~\ref{fig:SF_0p05}, where we
present the ``best'' and the ``worst'' fits. We see that even the
``worst'' fits are actually quite reasonable. In the same figure
we show $s$-wave and $d$-wave gaps corresponding to the largest
$\lambda_s$ and $\lambda_d$.


The tables show the same trends that we discussed in the previous
section, the most relevant one is the overall increase of all
interactions when SF component is added, and further increase of
electron-hole interaction in both $s$-wave and $d$-wave channels.
In this Section we focus on the features associated with the
dependence on doping. For briefness, we show only the case of full
interaction with SF component.

First, we see from Fig.~\ref{fig:SF_0p05} that for positive $\mu$
(when the electronic structure consists of 4FSs),  $s$-wave
solution with $\lambda_s >0$ has nodes on the electron FSs. The
$d$-wave solution with $\lambda_d >0$ has symmetry-related nodes
on the hole FSs, but no nodes on electron FSs. That $\lambda_d >0$
is the combination of the two effects: (i) the $d$-wave component
of the interaction between electron pockets is negative (${\tilde
u}_{ee} <0$), i.e., there is a direct $d$-wave attraction between
electron pockets, and (ii) $d$-wave components of electron-hole
interaction ${\tilde u}_{h_1e}$ and ${\tilde u}_{h_2e}$ are quite
large. Both effects give rise to $\lambda_d >0$ even if we set all
$\cos 2 \theta$ components of the interactions to zero.  Within
LAHA, we can check the relative importance  of the two effects by
artificially setting one of them to zero. We see from
Tables~\ref{tab:3} and \ref{tab:1} that ${\tilde u}_{ee}$ is
attractive, but very weak. The primary reason for $\lambda_d >0$
are large values of electron-hole interaction ${\tilde u}_{h_1e}$
and ${\tilde u}_{h_2e}$. Accordingly, the driving force for
$d$-wave attraction is strong $d$-wave component of the
pair-hopping between electron and hole pockets.  In this respect,
the mechanism is quite similar to that for sign-changing $s$-wave
gap.



\begin{figure*}[htp]
$\begin{array}{cc}
\includegraphics[width=2in]{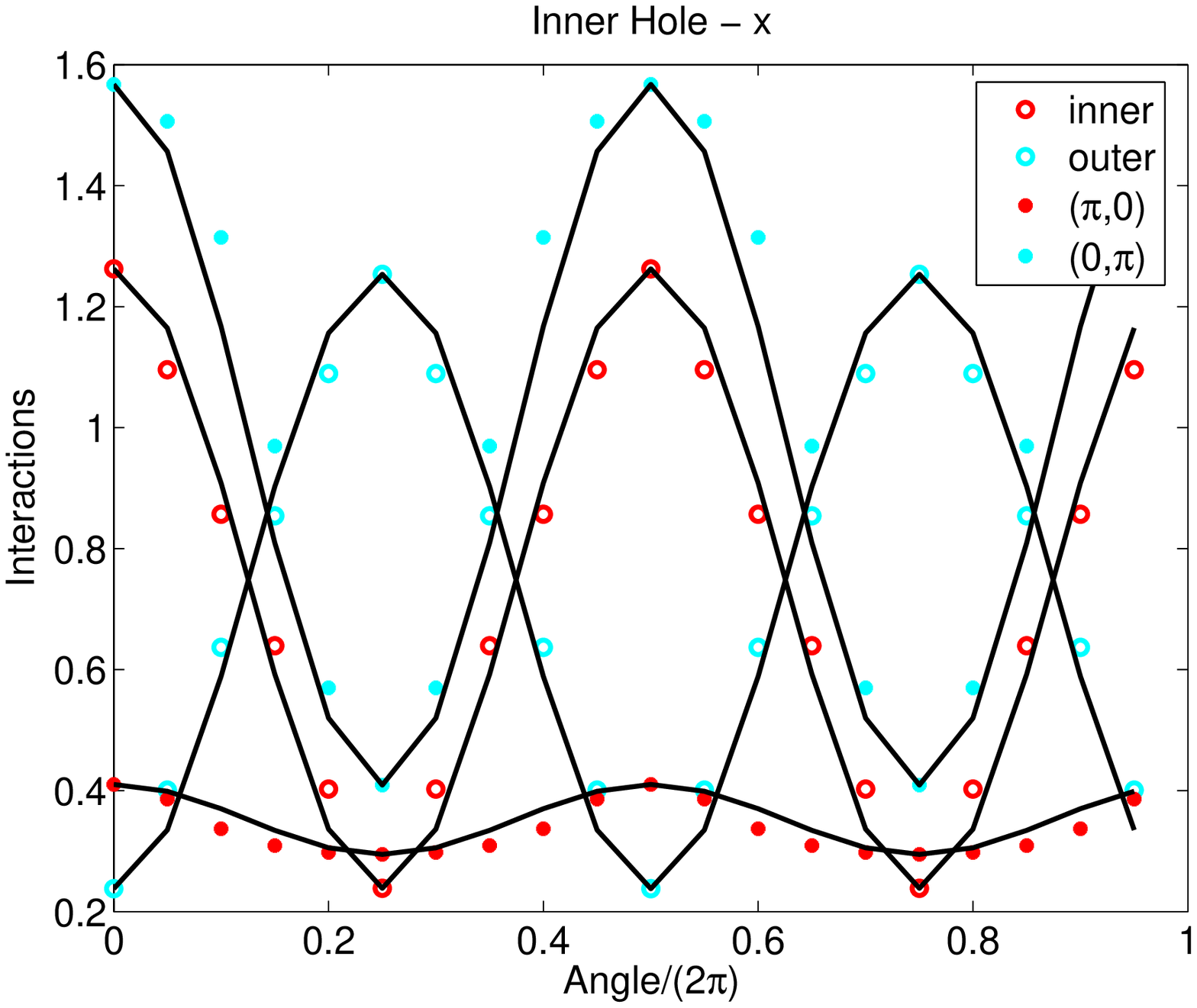}&
\includegraphics[width=2in]{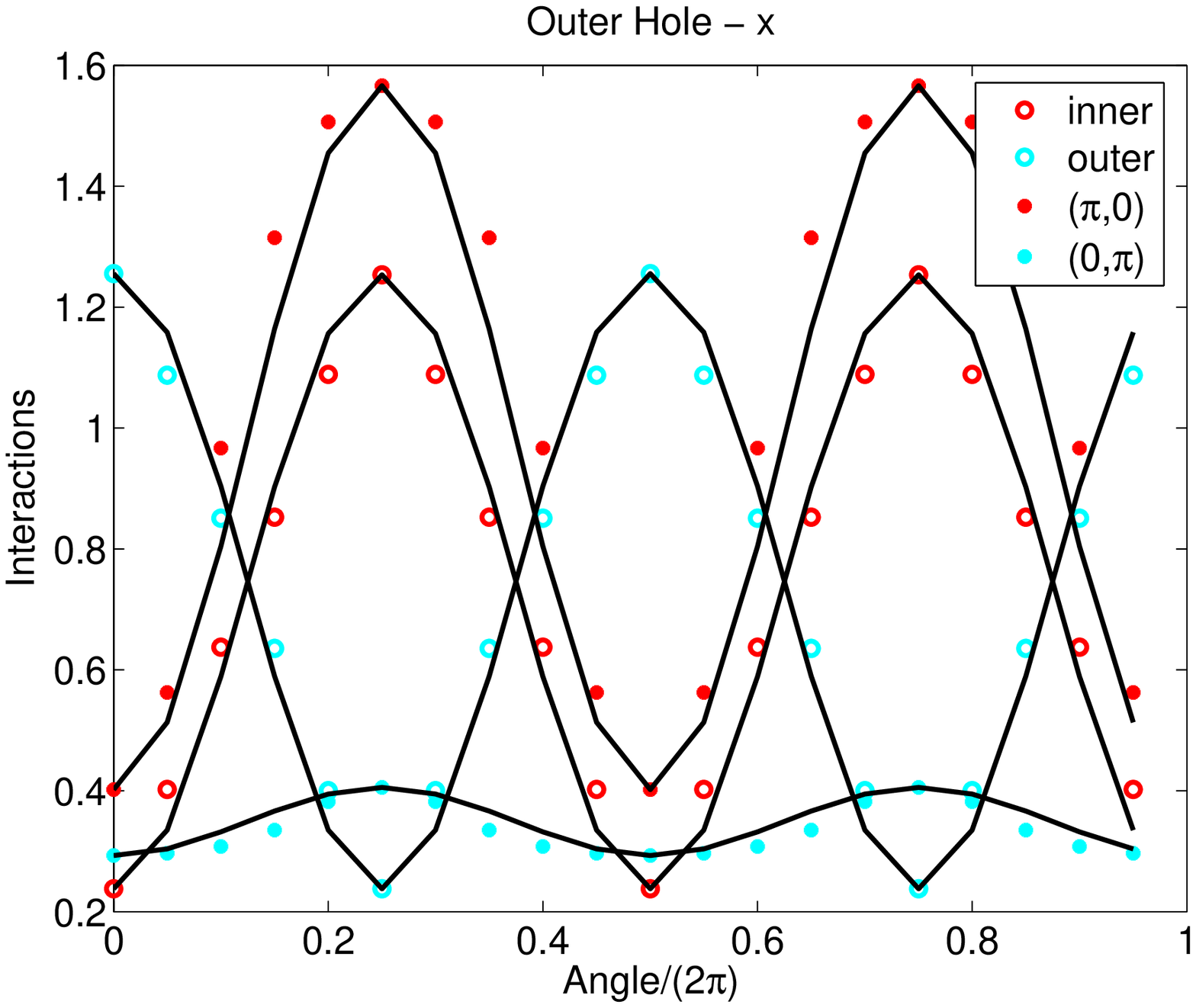}\\
\includegraphics[width=2in]{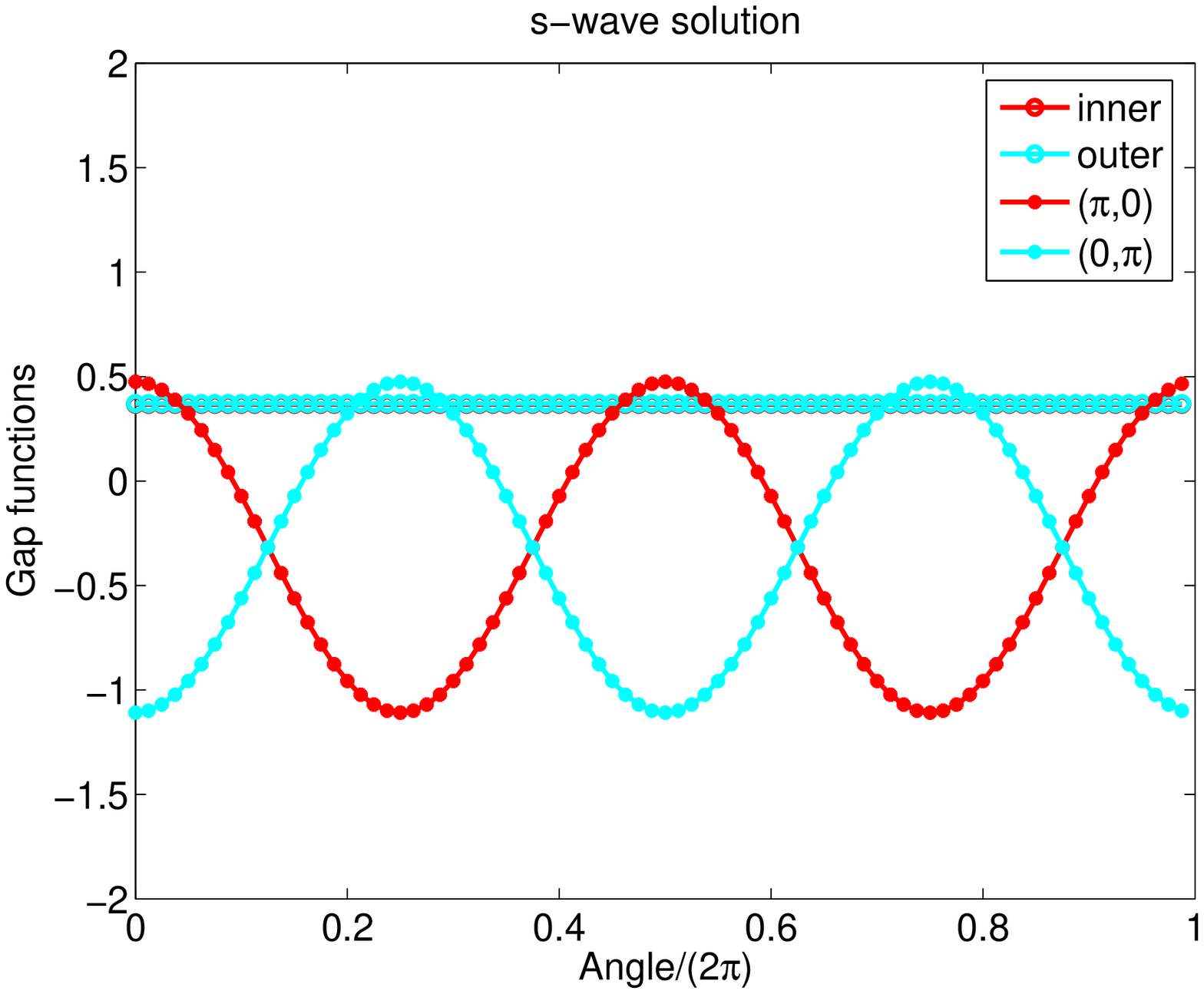}&
\includegraphics[width=2in]{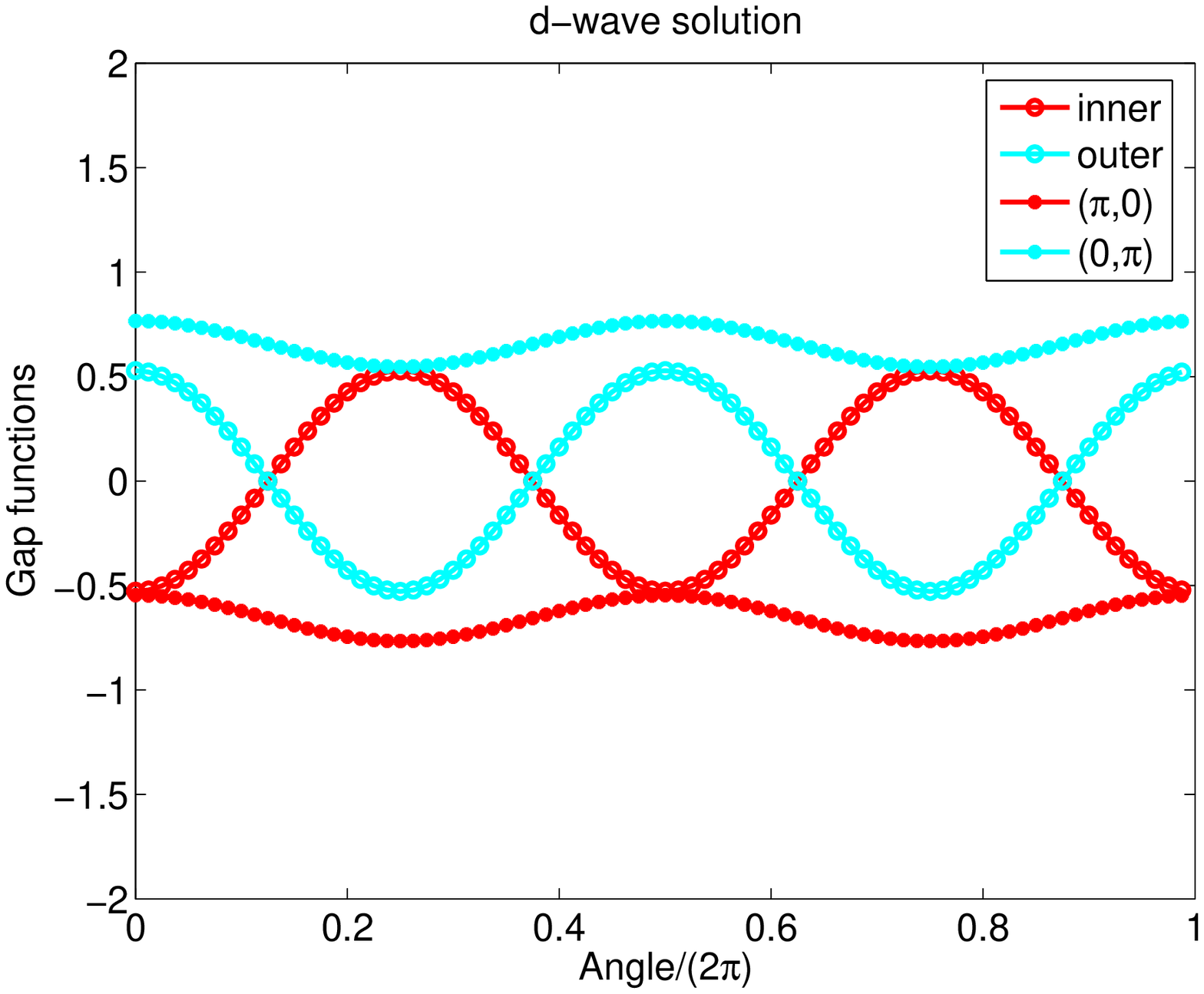}
\end{array}$
\caption{\label{fig:SF_0p18} The same as in Fig.
\ref{fig:SF_0p05}, but for different $\mu=0.18$. $\lambda_s =
0.21$, $\lambda_d = 0.35$. Again, $\lambda_d$ is larger, and the
difference between $\lambda_d$ and $\lambda_s$ is larger than for
$\mu=0.05$.}
\end{figure*}

How far in electron doping does this mechanism remain the leading
one? To analyze this, we also considered the case of a larger
$\mu=0.18$ ($n_e=6.23$), when hole pockets almost disappear. The
FS for this case is shown in Fig.~\ref{fig:FS_mu_0.05and0.18}, the
parameters extracted from the fit are shown in Table~\ref{tab:1},
and the fits and the gaps are presented in Fig.~\ref{fig:SF_0p18}.
A somewhat surprising result is that there is very little change
compared to the case of smaller $\mu=0.05$, when hole pockets are
much larger. Still, the dominant interaction in the $d$-wave
channel is the pair hopping between hole and electron FSs. The
$d$-wave component of electron-electron interaction is negative
(i.e., attractive), but it remains very small.

Note also that for both $\mu$'s $\lambda_d > \lambda_s$, and the
difference increases as electron doping increases. This does not
necessarily mean that $d$-wave is the leading instability because
$T_c$ for $s$-wave and $d$-wave superconducting instabilities have
different prefactors. Still, a larger value of $\lambda_d$ implies
that $d$-wave superconductivity is certainly a possibility in
electron-doped pnictides. A more exotic mixed $s+ id$ state is
also quite possible,~\cite{hanke} but to study it one obviously
needs to solve a non-linear gap equation, which is beyond the
scope of this work.

\begin{figure*}[htp]
$\begin{array}{cc}
\includegraphics[width=2in]{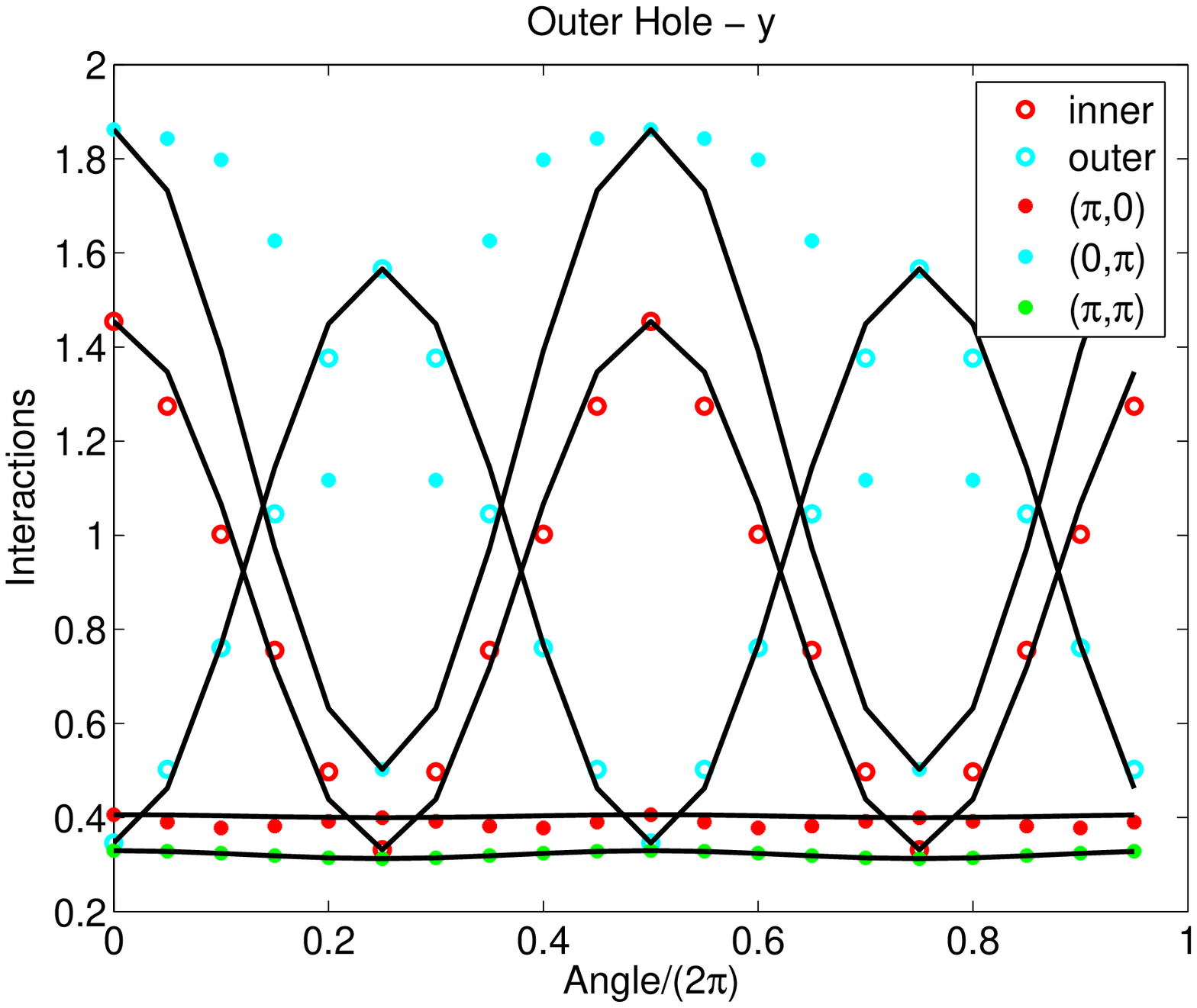}&
\includegraphics[width=2in]{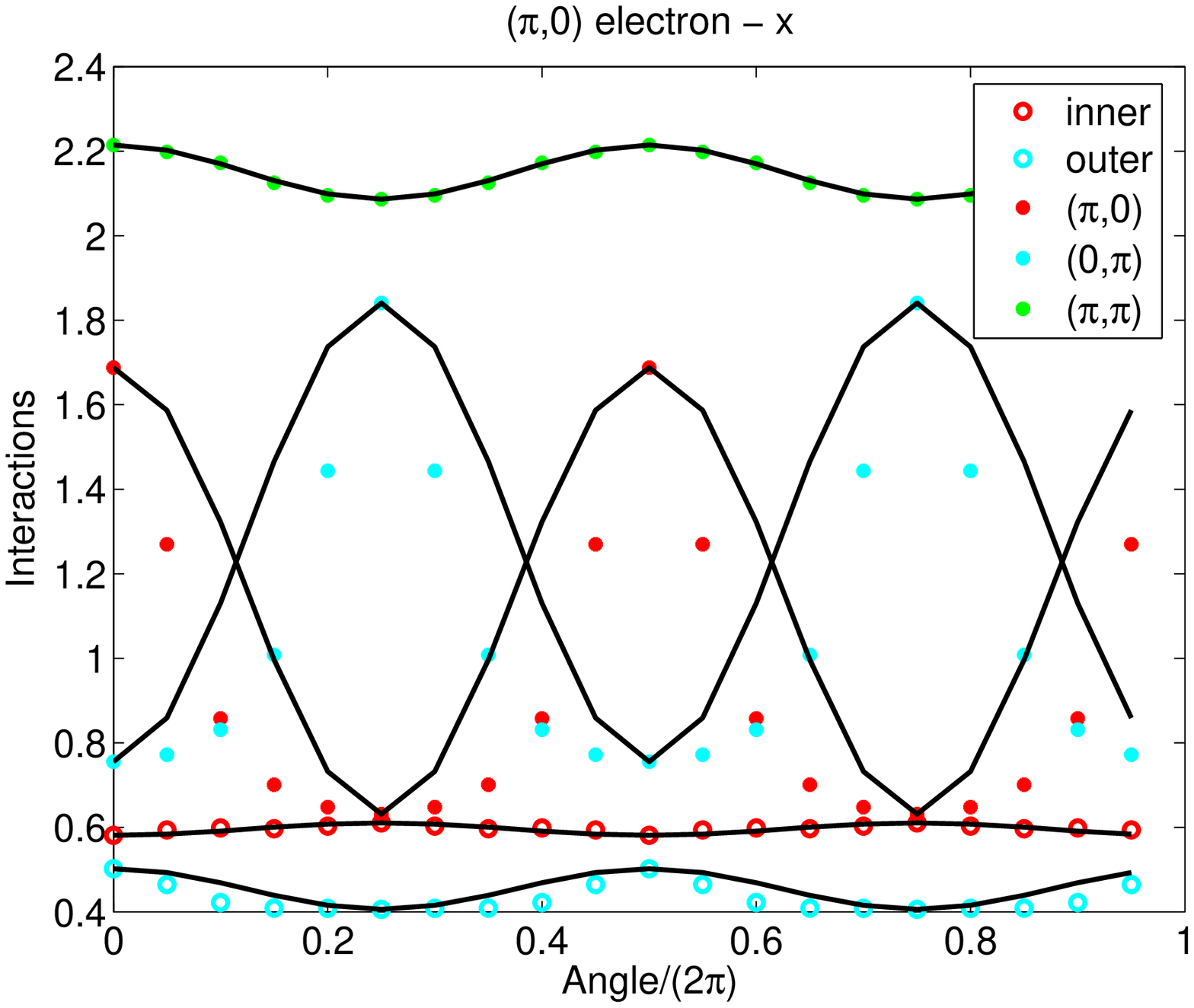}\\
\includegraphics[width=2in]{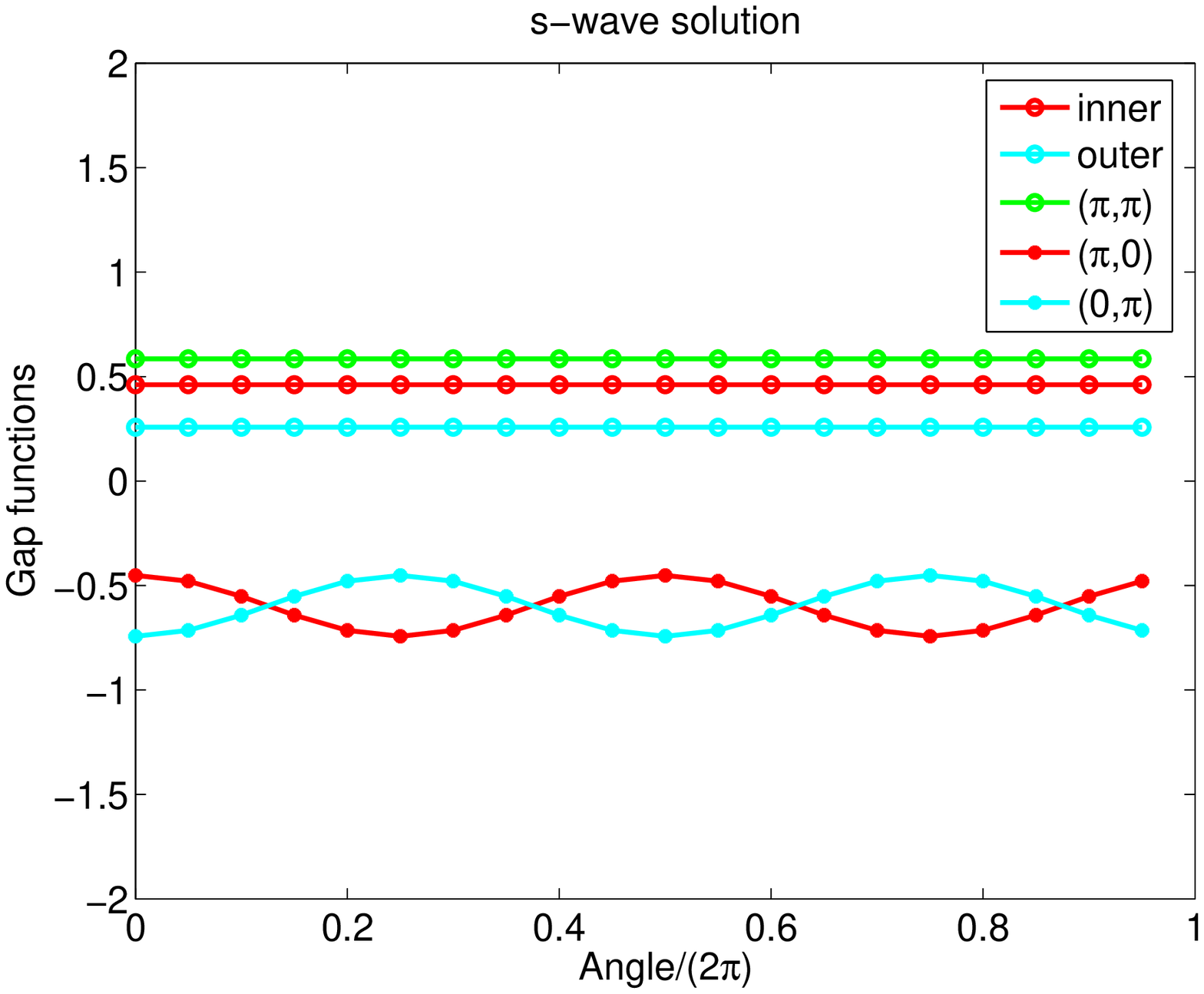}&
\includegraphics[width=2in]{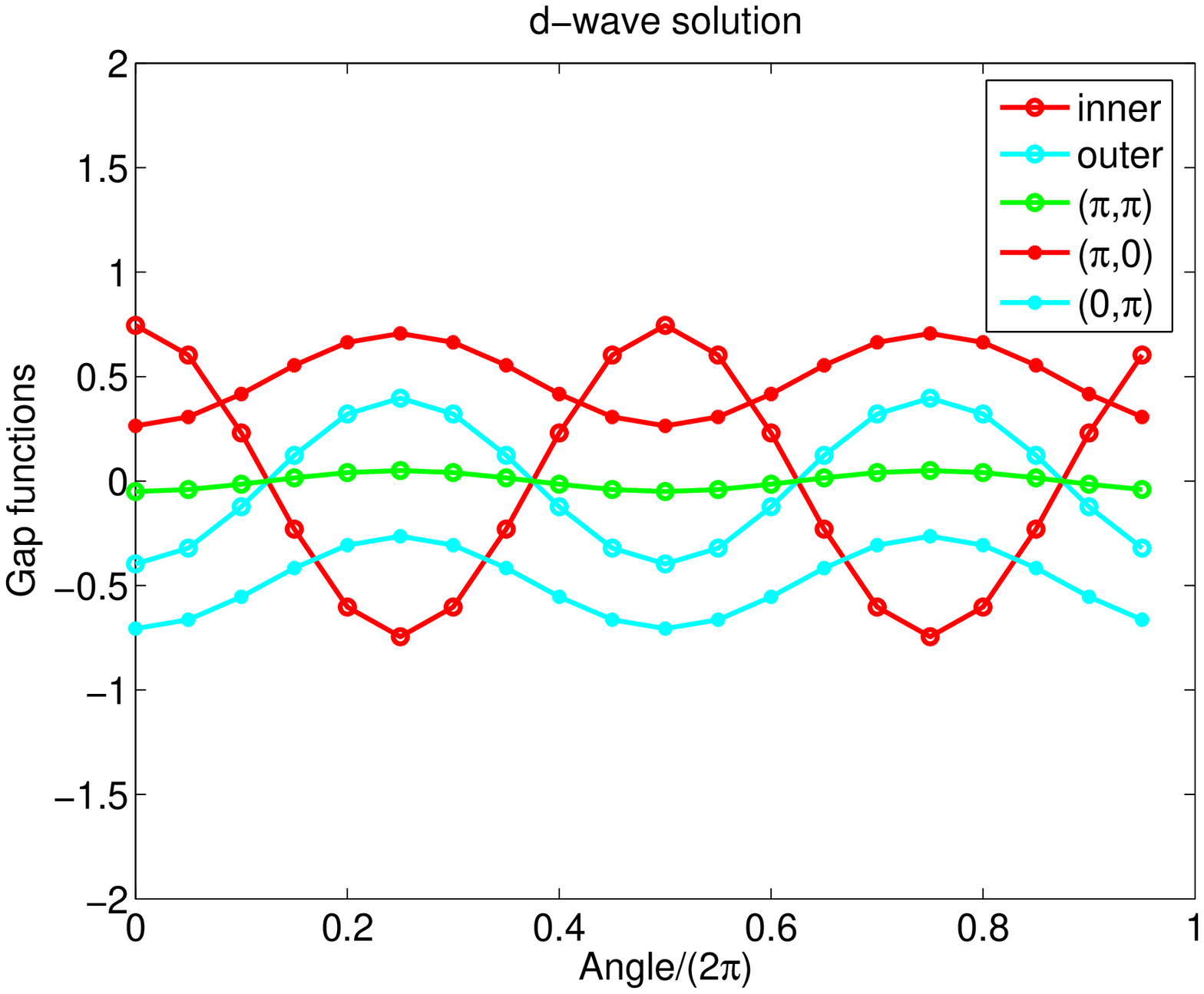}
\end{array}$
\caption{\label{fig:SF_m0p05} The same as in Fig.~\ref{fig:SF_0p05}, but for negative $\mu=-0.05$, when extra hole FS appears. $\lambda_s = 0.58$, $\lambda_d = 0.31$. Observe that now $\lambda_s > \lambda_d$ and $s^\pm$ gap has no nodes.}
\end{figure*}

\begin{figure*}[htp]
$\begin{array}{cc}
\includegraphics[width=2in]{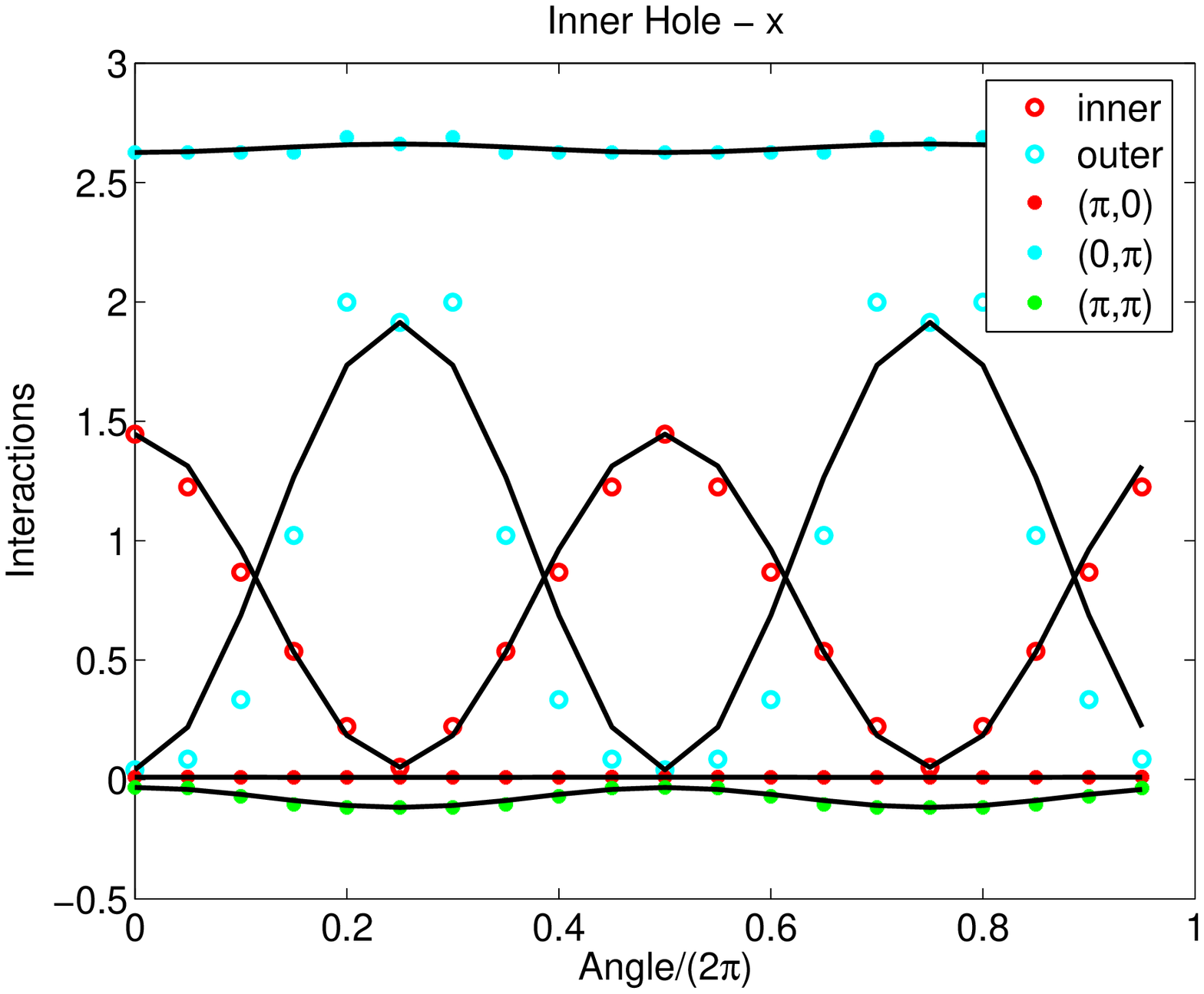}&
\includegraphics[width=2in]{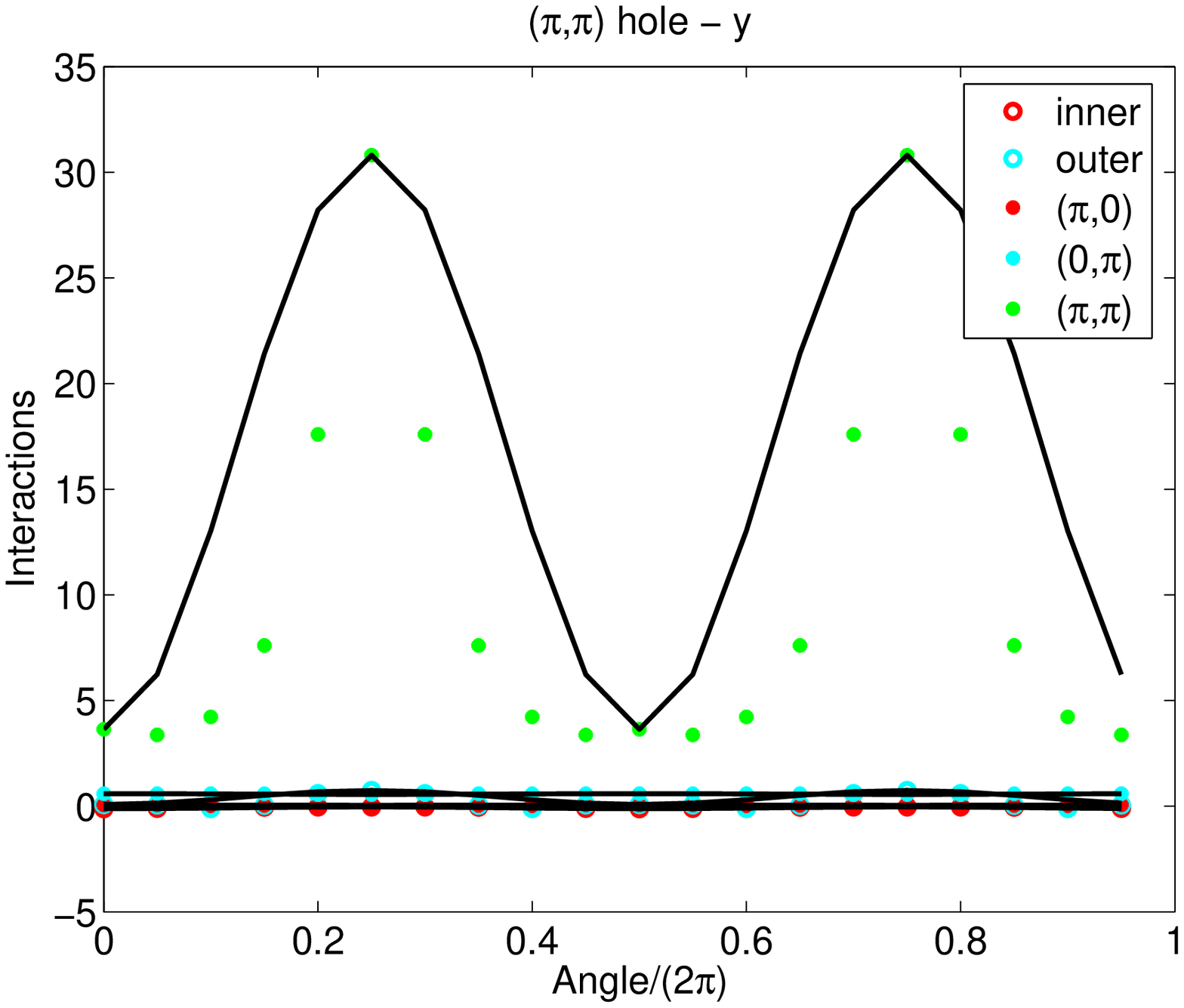}\\
\includegraphics[width=2in]{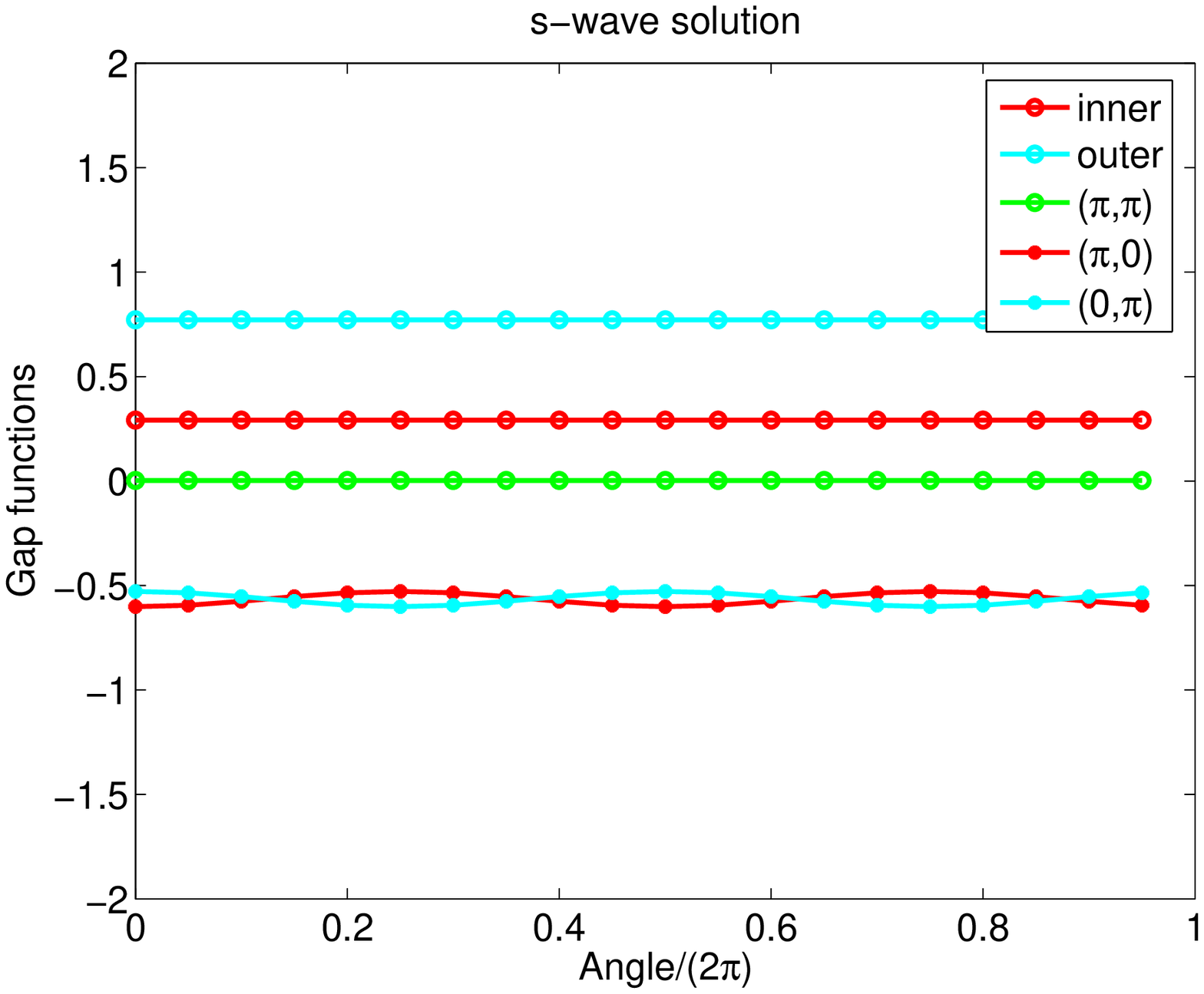}&
\includegraphics[width=2in]{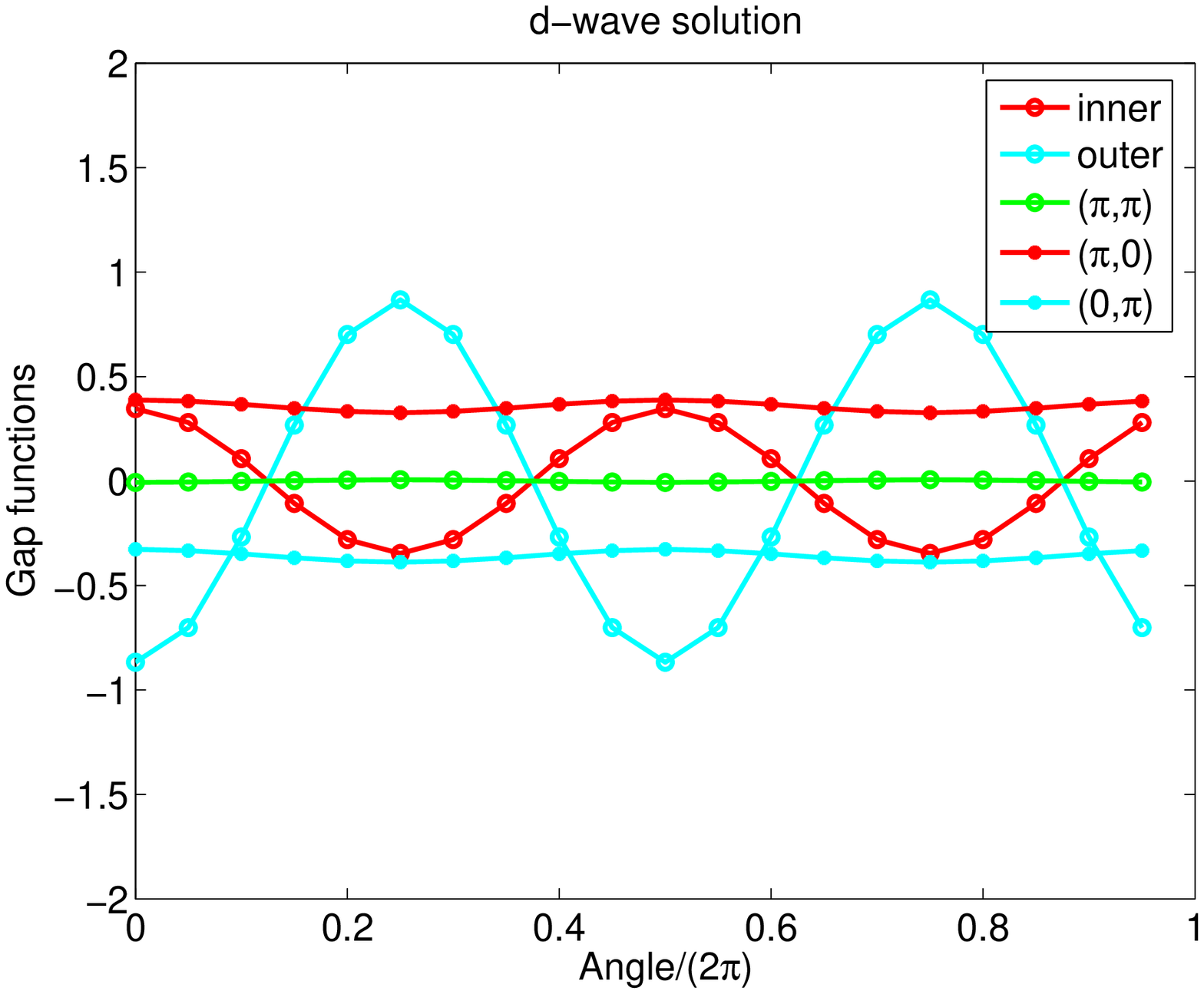}
\end{array}$
\caption{\label{fig:SF_m0p18} The same as in
Fig.~\ref{fig:SF_m0p05}, but for  $\mu=-0.18$ and $U=0.9, J=0.0,
V=0.9$ We obtain $\lambda_s = 1.8$, $\lambda_d = 1.2$. Observe
that, still, $\lambda_s > \lambda_d$ and $s^\pm$ gap has no
nodes.}
\end{figure*}

\subsection{Hole doping}

\begin{figure}[htp]
$\begin{array}{c}
\includegraphics[width=2in]{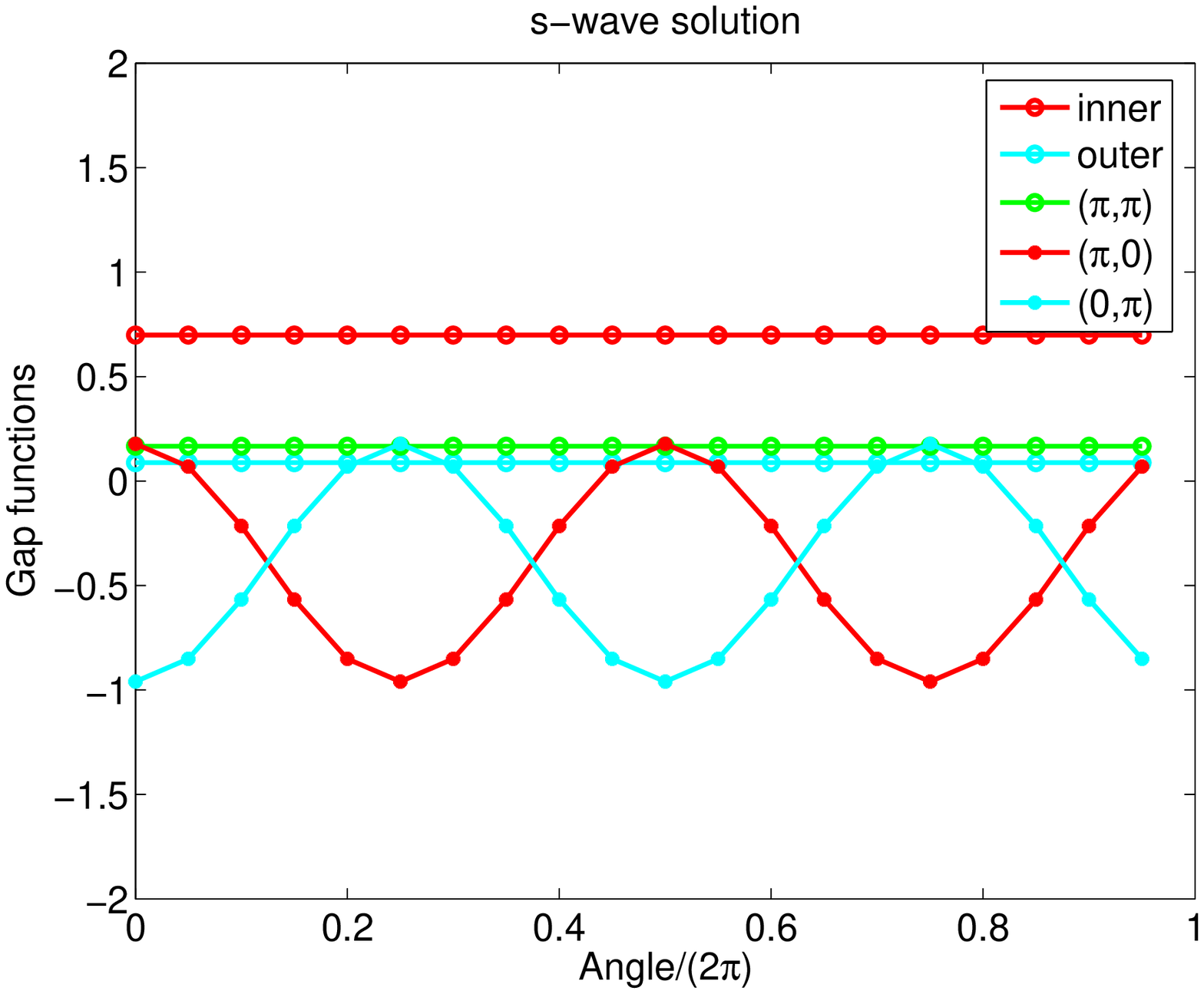}\\
\includegraphics[width=2in]{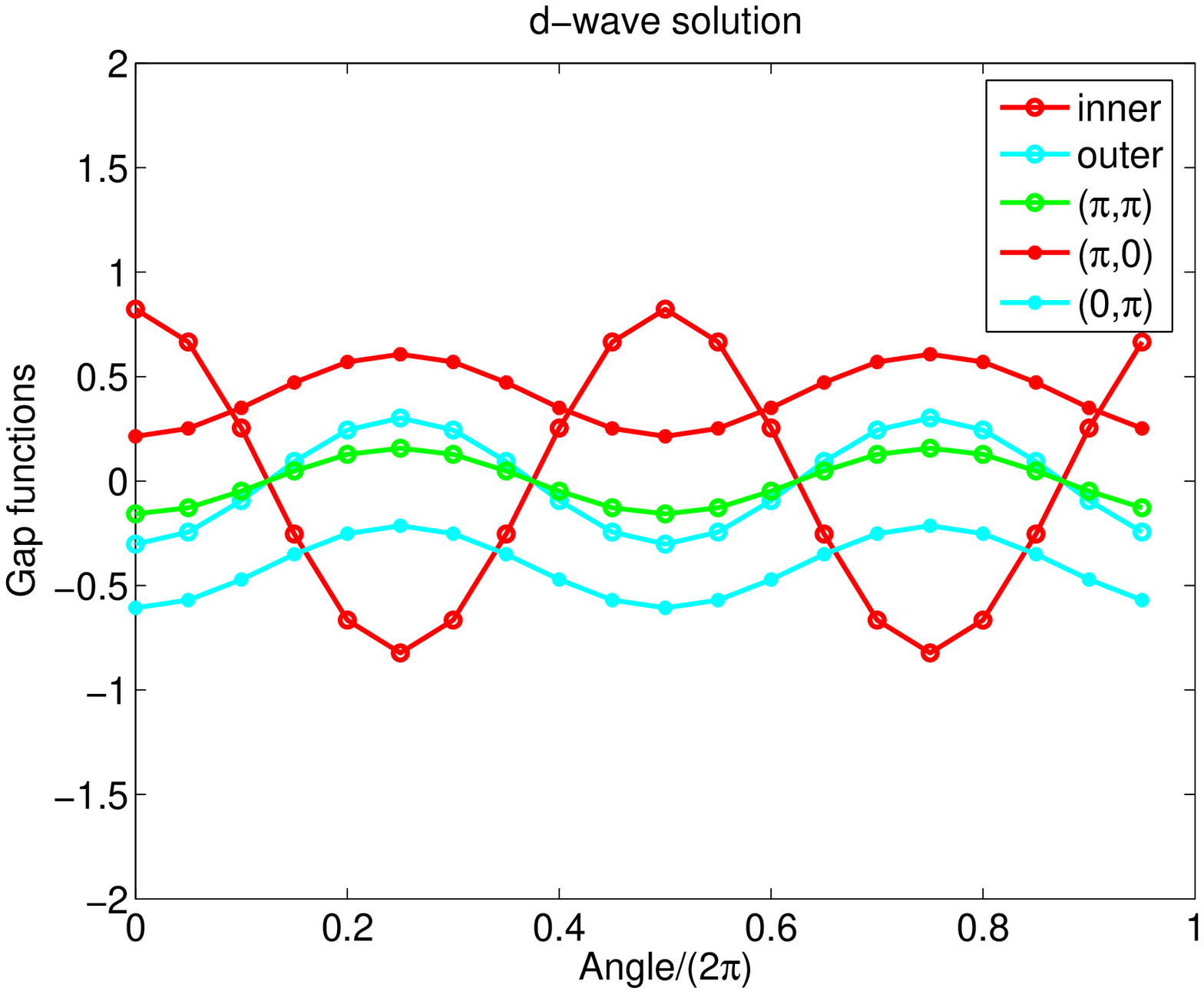}
\end{array}$
\caption{\label{fig:5FS_nodes} $s^\pm$ gap structure obtained
within LAHA  for the parameter set \#3(SF) for $\mu=-0.05$. The
coupling $\lambda_s=0.79$. The gap has nodes on electron FS
despite that there is the third Fermi surface at $(\pi,\pi)$. This
nodal solution has a strong $d$-wave competitor for which
$\lambda_d=0.8$.}
\end{figure}

We next consider representative cases of hole doping by setting
$\mu$ to  negative values, $\mu=-0.05$ ($n_e=5.95$) and
$\mu=-0.18$ ($n_e=5.53$). In Fig.~\ref{fig:FS_mu_0.05and0.18} we
show the FS for these two $\mu$'s. The FS now has an additional
hole pocket centered at $(\pi,\pi)$.  For $\mu=-0.05$ hole and
electron pockets are of comparable size, for $\mu=-0.18$, the
electron pockets almost disappear. The fits to LAHA for
$\mu=-0.05$ are presented in Fig.~\ref{fig:SF_m0p05} and the
parameters extracted from the fits are summarized in
Table~\ref{tab:5}. We see from Fig.~\ref{fig:SF_m0p05} that for
hole doping the situation is different in two aspects. First,  the
$s$-wave gap has no nodes; second, $\lambda_s$ is substantially
larger than $\lambda_d$, i.e., $s$-wave superconductivity is the
most likely scenario. To understand these differences,  compare
Table~\ref{tab:3} with Table~\ref{tab:5}). We see that the
interactions between two hole FSs at $(0,0)$ and two electron FSs
do not change substantially between electron-doped and hole-doped
cases, but for hole-doped case there appear additional hole-hole
and hole-electron interactions associated with the fifth hole
pocket. These additional interactions are weak and irrelevant for
the  $d$-wave component of $\Gamma_{ij} (\k_F, \k'_F)$, but are
quite strong for the $s$-wave component. We analyzed  the
$5\times5$ gap equation for the $s$-wave gap, and found that these
additional interactions effectively increase angle-independent
component of electron-hole interaction, which favors no-nodal
$s^\pm$ gap. As the consequence, the system develops an $s^\pm$
solution with relatively large $\lambda_s >0$, even if all
interactions are set to be angle independent. When we include  the
angle-dependent components, the gap acquires some momentum
dependence along electron FSs, but for given parameters this
dependence is weak and the gap remains nodeless. There is no
``theorem'', however, that the $s^\pm$  solution is always
nodeless for five FS pockets. In Fig.~\ref{fig:5FS_nodes} we show
the gap for the full interaction for the set \#3. We see that the
gap does in fact have nodes on the electron FSs.

We also note that, for the  $d$-wave case,  the electron-electron
interaction ${\tilde u}_{ee}$ is repulsive, and $d$-wave solution
with $\lambda_d >0$ is again the result of relatively strong
$d$-wave electron-hole interactions ${\tilde u}_{h_1e}$ and
${\tilde u}_{h_2e}$ involving the two hole pockets centered at
$(0,0)$.

The outcome of this study is quite simple -- no-nodal $s^\pm$
solution is the leading instability if the angle-independent part
of the electron-hole interaction $u_{he}$ (either the direct one
or  the effective one, in case of 5 FSs) is sufficiently large
compared to $(u_{hh} u_{ee})^{1/2}$.  In the opposite case,  the
$s$-wave solution has nodes on electron FSs, and $d_{x^2-y^2}$
pairing is a strong competitor.

Consider next how far in doping the nodeless $s^\pm$ gap remains
the leading instability. To analyze this, we turn to the case of
$\mu =-0.18$, when  the electron pockets almost disappear. The FS
for this case is shown in Fig.\ref{fig:FS_mu_0.05and0.18}, the
parameters extracted from the fit are shown in Table
\ref{tab:5_1}, and the fits and the gaps are presented in
Fig.~\ref{fig:SF_m0p18}.


On analyzing the structure of the interactions and the gaps  and
comparing them to the case of $\mu=-0.05$ we see that the key
features survive despite the small size of the electron pockets.
Namely,  the leading instability remains $s^\pm$, the gap has no
nodes, and the driving force for the pairing is the interaction
between hole and (still existing) electron pockets.  The $d$-wave
channel is a competitor ($\lambda_d >0$), but still, $s^\pm$ state
has larger $\lambda$.

The conclusion of this subsection is that, as long as both hole
and electron pockets are present, the pairing instability is
essentially driven by electron-hole interaction.  This should
obviously change at even larger hole or electron dopings, when
only one type of pockets remain and the pairing (if it exists)
should come from the interaction either between hole pockets or
between electron pockets.

\section{Overdoping \label{sec:Overdoping}}

Finally, we consider the case of strong electron or hole doping,
when hole FSs disappear and only electron ones at $(0,\pi)$ and
$(\pi,0)$ remain, and the case of strong hole doping, when
electron FSs disappear and only hole FSs at $(0,0)$ and
$(\pi,\pi)$ remain. We consider the two limits separately.

\subsection{Strong electron doping}

Strongly electron doped FeSC include recently discovered FeSe
superconductors AFe$_2$Se$_2$ (A = K, Rb, Cs)~\cite{exp:AFESE}.
$T_c$ in these materials is quite high and reaches almost $40K$.
The electronic structure of these materials in the folded
Brillouin zone is a bit more involved because (i) AFe$_2$Se$_2$
has body-centered tetragonal structure what makes the folding  of
the electron FSs a more complex procedure than just the mixing of
the two ellipses which would be the case for a simple tetragonal
structure~\cite{mazin,kontani}, and (ii) there is apparently a
small electron pocket at $(0,0)$~\cite{exp:AFESE_ARPES}. Because
our main intention is to understand what causes the pairing in the
absence of hole pockets, we follow Ref.~\onlinecite{graser_11} and
neglect the peculiarities of the folding procedure and potential
electron pocket at $(0,0)$.  Corrections to this approach have
been discussed in references~\onlinecite{mazin}
and~\onlinecite{kontani}.

RPA/SF and fRG studies~\cite{graser_11,das,dhl_AFESE} applied to
AFe$_2$Se$_2$  showed that the dominant instability is in the
$d$-wave channel. The issue we want to address is whether the
interaction between the two electron pockets alone is capable to
give rise to sizable $d$-wave attraction for large positive $\mu$.
We recall in this regard that  for the two cases of electron
doping that we considered before ($\mu=0.05$ and $\mu=0.18$),
the direct electron-electron interaction was attractive but very
small, and the attraction in  the $d$-wave channel was primarily
the result of strong $d$-wave component of electron-hole
interaction. Recall that at $\mu=0.18$  the hole FSs are already
very small. Hole excitations are visible in ARPES in AFe$_2$Se$_2$
above the gap of $60-100meV$ (Refs.~\onlinecite{exp:AFESE_ARPES}),
and it is certainly a possibility that the dominant mechanism of
$d$-wave attraction between fermions on electron FSs in
AFe$_2$Se$_2$ is virtual pair hopping to gapped hole states.

Let us elaborate on this point. When hole FSs are present, this
pair hopping give rise to an effective attractive $d$-wave
interaction between electron pockets with magnitude $u^2_{he} L$,
where $L = log E_F/T_c$ comes from integration over low-energy
hole states. This  interaction can well exceed a direct $u_{ee}$
because for relevant values of the logarithm $u_{he} L = O(1)$. [A
way to see this is to express $\Delta_h$ via $\Delta_e$ in the
matrix gap equation and re-write it solely as equation for
$\Delta_e$.] When hole excitations are gapped, the logarithm is
cut and scales as  $L = log \left[E_F/(T^2_c + E^2_0)\right]$,
where $E_0$ is the gap for hole states.  Because in AFe$_2$Se$_2$
$E_0 \leq E_F$ (Ref. \onlinecite{exp:AFESE_ARPES}) it is
not-apriori guaranteed that $u_{eh} L$ is now small and can be
neglected. fRG study of superconductivity in  AFe$_2$Se$_2$
(Ref.~\onlinecite{dhl_AFESE}) does include this contribution
because RG procedure incorporates renormalizations coming from
energies above $E_F$.

A way to verify whether the direct electron-electron interaction
or the pair hopping to gapped hole states is the dominant
mechanism of $d$-wave attraction in  AFe$_2$Se$_2$  is to use
RPA/SF approach which only considers the interaction between
fermions right at the FSs and neglects pairing interactions via
intermediate gapped states, and see whether $\lambda_d$ is large
enough (e.g., comparable to $\lambda_s$ for hole doping), and
whether  the $d$-wave gap structure is similar to that obtained in
fRG which includes virtual processes via gapped states.  If
$\lambda_d$ is not small and the gap structure is similar to that
in fRG, electron-hole interaction is likely irrelevant, and
$d$-wave attraction comes from the direct electron-electron
interaction.

\begin{figure*}[htp]
$\begin{array}{cc}
\includegraphics[width=2in]{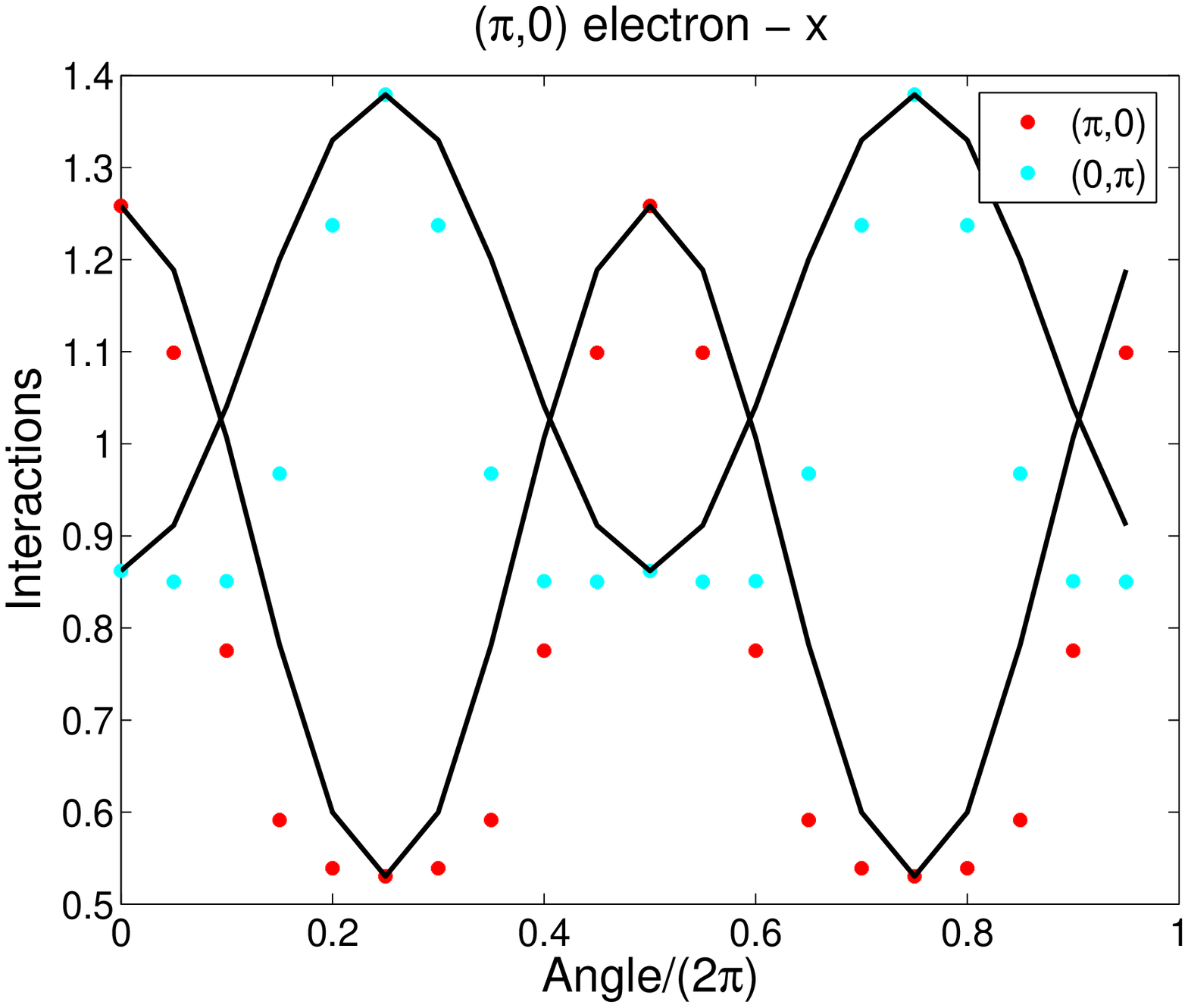}&
\includegraphics[width=2in]{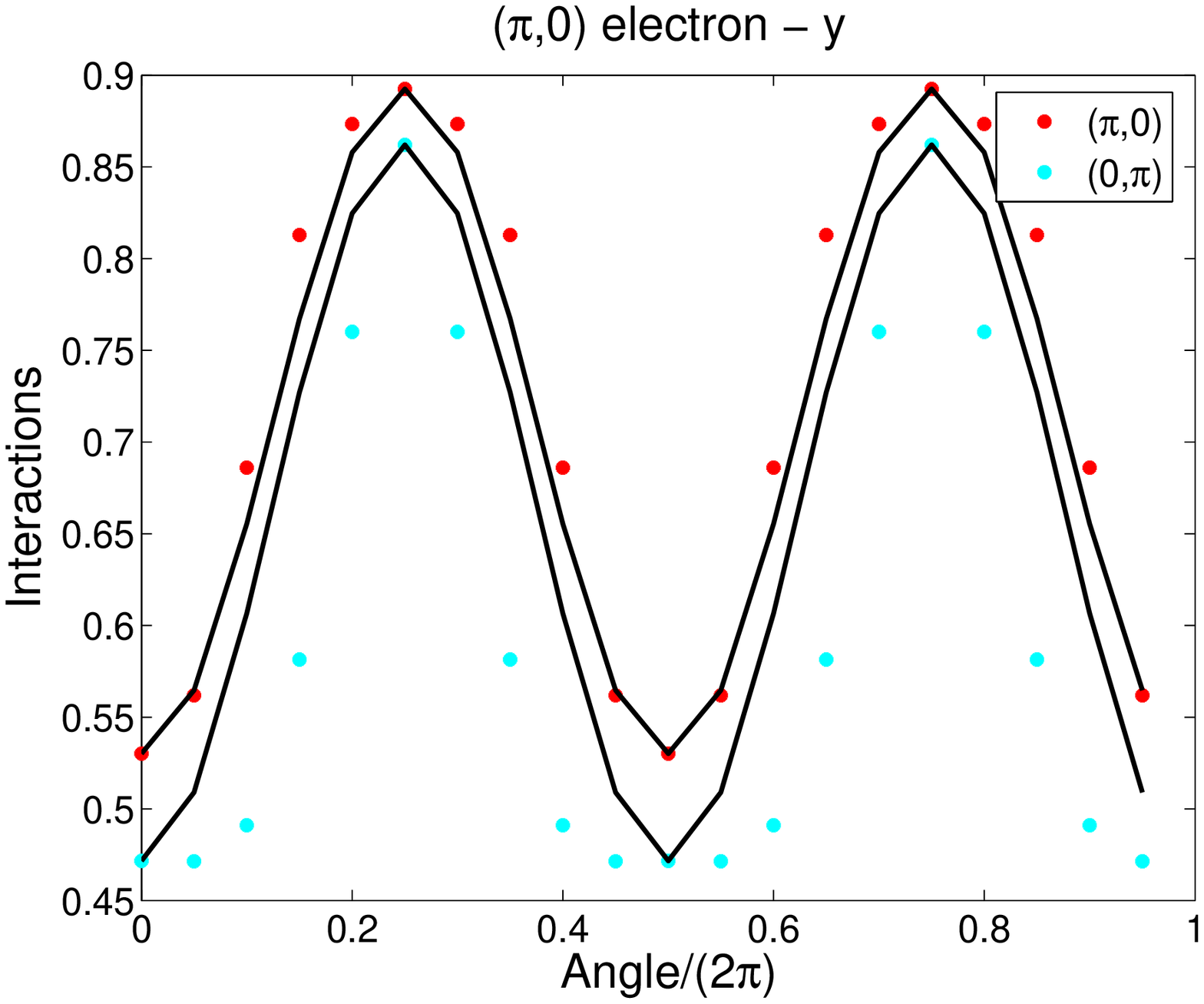}\\
\includegraphics[width=2in]{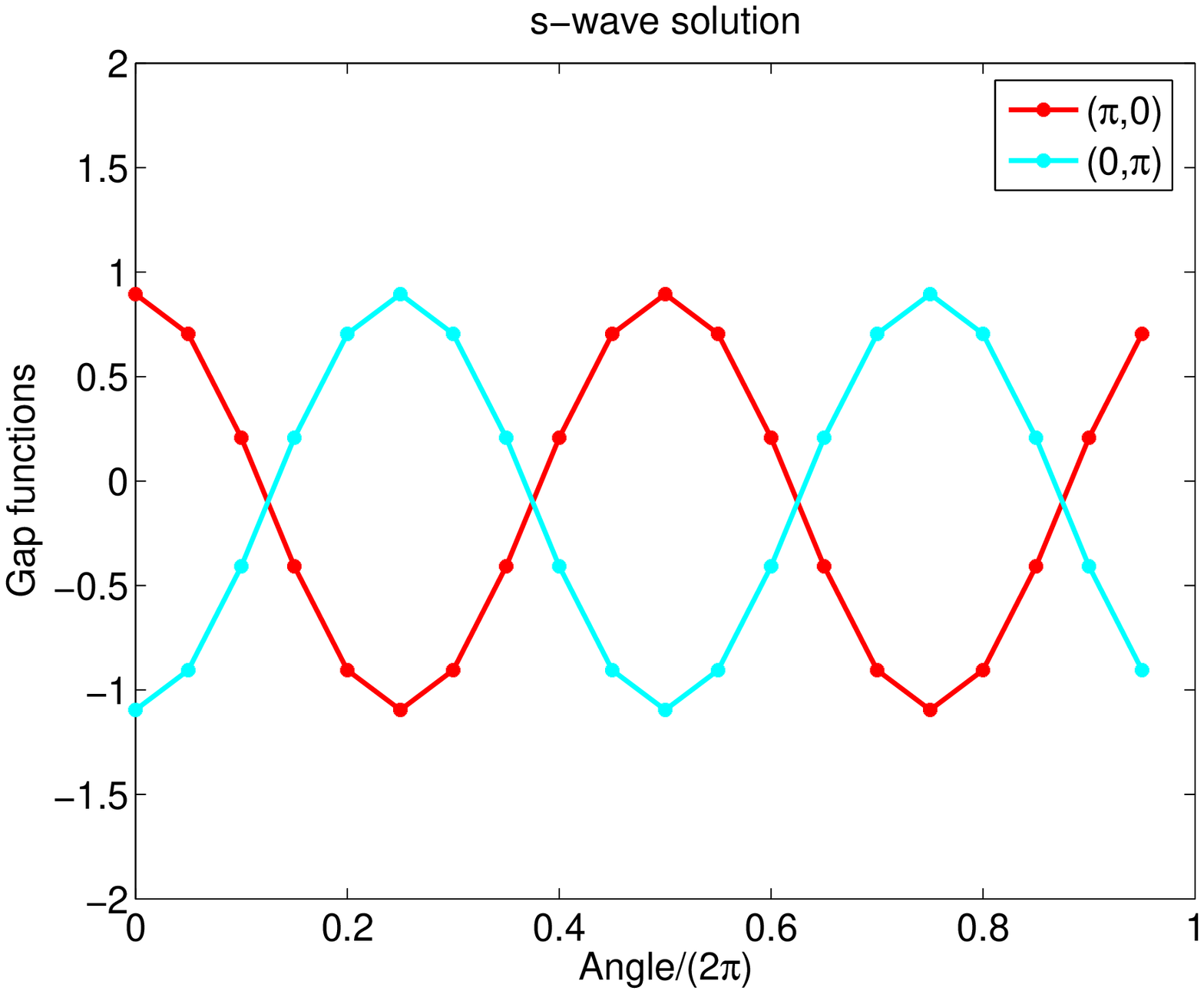}&
\includegraphics[width=2in]{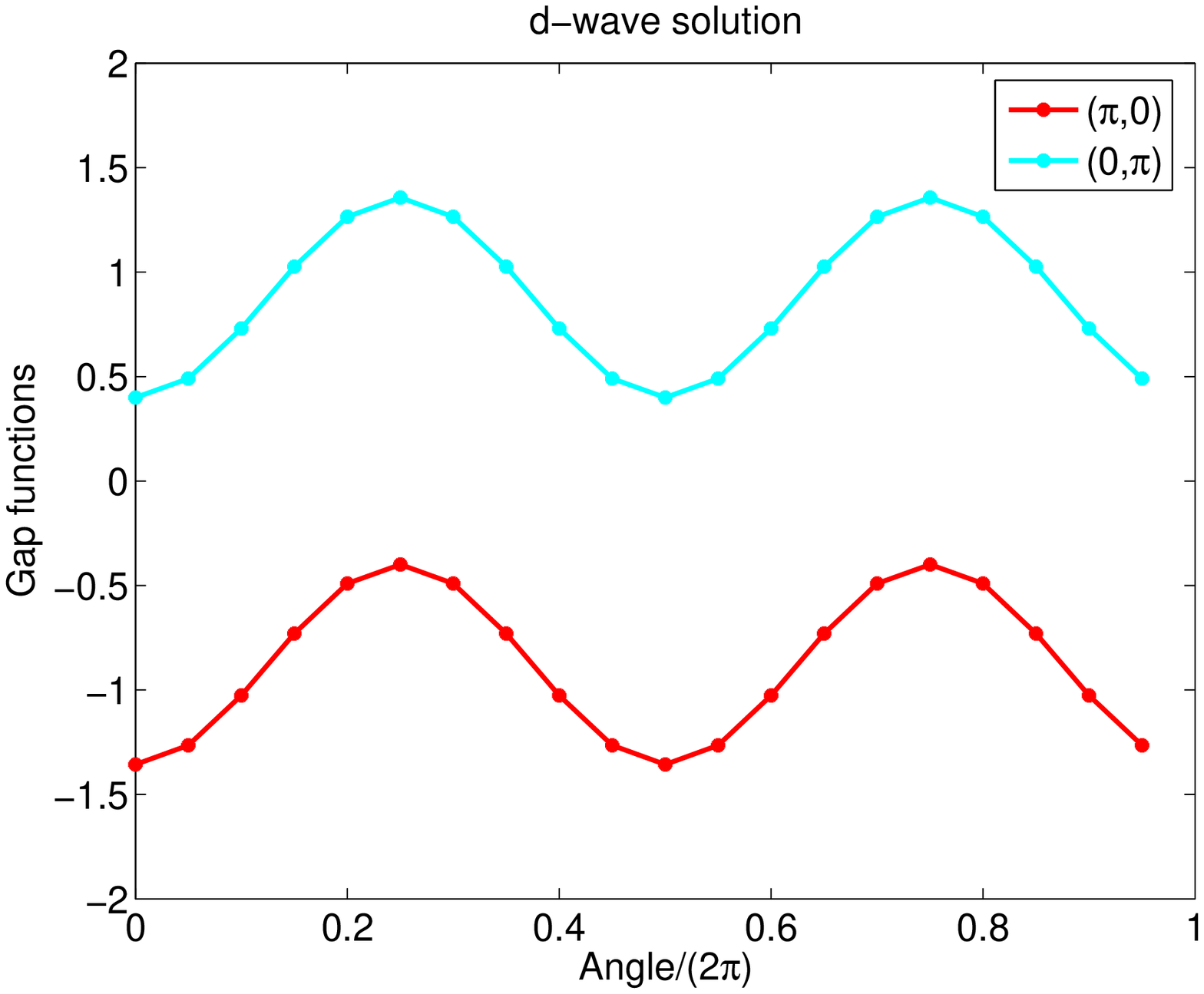}
\end{array}$
\caption{\label{fig:Max_Tom_KFeSe_fits} The fits of the
interactions and the structure of $s$-wave and $d$-wave gaps in
LAHA  for the case of heavy electron doping, when only electron
FSs are present. The $\lambda$s are $\lambda_s=-0.12$ and
$\lambda_d=0.13$. ($\mu=0.30$)}
\end{figure*}

\begin{figure*}[htp]
$\begin{array}{cc}
\includegraphics[width=2in]{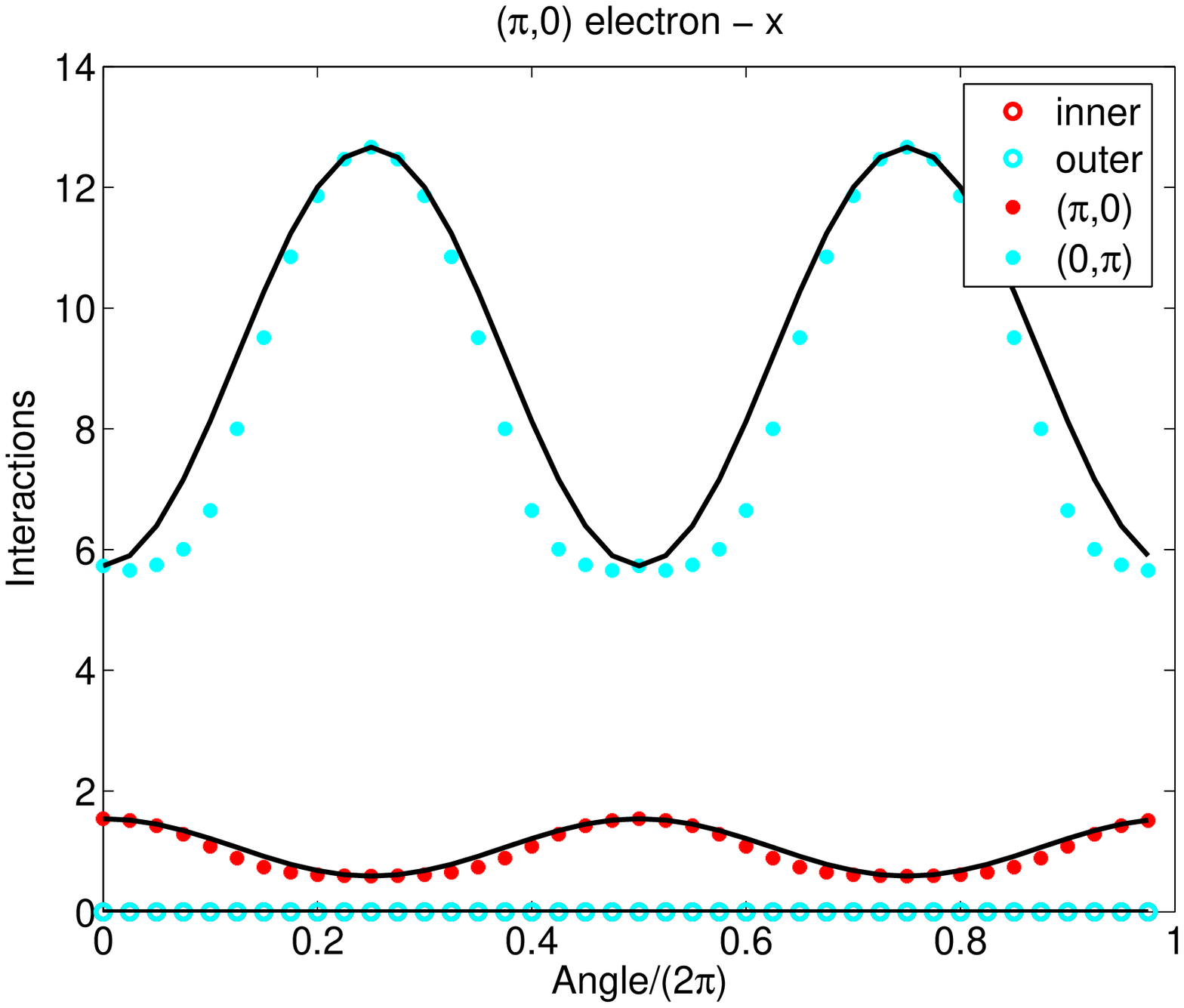}&
\includegraphics[width=2in]{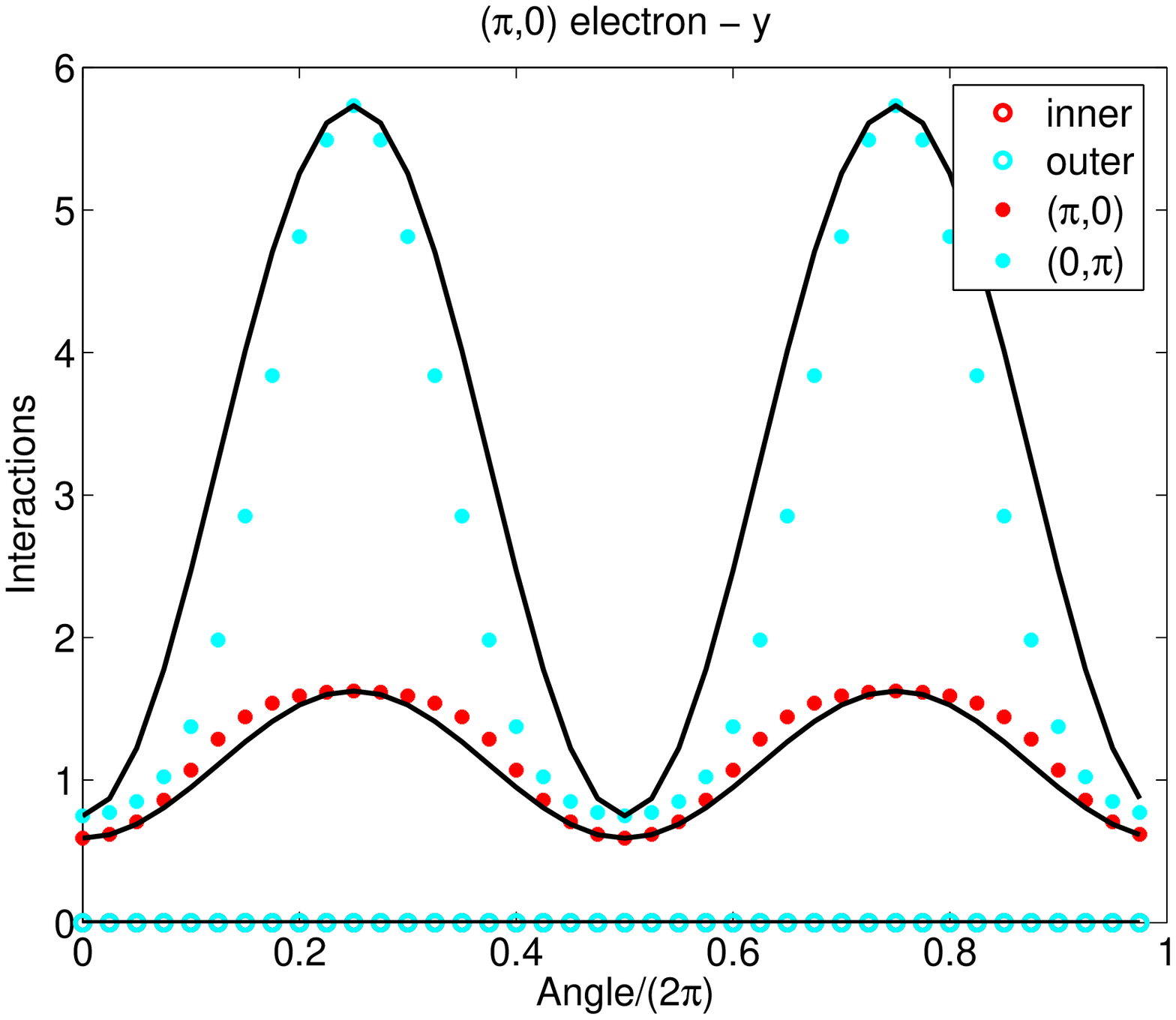}\\
\includegraphics[width=2in]{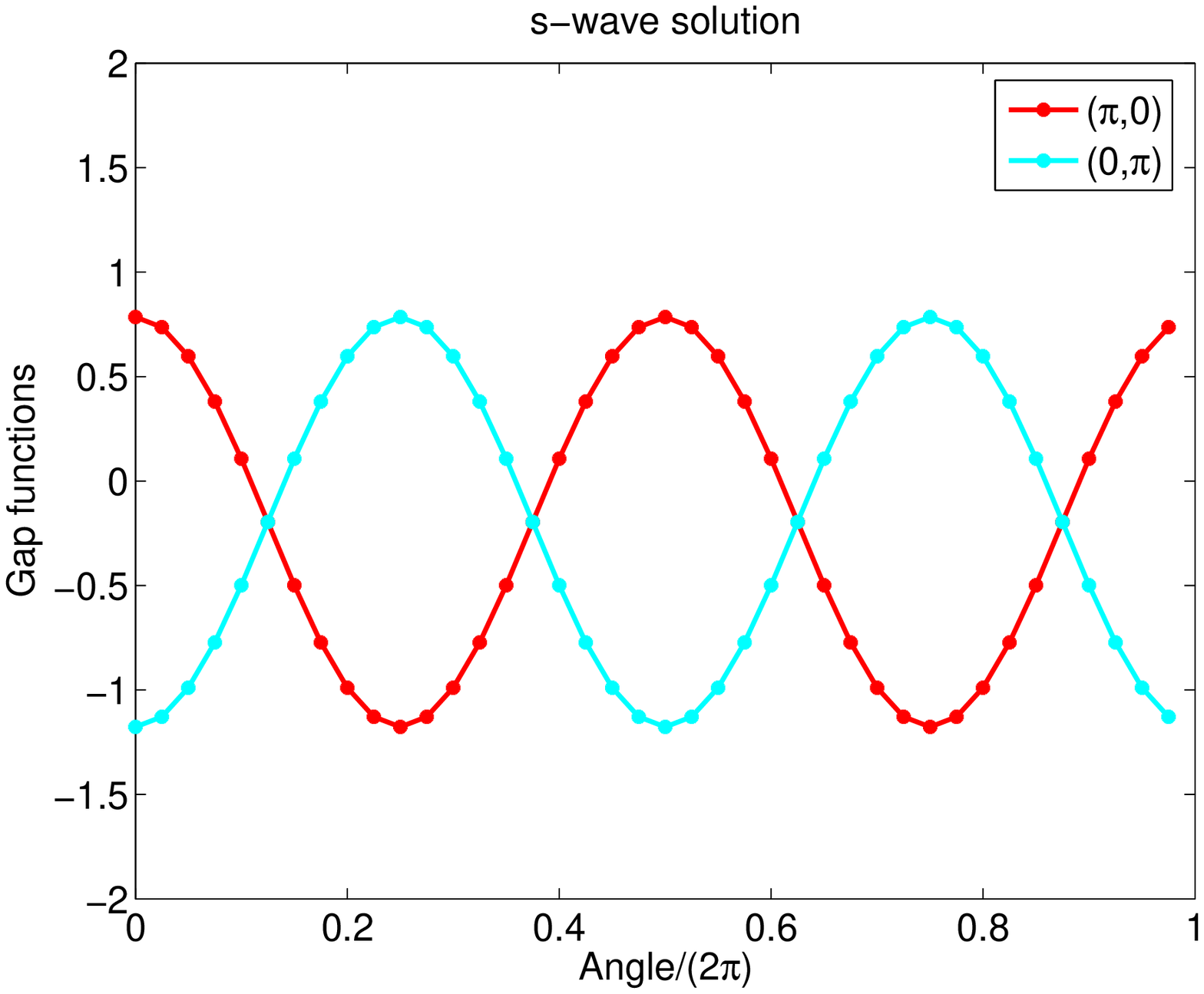}&
\includegraphics[width=2in]{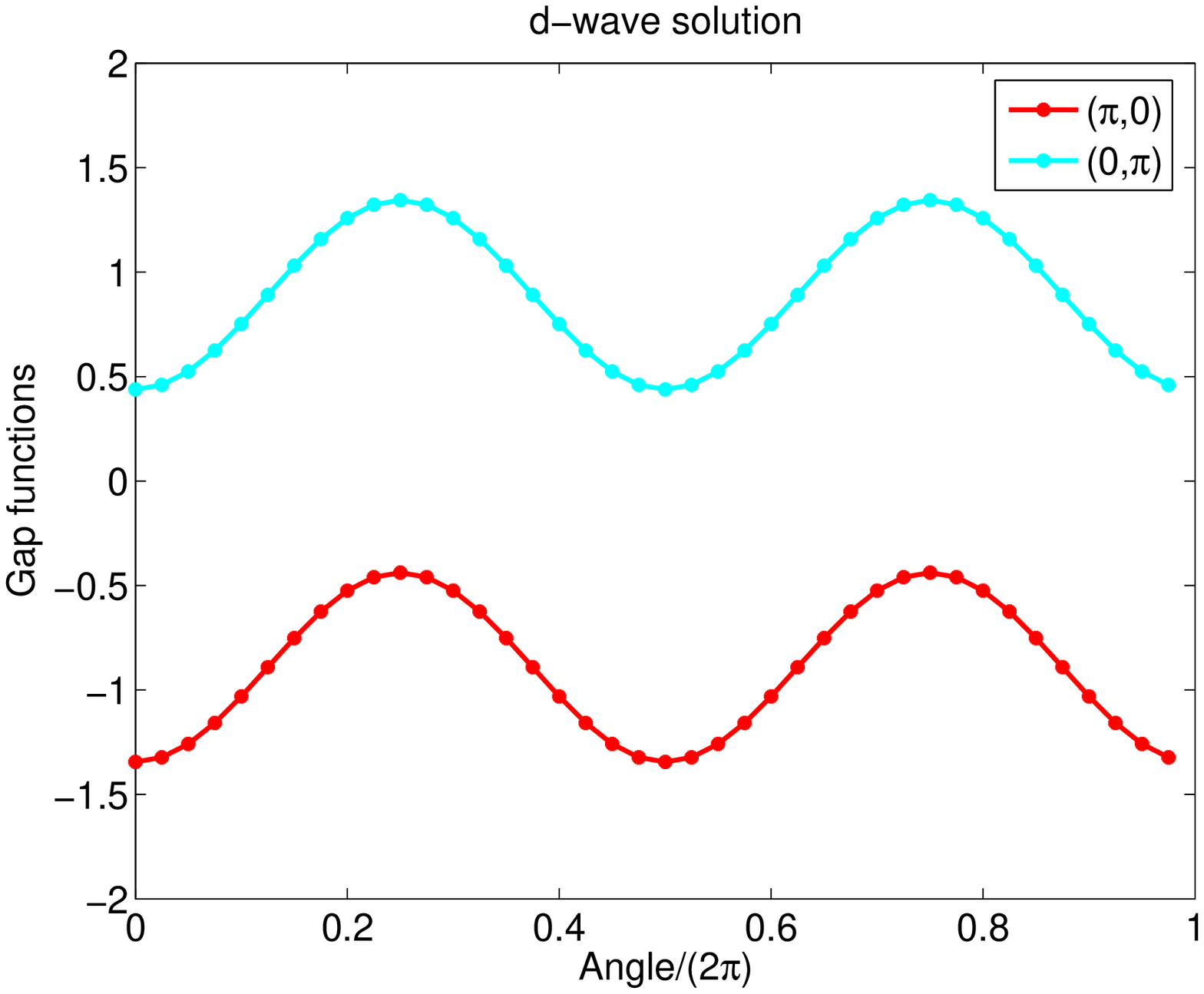}
\end{array}$
\caption{\label{fig:Tom_KFeSe_fits} The same as in Fig. \ref{fig:Max_Tom_KFeSe_fits} but for a different band structure, used in Ref.~\onlinecite{graser_11}.
The couplings are $\lambda_s=0.1$ and $\lambda_d=5.9$.}
\end{figure*}

In Fig.~\ref{fig:Max_Tom_KFeSe_fits} we show the fit to
electron-electron interactions obtained in RPA/SF formalism to Eq.~\ref{eq:interactions} in which we only kept electron-electron
interactions.  The parameters in $s$-wave and $d$-wave channels,
extracted from the fit are shown in Table~\ref{tab:Max_Tom_s and
d}.

\begin{table}[htp]
\caption{$s$- and $d$-wave parameters for the case of strong
electron doping, when there are no hole FSs. }
\label{tab:Max_Tom_s and d}
\begin{ruledtabular}
\begin{tabular}{ccccc}
$s$-wave&$u_{ee}$&$\alpha_{ee}$&$\beta_{ee}$&$\lambda_s$\\
&$0.84$&$0.09$&$0.04$&-0.12\\
\hline
$d$-wave&$\tilde{u}_{ee}$&$\tilde{\alpha}_{ee}$&$\tilde{\beta}_{ee}$&$\lambda_d$\\
&$-0.04$&0.88&-0.75&0.13
\end{tabular}
\end{ruledtabular}
\end{table}

\begin{table}[htp]
\caption{$s$- and $d$-wave parameters for the case of strong
electron doping, with band structure from Ref.
\onlinecite{graser_11}, when there are no hole FSs.}
\label{tab:KFeSe_s and d}
\begin{ruledtabular}
\begin{tabular}{ccccc}
$s$-wave&$u_{ee}$&$\alpha_{ee}$&$\beta_{ee}$&$\lambda_s$\\
&$3.65$&$0.20$&$0.03$&0.1\\
\hline
$d$-wave&$\tilde{u}_{ee}$&$\tilde{\alpha}_{ee}$&$\tilde{\beta}_{ee}$&$\lambda_d$\\
&$-2.57$&0.29&-0.0&5.9
\end{tabular}
\end{ruledtabular}
\end{table}

We solved $3 \times 3$ gap equations in $s$-wave and $d$-wave
channels with these parameters, and show the results in Fig.
\ref{fig:Max_Tom_KFeSe_fits}. We clearly see that $d$-wave
 eigenvalue is positive.

For completeness, we also computed  the d-wave eigenvalue
 for a different
band structure, used in Ref.~\onlinecite{graser_11},
 which still has only electron FSs remaining.
 The results are shown in Fig. \ref{fig:Tom_KFeSe_fits}
and Table~\ref{tab:KFeSe_s and d}. We find that the $d$-wave
eigenvalue is again positive and
 the d-wave gap structure  is quite similar to that in Fig.
\ref{fig:Max_Tom_KFeSe_fits}.  The magnitude of $\lambda_d$, however,
 depends on the choice of
 the band structure.

With LAHA fit, we are in position to analyze the pairing in more
detail and check if the $d$-wave attraction comes from
angle-independent or angle-dependent parts of electron-electron
interaction. In the absence of hole FSs, $\lambda_d$ is the
solution of  $2\times2$ gap equation and is given by \beq
\lambda_d = -{\tilde u}_{ee} \left(2 {\tilde \beta}_{ee} +1 \pm
\left((2{\tilde \beta_{ee}}-1)^2 + 8 {\tilde
\alpha}^2_{ee}\right)^{1/2} \right) \label{fr_1} \eeq In general,
$\lambda_d >0$ either because ${\tilde u}_{ee} <0$, or ${\tilde
u}_{ee} >0$ but ${\tilde \alpha}^2_{ee} > {\tilde \beta}_{ee}$. In
the first case, $d$-wave gap is due to a direct attraction between
fermions from the two electron pockets, much in analogy with ``hot
spot'' scenario for the cuprates, and the gap has only modest
variation along the electron FSs and no nodes. In the second case,
$d$-wave gap should have nodes.  We see from
Table~\ref{tab:Max_Tom_s and d} and~\ref{tab:KFeSe_s and d} that
${\tilde u}_{ee} <0$ , i.e.  $d$-wave attraction in AFe$_2$Se$_2$
is primarily due to the existence of a constant
(angle-independent)
 attractive $d-$wave interaction between the two electron pockets.

To understand why ${\tilde u}_{ee} <0$  we remind that ${\tilde
u}_{ee}$ is the difference between intra-pocket and inter-pocket
electron-electron interactions ${\tilde u}_{ee} = (N_F/2)
(\Gamma_{e_1e_1} - \Gamma_{e_1e_2})$ (see Eq.
(\ref{eq:interactions}). The bare values of $\Gamma_{e_1e_1}$ and
$\Gamma_{e_1e_2}$  are quite close.  For small electron doping,
magnetic fluctuations are predominantly peaked at $(0,\pi)$ and
$(\pi,0)$, and neither of these two interactions are selected for
relative increase. As a result, the full intra-pocket and
inter-pocket electron-electron interactions remain of nearly equal
magnitude, and the difference between the two is small. Once the
hole pockets disappear, the peak in magnetic susceptibility shifts
to near $(\pi,\pi)$, according to RPA~\cite{graser_11}. In this
situation, SF enhance the inter-pocket $\Gamma_{e_1e_2}$  compared
to the intra-pocket $\Gamma_{e_1e_1}$, and ${\tilde u}_{ee}$
becomes negative (attractive).

We also note that  $s$-wave solution  remains attractive for one choice of the
 band structure (see
Table~\ref{tab:KFeSe_s and d}), although $\lambda_s < \lambda_d$.
 The two solutions for
$\lambda_s$ are given by
 \beq
\lambda_s = -u_{ee} \left(2
\beta_{ee} +1 \pm \left((2\beta_{ee}-1)^2 + 8
\alpha^2_{ee}\right)^{1/2} \right) \label{fr_2}.
 \eeq

The interaction $u_{ee}$ is given by $u_{ee} = (N_F/2)
(\Gamma_{e_1e_1} + \Gamma_{e_1e_2})$ and is positive when both
intra-pocket $\Gamma_{e_1e_1}$ and inter-pocket $\Gamma_{e_1e_2}$
are positive,
 as in Tables \ref{tab:KFeSe_s and d} and \ref{tab:Max_Tom_s and d}).
 For $u_{ee} >0$ a positive
$\lambda_s$ appears when $\alpha^2_{ee} > \beta_{ee}$ which is satisfied
 for the parameters in Table~\ref{tab:KFeSe_s and d}).  Such a
solution is induced by angle-dependent electron-electron
interaction terms, and the $s$-wave gap should have nodes on the
electron FSs. This is consistent with Fig.
\ref{fig:Tom_KFeSe_fits}.

The structure of both $d-$wave and $s-$wave gap functions
 is also quite consistent with the
 solution obtained within fRG~\cite{dhl_AFESE}. The authors of that paper
 argued that the attraction in the $d-$wave channel is due to virtual hopping to
gapped hole states, which is captured in fRG.  We, on the contrary,
 believe that the good
agreement between fRG  and  LAHA/RPA approach, which does not include gapped hole states,
 indicates that the primary cause for a d-wave pairing is the direct interaction between the two electron pockets, present in both approaches.
 We have not investigated the influence of gapped hole states
directly, however.

The issue that we don't discuss in this paper is how the $d$-wave
gap evolves under the transformation into the folded zone. The
folding is a non-trivial procedure for the case of $d$-wave
pairing because the two electron FSs  do hybridize in the folded
zone, and this frustrates the $d$-wave gap which changes sign
between the two unhybridized FSs. Further complication is that
AFe$_2$As$_2$ has body-centered tetragonal lattice, and the two
electron FSs which eventually hybridize differ by  $k_z = \pi$ and
are rotated by $\pi/2$ before hybridization.
 There is an  argument that in this
situation the $d$-wave gap
must have nodes near $k_z = \pi/2$
(see Ref.~\onlinecite{mazin}). However, the issue of the pairing in the presence of the hybridization
is not settled at the moment and we will not dwell on it.

Finally, in our discussion we assumed that both intra-pocket
interaction $\Gamma_{e_1e_1}$ and inter-pocket interaction
$\Gamma_{e_1e_2}$ are positive. Then $u_{ee}$ is definitely
positive.  Intra-pocket interaction is certainly positive, but the
sign of interpocket interaction depends on the interplay between
$U, V$, $J$ and $J'$ in the same way as the sign of $u_{he}$
depends on the interplay between intra-orbital and inter-orbital
interactions.  If $\Gamma_{e_1e_2} <0$ the $d$-wave ${\tilde
u}_{ee} =  (N_F/2) (\Gamma_{e_1e_1} +|\Gamma_{e_1e_2}|)$ becomes
positive, while $s-$wave $u_{ee} =  (N_F/2) (\Gamma_{e_1e_1} -
|\Gamma_{e_1e_2}|)$ becomes negative, if $ |\Gamma_{e_1e_2}| >
\Gamma_{e_1e_1}$.  In this situation, the system develops an
$s-$wave pairing with equal sign of the gap on the two hole
pockets.  We make a conjecture that $\Gamma_{e_1e_2}$ is negative
if one uses orbital $J_1-J_2$ model with the interaction between
first and second neighbors in real space (and $J_2 >J_1$), instead
of the on-site orbital Hubbard/Hund model. This would explain why
both strong coupling~\cite{si_11,bernevig_11} and weak
coupling~\cite{bernevig_11} studies of the pairing in the orbital
$J_1-J_2$ model yielded an $s-$wave pairing.

\subsection{Strong hole doping}

Superconductivity at strong hole doping, when only hole FSs are
present, has been observed in
\KFA~\cite{KFeAs_exp,KFeAs_exp_nodal}, which is at the opposite
end from parent \BFA in the family of K$_{1-x}$Ba$_x$Fe$_2$As$_2$.
$T_c$ in this material is rather low, only $3K$, but the interest
in \KFA is fueled by the fact that penetration depth and thermal
conductivity measurements point to nodal behavior, much as in
\LFPO.  The  fRG study by Thomale \textit{et
al.}~\cite{fRG_KBaFeAs} shows that the leading instability  is in
$d$-wave channel, consistent with the observation of  nodes.
Interestingly,  Thomale \textit{et al.} have found that the
$d$-wave gaps on the three hole FSs (two centered at $(0,0)$ and
one at $(\pi,\pi)$) are ``in phase'' with each other, i.e.,
$\Delta_i (\phi) = \Delta_i \cos{2 \phi}$ ($i=1-3$), and all
$\Delta_i$ are of the same sign.  This is {\it not} the $d$-wave
solution with the largest $\lambda_d$ at smaller doping, when both
hole and electron pockets are present. In that solution, there is
a $\pi$ phase shift between the  gaps on the two FSs centered at
$(0,0)$ (see Figs.~\ref{fig:SF_m0p18} and~\ref{fig:SF_m0p05}). The
solution with ``in phase'' $d$-wave gaps is one of 5 $d$-wave
solutions in  the LAHA formalism, but with
 negative $\lambda_d$ at small/moderate hole doping.

Given this discrepancy, it is interesting to analyze strong hole
doping in our approach.  We set $\mu=-0.3$ ($n_e=4.88$) and
consider the same set of parameters as Thomale \textit{et al.} ($U
=3$, $J=0.6$, $V =1.85$), but rescale all interactions by $a =
0.25$ to avoid a magnetic instability in the RPA/SF analysis. The
magnetic susceptibility $\chi (q)$ for such large  negative $\mu$
has a  maximum at $q=0$, and affects mostly the interactions
either near the FSs at $(0,0)$ or at $(\pi,\pi)$, but not
intra-pocket interactions between $(0,0)$ and $(\pi,\pi)$.

\begin{figure*}[htp]
$\begin{array}{cc}
\includegraphics[width=2in]{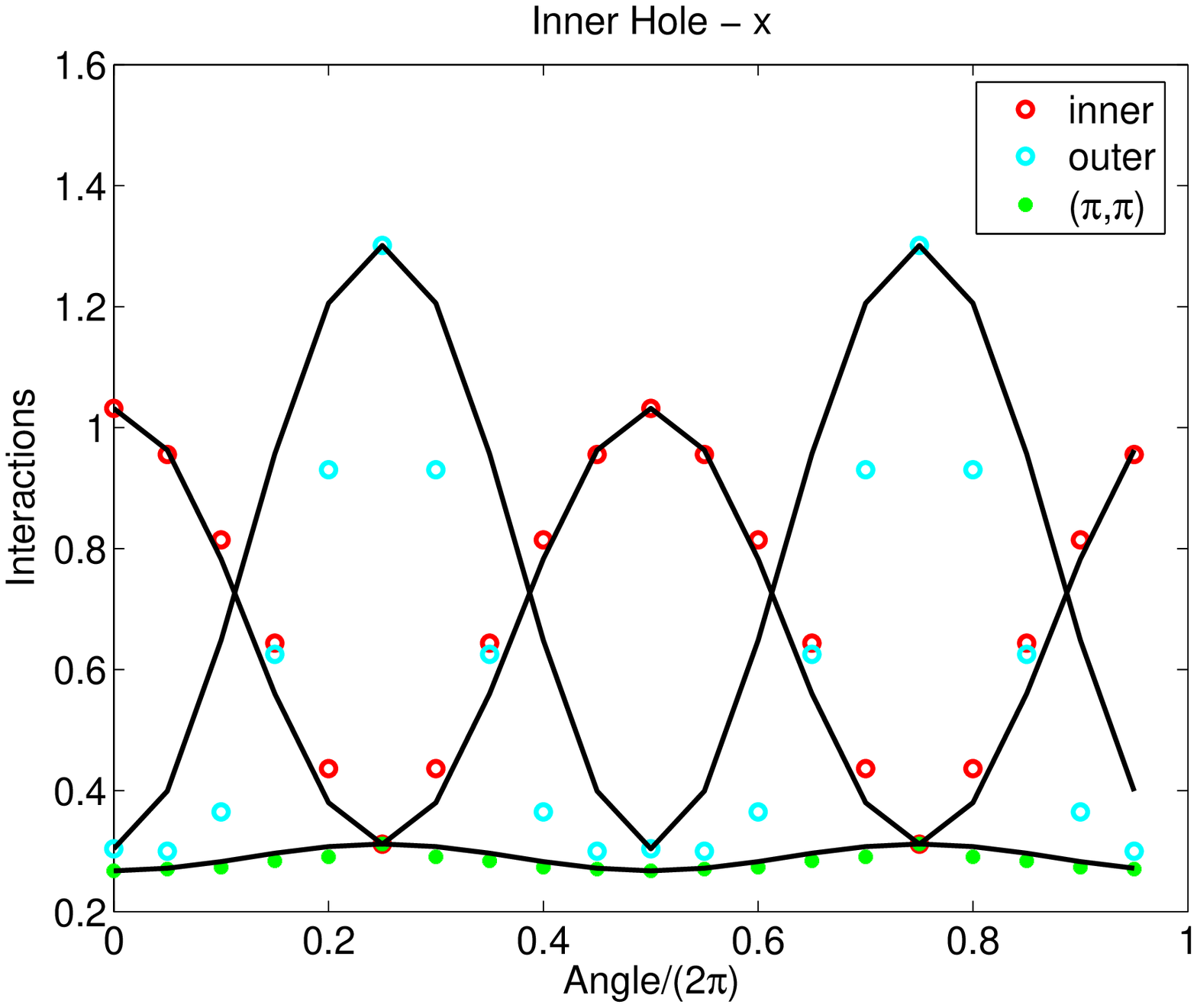}&
\includegraphics[width=2in]{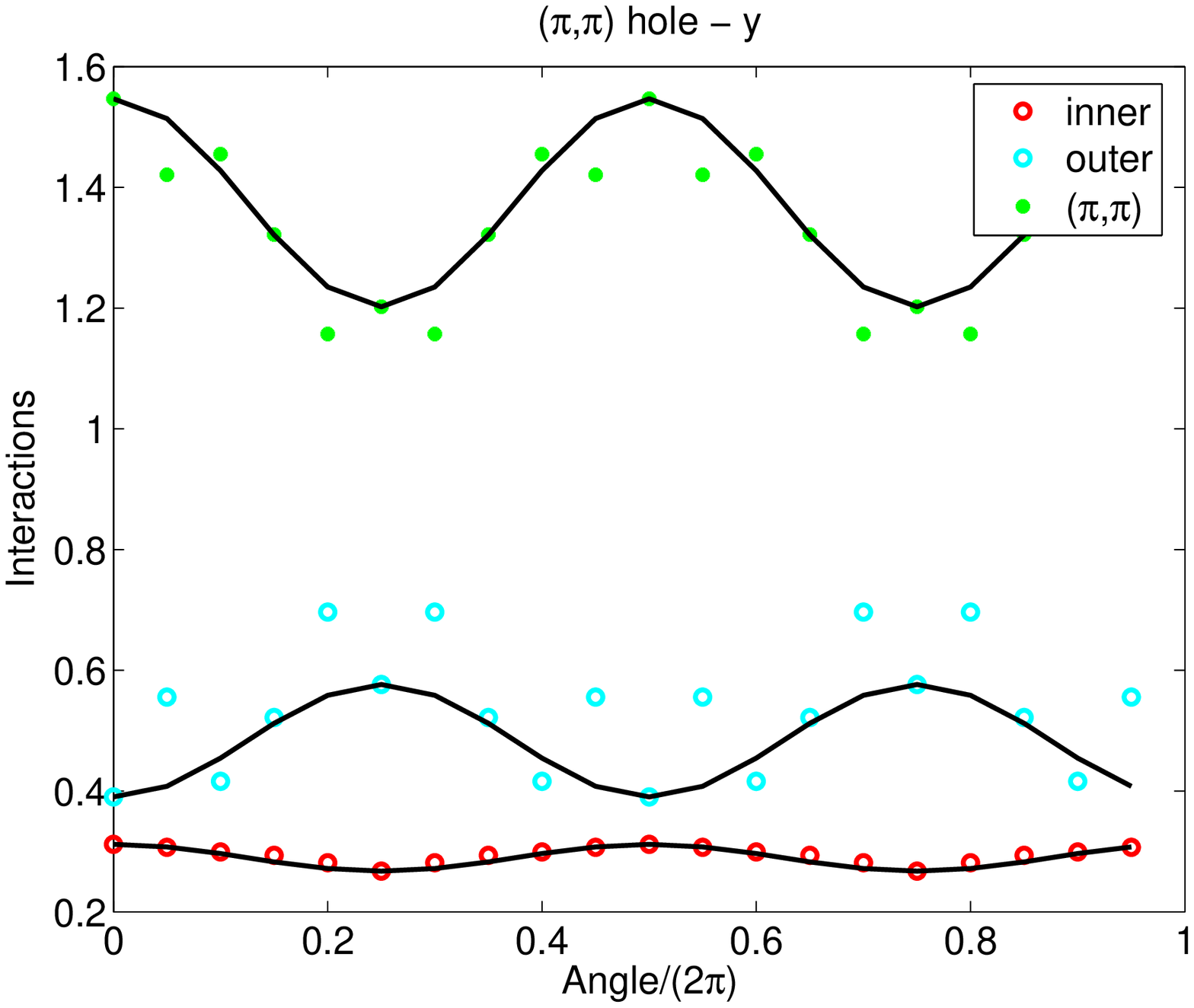}\\
\includegraphics[width=2in]{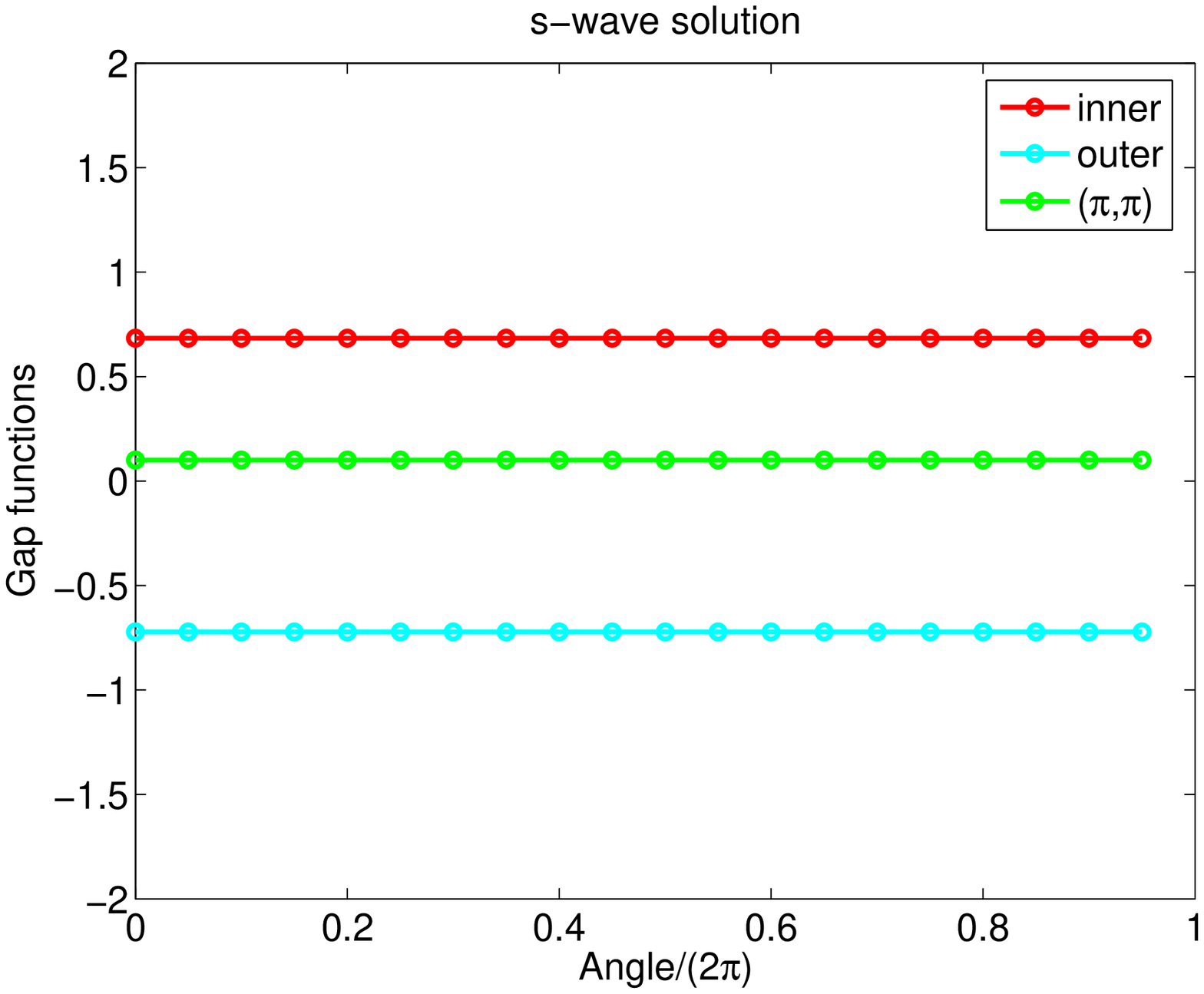}&
\includegraphics[width=2in]{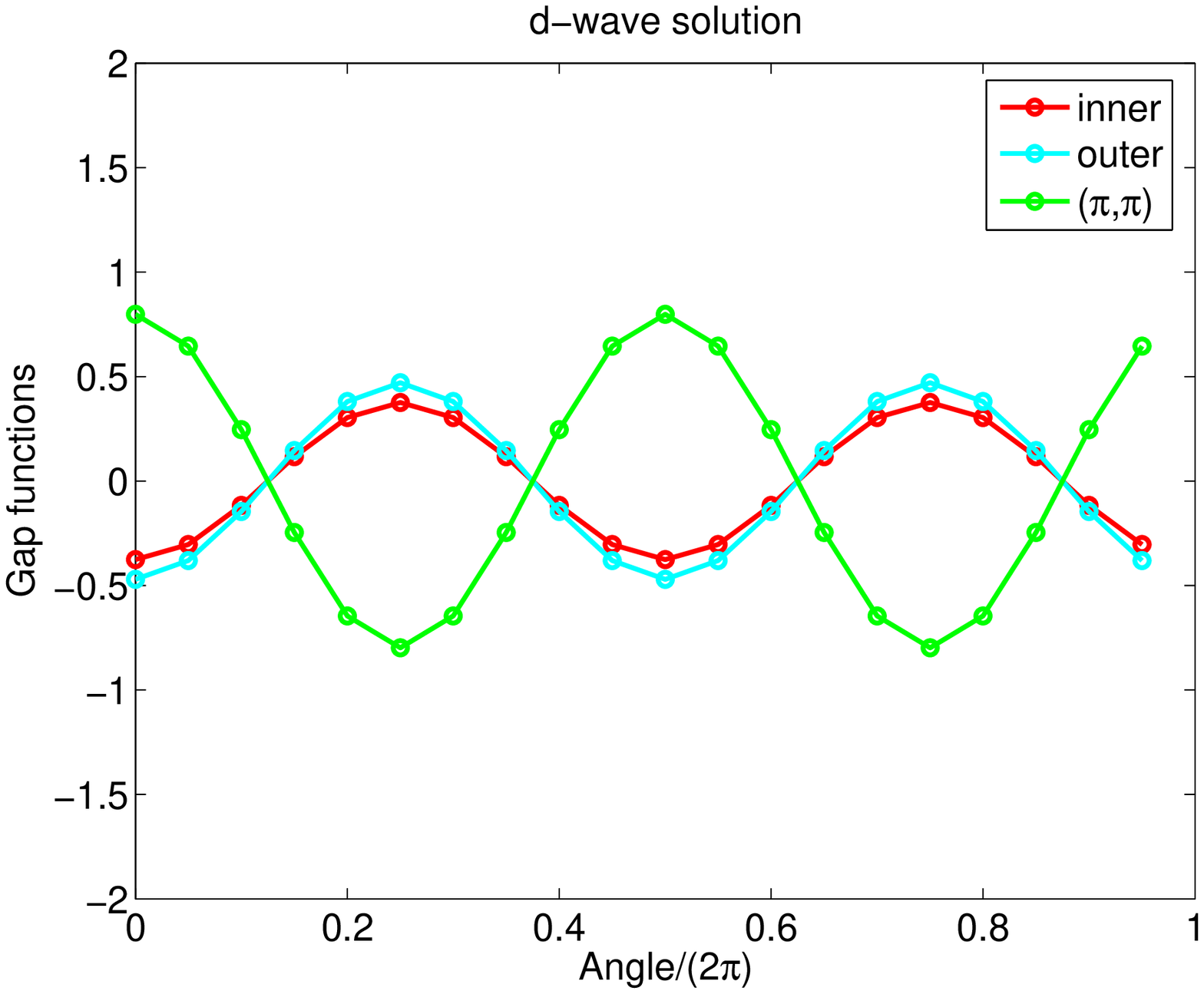}
\end{array}$
\caption{\label{fig:SF_m0p30} The fits of the interactions by LAHA
and the structure of $s$-wave and $d$-wave gaps for the case of
heavy hole doping, $\mu=-0.30$, when only hole FSs are present.
The parameters are $U=0.75, J=0.15, V=0.46$ (see text). The
eigenvalues are $\lambda_s=0.13$ and $\lambda_d=0.11$.}
\end{figure*}

The results for the fit of  the RPA/SF interactions to LAHA are
shown in Fig.~\ref{fig:SF_m0p30} and the interaction parameters
extracted from the fit are presented in Table~\ref{tab:KBaFeAs_s
and d}. We see that both $s$-and $d$-wave solutions are attractive
and both are different from $s-$wave and $d-$wave solutions at
smaller $|\mu|$, when  hole and electron FSs were present. In
particular, we see from Fig.~\ref{fig:SF_m0p30} that $s$-wave gap
now changes sign between the two hole FSs centered at $(0,0)$,
while in the $d$-wave solution the gaps on these two FSs are ``in
phase''. We remind the reader that at smaller $|\mu|$ the $s$-wave
gaps on the two hole FSs are of the same sign, while $d$-wave gaps
on these FSs have a phase shift of $\pi$ (see
Figs.~\ref{fig:SF_m0p18} and~\ref{fig:SF_m0p05}).

Consider first the $s-$wave channel. A simple analysis of the $3
\times 3$ equation shows that, in the absence of electron pockets,
the reason for the $s$-wave attraction is a strong $h_1-h_2$
interaction between the two pockets at $(0,0)$, which exceeds the
intra-pocket interaction for any of these two pockets (see
Table~\ref{tab:KBaFeAs_s and d}). Intra and inter-pocket
interactions are positive, and the solution with a positive
$\lambda$ obviously corresponds to a sign-changing gap between the
two hole pockets, in full analogy with the sign-changing solution
in a model with strong inter-pocket interaction between hole and
electron pockets. The presence of the third hole FS at $(\pi,\pi)$
is not a factor in this consideration because the intra-pocket
$s$-wave repulsion for this FS is quite strong, and the gap at
$(\pi,\pi)$ is only induced by  much weaker interactions with
$(0,0)$ pockets.

Consider next  the  $d$-wave channel. A straightforward analysis
of the $3\times 3$ $d$-wave gap equation shows that the solution
with $\lambda_d >0$ exists for two reasons, both  specific to the
case of no electron pockets.   First, we see from
Table~\ref{tab:KBaFeAs_s and d} that the $d$-wave intra-pocket
interaction within the $(\pi,\pi)$ pocket, ${\tilde u}_{h_3,h_3}$,
is now negative (i.e., attractive). Second,  intra-pocket
interaction  ${\tilde u}_{h_1,h_2}$ between the two pockets at
$(0,0)$ is  negative and larger in magnitude than repulsive
${\tilde u}_{h_1,h_1}$ and  ${\tilde u}_{h_2,h_2}$.  In
consequence, if we momentarily decouple the pocket at $(\pi,\pi)$
and the two pockets at $(0,0)$, we obtain two solutions with
positive $\lambda_d$. One corresponds to a gap only on $(\pi,\pi)$
pocket, another to in-phase gaps on the two pockets at $(0,0)$.
The third solution is the one in which there is the $\pi$ phase
shift between the two gaps at $(0,0)$. This solution has negative
$\lambda_d$ and is irrelevant. The residual, much weaker
interaction between the pockets at $(0,0)$ and $(\pi,\pi)$ couples
the two solutions with positive $\lambda_d$ and sets the phase
shift between the gaps at $(\pi,\pi)$ and $(0,0)$. In each of
these two coupled solutions, there is no phase shift between the
gaps at $(0,0)$, and the gap at $(\pi,\pi)$ is larger than the
gaps at $(0,0)$ simply because  ${\tilde u}_{h_3,h_3}$, is
attractive while  ${\tilde u}_{h_1,h_1}$ and  ${\tilde
u}_{h_2,h_2}$ are repulsive.  These features are not present in
the $d$-wave solution with the largest $\lambda_d$ for smaller
$|\mu|$, when both hole and electron FSs are present. For those
cases,  ${\tilde u}_{h_3,h_3}$, is repulsive, and there is a
$\pi$ phase shift between the two gaps at $(0,0)$ because
${\tilde u}_{h_1,h_2}$ is dominated by the interactions with
electron pockets ${\tilde u}_{h_1e}$ and   ${\tilde u}_{h_2e}$
which, by symmetry, are of different signs.  We see therefore that
the two $d$-wave solutions of the linearized gap equation which
give the two largest $\lambda_d$ are not the same as the solution
with the largest $\lambda_d$ at smaller $|\mu|$.

The larger value of $\Delta_{h_3}$ compared to $\Delta_{h_1}$ and
$\Delta_{h_2}$ and the in-phase structure of the gaps at $(0,0)$
are consistent with  the fRG  $d$-wave solution by Thomale
\textit{et al.}~\cite{fRG_KBaFeAs}. There is only one relatively
minor disagreement: for
 our parameters the solution with the
largest $\lambda_d$ is the one for which $(\pi,\pi)$ gap and
$(0,0)$ gaps have relative phase shift $\pi$, i.e., are  ``of
opposite sign'' (see Table~\ref{tab:Xtreme_hole(SF)}). Thomale
\textit{et al.} found the solution with the ``equal sign'' of all
three $d$-wave gap.  We verified, however, that the selection of
the phase between $(\pi,\pi)$ and $(0,0)$ gaps is sensitive to the
interplay between ${\tilde u}_{h_1,h_3}$ and ${\tilde
u}_{h_2,h_3}$, which are small in magnitude and have different
signs (see Table~\ref{tab:KBaFeAs_s and d}). Already a small
modification of these parameters makes $\lambda_d$ larger for the
solution with the same phase for the gaps on all three FSs, the
same as in fRG solution.

As about the comparative strength of s-wave and d-wave pairing
components, we found for the particular $\mu =-0.30$ that we
considered, that  $\lambda_s$ and $\lambda_d$ are comparable:
$\lambda_s =0.13$, $\lambda_d =0.11$. However,  the rate with
which $\lambda_d$ increases with the hole doping well exceeds that
for $\lambda_s$, and at larger dopings $d-$wave channel almost
certainly becomes the  most attractive one. This is also
consistent with the fRG analysis by Thomale \textit{et al.}  who
found that d-wave coupling becomes larger than s-wave coupling at
large enough hole dopings.

The pairing in heavily hole doped FeSCs was recently studied
within RPA for 5-band orbital model by Suzuki et al~\cite{suzuki}.
They found that the pairing is driven by incommensurate spin
fluctuations and s-wave and d-wave pairing amplitudes are of about
the same strength. This fully agrees with our analysis. There is
one difference, however -- Suzuki {\it et al} attributed
attraction in the s-wave channel to the still strong interaction
between hole states and gapped electron states near $(0,\pi)$ and
$(\pi,0)$, while in our case the attraction in s-wave channel is
due to strong interaction between inner and outer hole pockets
centered at $(0,0)$ (see Table \ref{tab:KBaFeAs_s and d}). The
authors of Ref.~\cite{suzuki} cited recent observation of
incommensurate spin fluctuations in KFe$_2$As$_2$ (Ref.
\cite{chlee}) as evidence for still strong interactions between
fermions from near $\Gamma$ and $(\pi,\pi)$ points and from near
$(0,\pi)$ and $(\pi,0)$. This is certainly a possibility, but we
point out that the interaction within the hole pocket centered at
$(\pi,\pi)$ also gives rise to incommensurate spin fluctuations at
rather large momenta, because of a large size of that pocket. We
recall that in  our theory, the magnetically-enhanced interaction
within the $(\pi,\pi)$ hole pocket is the driving force for the
d-wave pairing.

 \subsection{strong electron vs strong hole doping}

We see that in our theory  there is an attraction in $d$-wave channel
at both strong hole doping and strong
electron doping (at least, for the model and parameters which we considered). The two
limits are, however, quite different from a physics perspective.
 For the case of strong electron doping, the enhancement of the
spin susceptibility around $(\pi,\pi)$ unambiguously leads to a
attraction in the $d$-wave channel, i.e. to $\lambda_d >0$. For
strong hole doping, the susceptibility is peaked at $(0,0)$, which
affects ${\tilde u}_{h_1,h_1}$, ${\tilde u}_{h_1,h_2}$, ${\tilde
u}_{h_2,h_2}$, and ${\tilde u}_{h_3,h_3}$. The attractive $d$-wave
solution is the result of negative ${\tilde u}_{h_3,h_3}$ and a
larger value of ${\tilde u}^2_{h_1,h_3}$ compared to ${\tilde
u}_{h_1,h_1}$${\tilde u}_{h_2,h_2}$.  There is no fundamental
reason why it should be so except that for large non-circular
Fermi surfaces in 2D, the particle-hole susceptibility $\chi (q)$
is larger at $2k_F$ than at $q=0$.  and  the appearance of the
solution with a positive $\lambda_d$ at strong hole doping is very
likely accidental.  In other words, similar materials without
electron pockets could easily have $s$-wave or a different
$d$-wave pairing state.

\begin{table}[htp]
\caption{$s$- and $d$-wave parameters for the case of
strong hole doping (with SF component), when there are no electron FSs.}
\label{tab:KBaFeAs_s and d}
\begin{ruledtabular}
\begin{tabular}{cccccccc}
$s$-wave&$u_{h_1h_1}$&$u_{h_2h_2}$&$u_{h_3h_3}$&$u_{h_1h_2}$&$u_{h_1h_3}$&$u_{h_2h_3}$&$\lambda_s$\\
&0.67&0.69&1.37&0.80&0.29&0.48&0.13\\
\hline
$d$-wave&$\tilde{u}_{h_1h_1}$&$\tilde{u}_{h_2h_2}$&$\tilde{u}_{h_3h_3}$&$\tilde{u}_{h_1h_2}$&$\tilde{u}_{h_1h_3}$&$\tilde{u}_{h_2h_3}$&$\lambda_d$\\
&0.36&0.34&-0.17&-0.50&-0.02&0.09&0.11
\end{tabular}
\end{ruledtabular}
\end{table}

\begin{table}[htp]
\caption{The structure of $d$-wave gaps $\Delta_{h_i} (\phi_i) = \Delta_{h_i} \cos 2 \phi_i$ for $\mu=-0.30$ obtained by solving $3\times 3$ linearized gap equation (the gaps are in arbitrary units since only the ratios of the gaps matter).}
\label{tab:Xtreme_hole(SF)}
\begin{ruledtabular}
\begin{tabular}{cccc}
&sol. 1& sol. 2&sol. 3 \\
\hline
$\Delta_{h_1}$&0.38&0.60&0.71\\
$\Delta_{h_2}$&0.47&0.54&-0.70\\
$\Delta_{h_3}$&-0.80&0.60&-0.08\\
$\lambda$ & \textbf{0.11} &-0.06&-0.43
\end{tabular}
\end{ruledtabular}
\end{table}

\section{Conclusions \label{sec:Conclusions}}

In this paper we analyzed  the pairing symmetry and the structure
of the gap in FeSC by approximating the pairing interaction
between  low-energy fermions by leading angular harmonics. This
allowed us to decompose the  pairing interaction and study
separately contributions to pairing from scattering processes
between different FSs and the interplay between angle-independent
and angle-dependent parts of each interaction. The angular
dependence of the interactions are peculiar to FeSC because of the
multiorbital nature of low-energy excitations. [The interactions
in band representation are obtained by dressing up interactions in
orbital representation by angle-dependent coherence factors
associated with the hybridization of Fe $d$-orbitals]. We used the
band interaction obtained within RPA/SF formalism as an input,
fitted it within LAHA, verified that the fit is quite good for all
cases that we studied, and analyzed in detail how the pairing
interactions in $s$-wave and $d$-wave channels evolve with the
bare interaction and the one with extra SF component and between
hole and electron doping. We also analyzed the interplay between
$s$-wave and $d$-wave pairing.  Using the same procedure, we also
studied the pairing at large electron (hole) doping, when only
electron (hole) FSs are present. Throughout this paper  we treated
FeSC as quasi-2D systems and didn't address  potential new physics
associated with 3D effects.

The main conclusion of our study is that all pairing states
obtained so far at different dopings in FeSCs can be understood
within the same universal pairing scenario based on
spin-fluctuation exchange. We furthermore found that all these
pairing states appear naturally in the effective low-energy model
with small number of input parameters. We conjecture that the
approaches based on RPA (both analytical and  functional) and on
itinerant $J_1-J_2$ model reduce to this model at low energies,
however with different input parameters.

We used this effective model to study the doping evolution of the
pairing in hole and electron-doped FeSCs. We argue that the
pairing mechanisms at small/moderate  and large dopings are
qualitatively different -- when both hole and electron pockets are
present, the pairing is of Kohn-Luttinger type, driven by the
pair-hopping of fermions from hole to electron pockets, while at
larger hole or electron doping, the pairing is due to a direct
interaction between only hole or only electron pockets.  For
moderate hole dopings the leading pairing instability is towards
an $s\pm$ state with a nodeless gap. For moderate electron doping
a nodal $s\pm$ state is the leading instability, but $d$-wave is a
close competitor. For larger electron or hole dopings, when only
one type of FS is present, the leading pairing instability is
towards a d-wave state, which in case of strong electron doping is
nodeless, at least in 2D case.

We summarize below the detailed reasoning behind the observation
stated above by presenting the answers to the questions we posed
in the introduction-

\begin{itemize}

\item
{\em What is the origin of the strong angular dependence of the $s^\pm$
gap along the electron FSs?}

We found that the origin is different for bare and full
interactions. For bare interactions (no SF component), the
combination of intra- and inter-electron pocket repulsions are
stronger than the electron-hole interaction and the $s^\pm$
attractive solution for the gap is entirely due to the
angle-dependent parts of the electron-hole (electron-electron for
strong electron doped materials) interaction, much as was
anticipated in Refs. \onlinecite{tom_09} and \onlinecite{cvv_09}.
Namely, the system adjusts the magnitude of the angle-dependent,
$\pm \cos 2 \theta$ gap component along  the two electron FSs to
minimize the effect of the inter-electron-pocket repulsion.

For the full interaction, the electron-hole interaction is the
strongest, and the attractive $s^\pm$ solution exists even if all
interactions are taken to be angle-independent.  The
angle-dependent terms modify the $s^\pm$ gap by creating $\pm \cos
2 \theta$ gap components. Whether these components are large
enough to lead to nodes depends on details, but the generic trend
is that when the angle-independent part of the electron-hole
interaction is larger, the gap is less likely to have nodes.

\item
{\em Are the angular dependencies of all interactions relevant for the gap
structure, or can some interactions be safely approximated as
angle-independent?}

We found that the angle-dependent part of the electron-hole interaction is
the relevant one.  The angle-dependent parts of the electron-electron
interaction have little effect on the gap structure, at least for
the full interaction with the SF component.

\item
{\em Why do the $s^{\pm}$ solutions obtained within the RPA/SF and
fRG approaches have nodes for systems with two hole and two
electron FSs and no nodes for systems with three hole and two
electron FSs?}

We found that the angle-independent electron-hole interaction,
which favors a no-nodal $s^\pm$ gap, is further increased if the
third hole FS is present. For most of the parameter sets which we
analyzed, the $s^\pm$ gap has nodes in case of electron doping
(four FSs), but no-nodal solution is stabilized for hole doping
(five FSs). Kuroki \emph{et al}\cite{kuroki} have pointed out that
this can be traced back to the d$_{xy}$ orbital character of the
third hole pocket which interacts strongly with the d$_{xy}$
states at the tips of the electron pockets. We found, however,
that for some parameters nodal solutions survive in the presence
of the fifth FS, i.e., the disappearing of the nodes with the
appearance of the fifth FS is not a universal result. That aside,
the gap structure still evolves between nodal and no-nodal once we
change the magnitudes of angle-dependent parts of the
interactions.

\item
{\em What causes the pairing when only electron FSs are present?}

We found that the $d$-wave pairing is generally attractive and
competes with $s^\pm$ pairing for the electron-doped FeSC. At small
electron doping, the $d$-wave attraction is almost entirely due to
the $d$-wave electron-hole interaction, and the direct $d$-wave
interaction between electron pockets is weak.

For strong electron doping, when only electron FSs are present,
the situation is different. We found an {\it attractive} $d$-wave
interaction between electron pockets. The $d$-wave pairing is then
quite similar to the one in the magnetic  hot-spot pairing
scenario for the cuprates. In both cases, there is a $d$-wave
attraction between the FS sheets separated by $(\pi,\pi)$.

With regard to the subleading $s$-wave attraction, in our case it is due
to an angle-dependent $s$-wave component of the electron-electron interaction. We didn't consider the interaction via gapped hole states (another potential reason for $s$-wave attraction), but the similarity between our gap structure and the one obtained in the fRG study~\cite{dhl_AFESE}, which includes both interactions, indicates that the likely origin of the $s$-wave attraction is the angle-dependence of electron-electron interaction.

\item
{\em What causes the pairing when only hole FSs are present?}

We found that both $s$-wave and $d$-wave channels are attractive,
with comparable $\lambda_{s,d}>0$. Which pairing instability is
stronger depends on detail.

The reason for the $s$-wave instability is a strong repulsive
inter-pocket ($h_1-h_2$) interaction between the two pockets at
$(0,0)$, which exceeds the intra-pocket repulsion. This leads to a
sign-changing $s$-wave gap between the two hole pockets, in full
analogy with the sign-changing solution in a model with strong
inter-pocket interaction between hole and electron pockets.

The reason for the competing  $d$-wave instability is two-fold.
First, the $d$-wave component of the intra-pocket interaction
within the hole pocket at $(\pi,\pi)$ is negative (i.e.,
attractive), second, there is strong attractive $d$-wave
interaction between the two hole pockets at $(0,0)$. The
combination of these two reasons leads to a $d_{x^2-y^2}$ solution
with positive $\lambda_d$, in which the magnitude of the gap is
the largest on $(\pi,\pi)$ pocket, and the two gaps at $(0,0)$
have zero phase shift.

Both $s$-wave and $d$-wave solutions are different from the ones
at smaller hole dopings, when hole and electron FSs are present
(e.g., the $d$-wave solution with the largest $\lambda_d$ at
smaller $|\mu|$ is the one with the $\pi$ phase shift between the
two gaps at $(0,0)$.

\item
{\em How is the structure of the pairing interaction affected when the spin-fluctuation component is added to the direct fermion-fermion
interaction?}

We found that the SF interaction primarily changes the overall
magnitude of the interaction, while its angular dependence remains
nearly unchanged.  All components of the  pairing interaction
increase when the SF term is added. On top of this, there is an
additional increase of the hole-electron inter-pocket interaction,
both in the $s$-wave and in the $d$-wave channels. This additional
increase makes both $s$-wave and $d$-wave solutions attractive.
\end{itemize}

We have only studied the strictly 2D case thus far, and neglected
aspects of the 3D I4/mmm crystal symmetry characteristic of 122
materials and  the hybridization of electron pockets in the folded
zone. We nevertheless believe that the general evolution of
interactions and gap symmetry discussed here will be generic to
the FeSCs. The approach developed here can be easily modified to
study  superconductivity on hybridized FSs and
 can also be used to study in great detail
 SDW
instability in multi-orbital systems~\cite{ashvin}.

\section*{Acknowledgements}

We acknowledge helpful discussions with L. Benfatto, R. Fernandes,
W. Hanke, I. Eremin,
H. Kontani, K. Kuroki,
 Y. Matsuda, I. Mazin, R. Prozorov, D.
Scalapino, J. Schmalian, Z. Tesanovic, R. Thomale, M. Vavilov, and
A. Vorontsov. This work was supported by NSF-DMR-0906953 (S. M and
A.V.C). Partial support from MPI PKS (Dresden) (S.M. and A.V.C),
and Humboldt foundation (A.V.C) is gratefully acknowledged. T.A.M.
acknowledges support from the Center of Nanophase Materials
Sciences, which is sponsored at Oak Ridge National Laboratory by
the Office of Basic Energy Sciences, U.S. Department of Energy.
M.M.K and P.J.H. acknowledge support from DOE DE-FG02-05ER46236.
M.M.K. is grateful for support from RFBR (grant 09-02-00127),
Presidium of RAS program N5.7, Russian FCP (GK P891), and
President of Russia (grant MK-1683.2010.2).

\end{document}